\documentclass{lmcs} 

\keywords{Generalized automata, Myhill-Nerode theorem, minimal automaton, Burrows-Wheeler Transform, FM-index.}

\theoremstyle{plain} 

\usepackage{tikz}
\usetikzlibrary{arrows,automata,positioning}

\usepackage{amsfonts}
\usepackage{graphicx}
\usepackage{subcaption}
\usepackage{amssymb} 
\usepackage{amsmath}
\usepackage[normalem]{ulem}
\usepackage{lineno}

\usepackage{tabularx}

\usepackage{hyperref}
\hypersetup{hypertexnames=false}

\DeclareMathOperator{\Pref}{Pref}

\DeclareMathOperator{\BWT}{BWT}

\begin{document}


\title[A Myhill-Nerode Theorem for Generalized Automata, with Applications]{A Myhill-Nerode Theorem for Generalized Automata, with Applications to Pattern Matching and Compression}
\titlecomment{{\lsuper*}A preliminary version~\cite{cotumaccio2024stacs} of this article  appeared in the proceedings of the 41st Symposium on Theoretical Aspects of Computer Science (STACS 2024)}
\thanks{}	

\author[N.~Cotumaccio]{Nicola Cotumaccio\lmcsorcid{0000-0002-1402-5298}}

\address{Department of Computer Science, University of Helsinki, Pietari Kalmin katu 5, P.O. Box 68, Helsinki, 00014, Finland}	
\email{nicola.cotumaccio@helsinki.fi}  





\begin{abstract}
  \noindent The model of generalized automata, introduced by Eilenberg in 1974, allows representing a regular language more concisely than conventional automata by allowing edges to be labeled not only with characters, but also strings. Giammarresi and Montalbano introduced a notion of determinism for generalized automata [STACS 1995]. While generalized deterministic automata retain many properties of conventional deterministic automata, the uniqueness of a minimal generalized deterministic automaton is lost.

In the first part of the paper, we show that the lack of uniqueness can be explained by introducing a set $ \mathcal{W(A)} $ associated with a generalized automaton $ \mathcal{A} $. If $ \mathcal{A} $ is a conventional automaton, the set $ \mathcal{W(A)} $ is always trivially equal to the set of all prefixes of the language recognized by the automaton, but this need not be true for generalized automata. By fixing $ \mathcal{W(A)} $, we can derive for the first time a full Myhill-Nerode theorem for generalized automata, which contains the textbook Myhill-Nerode theorem for conventional automata as a degenerate case.

In the second part of the paper, we show that the set $ \mathcal{W(A)} $ leads to applications for pattern matching and data compression. Wheeler automata [TCS 2017, SODA 2020] are a popular class of automata that can be compactly stored using $ e \log \sigma (1 + o(1)) + O(e) $ bits ($ e $ being the number of edges, $ \sigma $ being the size of the alphabet) in such a way that pattern matching queries can be solved in $ O(m \log \log \sigma) $ time  ($ m $ being the length of the pattern). In the paper, we show how to extend these results to generalized automata. More precisely, a Wheeler generalized automaton can be stored using $ \mathfrak{e} \log \sigma (1 + o(1)) + O(e) $ bits so that pattern matching queries can be solved in $ O(m \log \log \sigma) $ time, where $ \mathfrak{e} $ is the total length of all edge labels.
\end{abstract}

\maketitle


\section{Introduction}\label{sec:introduction}

The class of regular languages can be defined starting from \emph{non-deterministic finite automata (NFAs)}. In his monumental work~\cite{eilenberg1974} on automata theory (which dates back to 1974), Eilenberg proposed a natural generalization of NFAs where edges can be labeled not only with characters but with (possibly empty) finite strings, the so-called \emph{generalized non-deterministic finite automata (GNFAs)}. While classical automata are only a special case of generalized automata, it is immediate to realize that generalized automata can only recognize regular languages, because it is well-known that epsilon transitions do not add expressive power~\cite{hopcroft2006}, and a string-labeled edge can be decomposed into a path of edges labeled only with characters. However, generalized automata can represent regular languages more concisely than classical automata. A standard measure of the complexity of a regular language is the minimum number of states of some automaton recognizing the language, and generalized automata may have fewer states than conventional automata. In generalized automata, we assume that both the number of states and the number of edges are finite, but the number of edges cannot be bounded by some function of the number of states and the size of the finite alphabet (and so edge labels may be arbitrarily long). As a consequence, in principle it is not clear whether the problem of determining the minimum number of states of some generalized automaton recognizing a given language is decidable. In~\cite{hashiguchi1991}, Hashiguchi showed that the problem is decidable by proving that there must exist a state-minimal generalized automaton for which the lengths of edge labels can be bounded by a function that only depends on the size of the syntactic monoid recognizing the language.

An NFA is a \emph{deterministic finite automaton (DFA)} if no state has two distinct outgoing edges with the same label. This local notion of determinism extends to global determinism, that is, given a string $ \alpha $, there exists at most one path labeled $ \alpha $ that can be followed starting from the initial state. However, this is not true for generalized automata (see Figure~\ref{fig:determinismhowtogeneralize}). When considering generalized automata, we must add the additional requirement that no state has two distinct outgoing edges such that one edge label is a prefix of the other edge label. By adding this requirement, we retrieve global determinism, thus obtaining \emph{generalized deterministic finite automata (GDFAs)}.

\begin{figure}[h!]
\centering
\scalebox{0.8}{
\centering
\begin{tikzpicture}[->,>=stealth', semithick, auto, scale=1]
\node[state, initial] (1)    at (0,0)	{$ 1 $};
\node[state] (2)    at (4,0)	{$ 2 $};
\node[state] (3)    at (0,-2)	{$ 3 $};
\node[state, accepting] (4)    at (4,-2)	{$ 4 $};

\draw (1) edge [] node [] {$ ab $} (2);
\draw (2) edge [] node [] {$ c $} (4);
\draw (1) edge [] node [] {$ a $} (3);
\draw (3) edge [] node [] {$ bc $} (4);

\end{tikzpicture}
}
 	\caption{No state has two distinct outgoing edges with the same label, but there are two distinct paths labeled $ abc $ from the initial state.}
    \label{fig:determinismhowtogeneralize}
\end{figure}

For every regular language, up to isomorphism there exists a \emph{unique} deterministic automaton recognizing the language and having the minimum number of states among all deterministic automata recognizing the language, the \emph{minimal DFA} of the language. More generally, a classical textbook result in automata theory is the \emph{Myhill-Nerode theorem}~\cite{myhill1957finite, nerode1958linear}. Let $ \Pref(\mathcal{L}) $ be the set of all strings prefixing at least one string in the language $ \mathcal{L} $. Recall that an equivalence relation $ \sim $ on $ \Pref (\mathcal{L}) $ (i) is \emph{right-invariant} if for every $ \alpha, \beta \in \Pref (\mathcal{L}) $ and for every $ \phi \in \Sigma^* $ we have $ \alpha \phi \in \Pref (\mathcal{L}) \Longleftrightarrow \beta \phi \in \Pref (\mathcal{L}) $ and $ \alpha \phi \in \Pref (\mathcal{L}) \Longrightarrow \alpha \phi \sim \beta \phi $, and (ii) has \emph{finite index} if the number of equivalence classes is finite. We have the following result\footnote{The classical formulation of the Myhill-Nerode theorem is slightly different from the statement of Theorem~\ref{theor:MyhillNerodetextbook} (see Remark~\ref{rem:standardmyhill}).}.

\begin{thm}[Myhill-Nerode theorem]\label{theor:MyhillNerodetextbook}
Let $ \mathcal{L} \subseteq \Sigma^* $. The following are equivalent:
\begin{enumerate}
    \item $ \mathcal{L} $ is recognized by an NFA.
    \item The Myhill-Nerode equivalence $ \equiv_{\mathcal{L}} $ has finite index.
    \item There exists a right-invariant equivalence relation $ \sim $ on $ \Pref (\mathcal{L}) $ of finite index such that $ \mathcal{L} $ is the union of some $ \sim $-classes.
    \item $ \mathcal{L} $ is recognized by a DFA.

\end{enumerate}
Moreover, if one of the above statements is true (and so all the above statements are true), then there exists a unique minimal DFA recognizing $ \mathcal{L} $ (that is, two DFAs recognizing $ \mathcal{L} $ having the minimum number of states among all DFAs recognizing $ \mathcal{L} $ must be isomorphic).
\end{thm}

The problem of studying the notion of determinism in the setting of generalized automata was approached by Giammarresi and Montalbano~\cite{giammaresi1999journal, giammaresi1995conference}. The notion of isomorphism can be naturally extended to GDFAs (intuitively, two GDFAs are isomorphic if they are the same GDFA up to renaming the states), and the natural question is whether every regular language admits a unique minimal GDFA up to isomorphism. This is not true in general: there can exist two or more non-isomorphic state-minimal GDFAs recognizing a given regular language. Consider the two distinct GDFAs in Figure~\ref{fig:nouniqueminimalGDFA}. It is immediate to check that the two (non-isomorphic) GDFAs recognize the same language, and it can be shown that no GDFA with fewer than three states can recognize the same language~\cite{giammaresi1999journal}.

\begin{figure}[h!]
     \centering
     \begin{subfigure}[b]{0.49\textwidth}
        \centering
        \scalebox{.8}{
\begin{tikzpicture}[->,>=stealth', semithick, auto, scale=1]
\node[state, initial] (1)    at (0,0)	{$ 1 $};
\node[state] (2)    at (2,0)	{$ 2 $};
\node[state, accepting] (3)    at (4, 0)	{$ 3 $};

\draw (1) edge [] node [] {$ a^3, ba^2 $} (2);
\draw (2) edge [loop below] node [] {$ ba^2, aba $} (2);
\draw (2) edge [] node [] {$ a^2 $} (3);
\end{tikzpicture}
}
\end{subfigure}
     \begin{subfigure}[b]{0.49\textwidth}
        \centering
               \scalebox{.8}{
\begin{tikzpicture}[->,>=stealth', semithick, auto, scale=1]
\node[state, initial] (1)    at (0,0)	{$ 1 $};
\node[state] (2)    at (2,0)	{$ 2 $};
\node[state, accepting] (3)    at (4, 0)	{$ 3 $};

\draw (1) edge [] node [] {$ a^2, ba $} (2);
\draw (2) edge [loop below] node [] {$ aba, a^2b $} (2);
\draw (2) edge [] node [] {$ a^3 $} (3);
\end{tikzpicture}
}
\end{subfigure}
 	\caption{Two state-minimal GDFAs recognizing the same regular language.}
    \label{fig:nouniqueminimalGDFA}
\end{figure}

The non-uniqueness of a state-minimal GDFAs seems to imply a major difference in the behavior of generalized automata compared to conventional automata, so it looks like there is no hope to derive a structural result like the Myhill-Nerode theorem in the model of generalized automata. It is natural to wonder whether the lack of uniqueness should be interpreted as a weakness of the model of generalized automata, or rather as a consequence of some deeper property. Consider a conventional automaton $ \mathcal{A} $ that recognizes a language $ \mathcal{L} $. As is typical in automata theory, we can assume that all states are reachable from the initial state, and all states are either final or allow reaching a final state. Then, the set $ \mathcal{W(A)} $ of all strings that can be read starting from the initial state and reaching some state is exactly equal to $ \Pref (\mathcal{L}) $. However, this is no longer true in the model of generalized automata: typically, we do not have $ \mathcal{W(A)} = \Pref (\mathcal{L}) $, but only $ \mathcal{W(A)} \subseteq \Pref (\mathcal{L}) $. For example, consider Figure~\ref{fig:nouniqueminimalGDFA}. In both automata we have $ a^3 \in \Pref (\mathcal{L}) $, but we have $ a^3 \in \mathcal{W(A)} $ only for the GDFA on the left.

Given $ \mathcal{W} \subseteq \Pref (\mathcal{L}) $, we say that a GNFA $ \mathcal{A} $ recognizing $ \mathcal{L} $ is a $ \mathcal{W} $-GNFA if $ \mathcal{W(A)} = \mathcal{W} $. We will show that, if $ \mathcal{L} $ is recognized by a $ \mathcal{W} $-GDFA, then there exists a \emph{unique} state-minimal $ \mathcal{W} $-GDFA recognizing $ \mathcal{L} $. In particular, our result will imply the uniqueness of the minimal automaton for standard DFAs, because for DFAs it must necessarily be $ \mathcal{W} = \Pref (\mathcal{L}) $.

We will actually prove much more. We will show that, once we fix $ \mathcal{W} $, then nondeterminism and determinism still have the same expressive power (thus extending the corresponding result for conventional automata, first proved by Rabin and Scott through the \emph{powerset construction}~\cite{rabin1959finite}), and it is possible to derive a characterization in terms of equivalence relations. In other words, we will prove a \emph{Myhill-Nerode theorem for generalized automata}. To this end, we will introduce the notion of \emph{locally bounded set} (Section~\ref{sec:preliminary}), which we can use to prove the following result.

\begin{thm}[Myhill-Nerode theorem for generalized automata]\label{theor:ng}
Let $ \mathcal{L} \subseteq \Sigma^* $ and let $ \mathcal{W} \subseteq \Sigma^* $ be a locally bounded set such that $ \mathcal{L} \cup \{\epsilon \} \subseteq \mathcal{W} \subseteq \Pref (\mathcal{L}) $. The following are equivalent:
\begin{enumerate}
    \item $ \mathcal{L} $ is recognized by a $ \mathcal{W} $-GNFA.
    \item The Myhill-Nerode equivalence $ \equiv_{\mathcal{L}, \mathcal{W}} $ has finite index.
    \item There exists a right-invariant equivalence relation $ \sim $ on $ \mathcal{W} $ of finite index such that $ \mathcal{L} $ is the union of some $ \sim $-classes.
    \item $ \mathcal{L} $ is recognized by a $ \mathcal{W} $-GDFA.    
\end{enumerate}
Moreover, if one of the above statements is true (and so all the above statements are true), then there exists a unique minimal $ \mathcal{W} $-GDFA recognizing $ \mathcal{L} $ (that is, two $ \mathcal{W} $-GDFAs recognizing $ \mathcal{L} $ having the minimum number of states among all $ \mathcal{W} $-GDFAs recognizing $ \mathcal{L} $ must be isomorphic).
\end{thm}

In particular, we will show that there is no loss of generality in assuming that $ \mathcal{W} \subseteq \Sigma^* $ is a locally bounded set such that $ \mathcal{L} \cup \{\epsilon \} \subseteq \mathcal{W} \subseteq \Pref (\mathcal{L}) $, because these are \emph{necessary} conditions for the existence of a $ \mathcal{W} $-GNFA (Lemma~\ref{lem:prooflocallybounded}). We conclude that our Myhill-Nerode theorem for GNFAs provides the first structural result for the model of generalized automata.

In the second part of the paper, we show that the set $ \mathcal{W(A)} $ sheds new light on the \emph{String Matching in Labeled Graphs (SMLG)} problem. The SMLG problem has a fascinating history that dates back more than thirty years. Loosely speaking, the SMLG problem can be defined as follows: given a directed graph whose nodes or edges are labeled with nonempty strings and given a pattern string, decide whether the pattern can be read by following a path on the graph and concatenating the labels. The SMLG problem is a natural generalization of the classical pattern matching problem on texts (which requires determining whether a pattern occurs in a text) because texts can be seen as graphs consisting of a single path. The pattern matching problem on text can be efficiently solved in $ O(n + m) $ time ($ n $ being the length of the text, $ m $ being the length of the pattern) by using the Knuth-Morris-Pratt algorithm~\cite{knuth1977}. The SMLG problem is more challenging, and the complexity can be affected by the specific variant of the pattern matching problem under consideration or the class of graphs to which the problem is restricted. For example, in the (approximate) variant where one allows errors in the graph, the problem becomes NP-hard~\cite{amir2000}, so generally errors are only allowed in the pattern. The SMLG problem was studied extensively during the nineties~\cite{manber1992, tasuya1993, park1995, amir2000, rautiainen2017, navarro2000}; Amir et al. showed how to solve the (exact) SMLG problem on arbitrary graphs in $ O(\mathfrak{e} + me) $ time~\cite{amir2000}, where $ e $ is the number of edges in the graph, $ m $ is the length of the pattern, and $ \mathfrak{e} $ is the total length of all labels in the graph. Recently, the SMLG problem has been back in the spotlight. Equi et al.~\cite{equi2023} showed that, \emph{on arbitrary graphs}, for every $ \epsilon > 0 $ the SMLG problem cannot be solved in $ O(me^{1 - \epsilon}) $ or $ O(m^{1 - \epsilon}e) $ time, unless the Orthogonal Vectors hypothesis fails. In applications (especially in bioinformatics) we often need faster algorithms, so
the SMLG problem has been restricted to classes of graphs on which it can be solved more efficiently. For example, \emph{Elastic Founder graphs} can be used to represent multiple sequence alignments (MSA), a central model of biological evolution, and on Elastic Founder graphs the SMLG problem can be solved in linear time under a number of assumptions which only have a limited impact on the generality of the model~\cite{equiefg2023, rizzo2022}.

The pattern matching problem on texts has been revolutionized by the invention of the Burrows-Wheeler Transform (BWT)~\cite{burrows1994} and the FM-index~\cite{ferragina2000, ferraginajacm2005}. The BWT of a string is a permutation of the string characters that allows \emph{compressing} the original string, and the FM-index is a data structure for solving pattern matching queries directly on the BWT-representation of the string. The Burrows-Wheeler Transform and the FM-index have established a new paradigm in bioinformatics, where the huge increase in genomic data requires the development of \emph{space-efficient} algorithms~\cite{simpson2010}. 

Recently, these ideas were extended to NFAs. In particular, \emph{Wheeler NFAs} are a popular class of automata on which the SMLG problem can be solved in linear time, while only storing a \emph{compact} representation of the Wheeler NFA~\cite{gagie2017, alanko2020}. A special case of Wheeler NFAs are de Bruijn graphs~\cite{bowe2012}, which are used to perform Eulerian sequence assembly~\cite{idury1995, pavel2001, bankevich2012}. Wheeler NFAs are also of relevant theoretical interest: for example, the powerset construction applied to a Wheeler NFA leads to a linear blow-up in the number of states of the equivalent DFA, and the equivalent DFA is Wheeler~\cite{alanko2021}; on arbitrary NFAs, the blow-up can be exponential.

The missing step is to determine whether it is possible to generalize the Burrows-Wheeler Transform and the FM-index to generalized automata, so that the resulting data structures can also be applied to Elastic Founder graphs and other classes of graphs where labels can be arbitrary strings. Indeed, in data compression it is common to consider edge-labeled graphs where one compresses unary paths in the graph to save space, and each path is replaced by a single edge labeled with the concatenation of all labels. For example, some common data structures that are stored using this mechanism are Patricia trees, suffix trees and pangenomes~\cite{navarro2016, baaijens2022, mäkinen2023}.

We will mainly focus on GDFAs (the extension to GNFAs is discussed in Appendix~\ref{app:from gdfas to gnfas}). We say that a GDFA is an \emph{$ r $-GDFA} if all edge labels have length at most $ r $ (so a GDFA is a conventional DFA if and only if it is a $ 1 $-GDFA). Let $ m $, $ e $ and $ \mathfrak{e} $ as above and let $ \sigma = |\Sigma| $. In Section~\ref{sec:WheelerGDFAs}, we will extend the notion of Wheelerness to GDFAs. The key ingredient will be the same set $ \mathcal{W(A)} $ that we use in our Myhill-Nerode theorem: we will consider a partial order $ \preceq_\mathcal{A} $ which sorts the set of all states with respect to the \emph{co-lexicographic order} of the strings in $ \mathcal{W(A)} $ (See Definition~\ref{def:WheelerGDFA}). We will then prove the following result.

\begin{thm}[FM-index of Wheeler GDFAs]\label{theor:fmindexintroduction}
    Let $ \mathcal{A} $ be a Wheeler $ r $-GDFA, with $ \sigma \le \mathfrak{e}^{O(1)} $ and $ r = O(1) $. Then, we can encode $ \mathcal{A} $ by using $ \mathfrak{e} \log \sigma (1 + o(1)) + O(e) $ bits so that later on, given a pattern $ \alpha \in \Sigma^* $ of length $ m $, we can solve the SMLG problem on $ \mathcal{A} $ in $ O(m \log \log \sigma) $ time. Within the same time bound, we can also decide whether $ \alpha $ is recognized by $ \mathcal{A} $. 
\end{thm}

In particular, if $ r = 1 $ (that is, if $ \mathcal{A} $ is a conventional Wheeler DFA), then $ \mathfrak{e} = e $, and we retrieve the time and space bounds already known for Wheeler automata~\cite{gagie2017, cotumacciojacm}. If $ r = O(1) $, we can still solve pattern matching queries in linear time (for constant alphabets), thus breaking the lower bound by Equi et al. while only storing a compressed representation of $ \mathcal{A} $.

\section{Preliminaries}\label{sec:preliminary}

Let $ \Sigma $ be a finite alphabet, and let $ \Sigma^* $ be the set of all finite strings on $ \Sigma $. We denote by $ \epsilon $ the empty string and by $ \Sigma^+ $ the set $ \Sigma^* \setminus \{\epsilon \} $ of all nonempty finite strings on $ \Sigma $. If $ \mathcal{L} \subseteq \Sigma^* $, let $ \Pref (\mathcal{L}) $ be the set of all prefixes of some string in $ \mathcal{L} $. Note that if $ \mathcal{L} \not = \emptyset $, then $ \epsilon \in \Pref (\mathcal{L)} $.

Let us introduce some definitions that will be useful for Theorem~\ref{theor:ng}. We say that $ \mathcal{L} \subseteq \Sigma^* $ is \emph{prefix-free} if no string in $ \mathcal{L} $ is a strict prefix of another string in $ \mathcal{L} $. Note that if $ \mathcal{L} $ is prefix-free and $ \epsilon \in \mathcal{L} $, then $ \mathcal{L} = \{\epsilon \} $. If $ \mathcal{L} \subseteq \Sigma^* $, the \emph{prefix-free kernel} of $ \mathcal{L} $ is the set $ \mathcal{K}(\mathcal{L}) $ of all strings in $ \mathcal{L} $ whose strict prefixes are all not in $ \mathcal{L} $. Note that $ \mathcal{K}(\mathcal{L}) $ is always prefix-free, and $ \mathcal{L} $ is prefix-free if and only if $ \mathcal{L} = \mathcal{K}(\mathcal{L}) $. For every $ \mathcal{L} \subseteq \Sigma^* $ and for every $ \alpha \in \mathcal{L} $, let $ \mathcal{L}_\alpha = \{\rho \in \Sigma^+ \;|\; \alpha \rho \in \mathcal{L} \} $. If $ \mathcal{L} \subseteq \Sigma^* $ and $ \mathcal{K}(\mathcal{L}_\alpha) $ is finite for every $ \alpha \in \mathcal{L} $, then we say that $ \mathcal{L} $ is \emph{locally bounded}.

We now recall how to define generalized automata~\cite{giammaresi1995conference, giammaresi1999journal}.

\begin{defi}\label{def:SNFAs}
A \emph{generalized non-deterministic finite automaton (GNFA)} is a 4-tuple $ \mathcal{A} = (Q, E, s, F) $, where $ Q $ is a finite set of states, $ E \subseteq Q \times Q \times \Sigma^* $ is a finite set of string-labeled edges, $ s \in Q $ is the initial state and $ F \subseteq Q $ is a set of final states. Moreover, we assume that, for every $ u \in Q $, (i) $ u $ is reachable from the initial state and (ii) $ u $ is co-reachable, that is, $ u $ is either final or allows reaching a final state.

A \emph{generalized deterministic finite automaton (GDFA)} is a GNFA such that, for every $ u \in Q $, (i) no edge leaving $ u $ is labeled with $ \epsilon $, (ii) distinct edges leaving $ u $ have distinct labels, and (iii) the set of all strings labeling some edge leaving $ u $ is prefix-free.
\end{defi}

The assumption that every state is reachable and co-reachable is standard in automata theory because, if a state is not reachable or it is not co-reachable, then it can be removed without changing the recognized language (as long as the set of states
does not become empty)\footnote{Since we assume that the set of states is non-empty (because we assume the existence of the initial state), under our definition the empty language is not recognized by any automaton. If we insist on including the empty language, we may introduce the additional notion of \emph{empty automaton}, and Theorem~\ref{theor:MyhillNerodetextbook} and Theorem~\ref{theor:ng} may be extended accordingly by considering the (unique) equivalence relation on the empty set.}. A conventional NFA (DFA, respectively) is a GNFA (GDFA, respectively) where all edges are labeled with characters from $ \Sigma $. Note that we explicitly require a GNFA to have finitely many edges (in conventional NFAs, the finiteness of the number of states automatically implies the finiteness of the number of edges because the alphabet is finite). If we allowed a GNFA to have infinitely many edges, then any nonempty (possibly non-regular) language would be recognized by a GNFA with two states, where the first state is initial, the second state is final, all edges go from the first state to the second state, and a string labels an edge if and only if it is in the language. By requiring a GNFA to have finitely many edges, the class of recognized languages is exactly the class of regular languages, because it is easy to transform a GNFA into a NFA with $ \epsilon $-transitions (that is, an NFA where edges can also be labeled with the empty string $ \epsilon $) that recognizes the same language by proceeding as follows: for every edge $ (u', u, \rho) \in E $, with $ \rho = r_1, \dots, r_{|\rho|} \in \Sigma^+ $, where $ r_1, \dots, r_{|\rho|} \in \Sigma $ and $ |\rho| \ge 2 $, we delete the edge $ (u', u, \rho) $, we add $ |\rho | - 1 $ new states $ z_1, \dots, z_{|\rho| - 1} $, and then we add the edges $ (u', z_1, r_1) $, $ (z_1, z_2, r_2) $, $ \dots $, $ (z_{|\rho| - 1}, u, r_{|\rho|}) $ (none of the new states is made initial or final).

Let us introduce some notation that will be helpful in the paper.

\begin{defi}\label{def:usual}
Let $ \mathcal{A} = (Q, E, s, F) $ be a GNFA.
\begin{itemize}
    \item For every $ \alpha \in \Sigma^* $, let $ J_\alpha $ be the set of all states that can be reached from the initial state by following edges whose labels, when concatenated, yield $ \alpha $. In other words, for every $ u \in Q $ we have $ u \in J_\alpha $ if and only if there exist $ t \ge 0 $, $ u_1, u_2, \dots, u_t \in Q $ and $ \alpha_1, \alpha_2, \dots, \alpha_t \in \Sigma^* $ such that (i) $ (s, u_1, \alpha_1), (u_1, u_2, \alpha_2), (u_2, u_3, \alpha_3), \dots, (u_{t - 1}, u_t, \alpha_t) \in E $, (ii) $ \alpha = \alpha_1 \alpha_2 \alpha_3 \dots \alpha_t $ and (iii) $ u_t = u $. Note that $ s \in J_\epsilon $.
    \item Let $ \mathcal{L(A)} $ be the language recognized by $ \mathcal{A} $, that is, $ \mathcal{L(A)} = \{\alpha \in \Sigma^* \;|\; J_\alpha \cap F \not = \emptyset \} $.
    \item For every $ u \in Q $, let $ I_u $ be the set of all strings that can be read from the initial state to $ u $ by concatenating edge labels, that is, $ I_u = \{\alpha \in \Sigma^* \;|\; u \in J_\alpha \} $. Note that for every $ u \in Q $ we have $ \emptyset \subsetneqq I_u \subseteq \Pref (\mathcal{L(A)}) $ because every state is reachable and co-reachable.
\end{itemize}
When $ \mathcal{A} $ is not clear from the context, we write $ J_\alpha^\mathcal{A} $ and $ I_u^\mathcal{A} $.
\end{defi}

\section{Generalized Automata: The Myhill-Nerode Theorem}\label{sec:nerodeGNFAs}

The Myhill-Nerode theorem for conventional automata (Theorem~\ref{theor:MyhillNerodetextbook}) provides some algebraic properties that $ \Pref(\mathcal{L}) $ must satisfy for $ \mathcal{L} \subseteq \Sigma^* $ to be a regular language. Intuitively, the link between the algebraic characterization and the automata characterization of regular languages is that, given an NFA $ \mathcal{A} = (Q, E, s, F) $ that recognizes $ \mathcal{L} $, we have $ \bigcup_{u \in Q} I_u = \Pref (\mathcal{L}) $: indeed, if $ \alpha \in \Pref (\mathcal{L}) $, one can read $ \alpha $ on $ \mathcal{A} $ starting from the initial state. However, if $ \mathcal{A} = (Q, E, s, F) $ is a GNFA that recognizes $ \mathcal{L} $, then we only have $ \bigcup_{u \in Q} I_u \subseteq \Pref (\mathcal{L)} $, because if $ \alpha \in \Pref (\mathcal{L}) $, then we can read $ \alpha $ on $ \mathcal{A} $ starting from the initial state, but possibly we do not reach the end of the last edge label.

Let us give the following definition.

\begin{defi}
Let $ \mathcal{A} = (Q, E, s, F) $ be a GNFA. Define:
\begin{equation*}
    \mathcal{W (A)} = \bigcup_{u \in Q} I_u.
\end{equation*}
We say that $ \mathcal{A} $ is a $ \mathcal{W(A)} $-GNFA.
\end{defi}

Note that for every $ \alpha \in \Sigma^* $ we have $ J_\alpha \not = \emptyset $ if and only if $ \alpha \in \mathcal{W(A)} $.

To obtain the Myhill-Nerode theorem for generalized automata (Theorem~\ref{theor:nerodegeneralized}), we need to characterize the sets $ \mathcal{W} \subseteq \Sigma^* $ for which there exists a Wheeler GNFA such that $ \mathcal{W}(\mathcal{A}) = \mathcal{W} $. Let us start with some basic properties.

\begin{lem}\label{lem:prooflocallybounded}
    Let $ \mathcal{A} = (Q, E, s, F) $ be a GNFA. Then:
    \begin{enumerate}
        \item $ \mathcal{L(A)} \cup \{\epsilon \} \subseteq \mathcal{W(A)} \subseteq \Pref (\mathcal{L(A)}) $.
        \item $ \mathcal{W}(\mathcal{A}) $ is locally bounded.
    \end{enumerate}
\end{lem}

\begin{proof}
    \begin{enumerate}
        \item We have $ \mathcal{L(A)} \subseteq \mathcal{W(A)} $ because if $ \alpha \in \mathcal{L(A)} $, then $ J_\alpha \cap F \not = \emptyset $ and in particular $ \alpha \in \mathcal{W(A)} $. Moreover, we have $ \epsilon \in \mathcal{W(A)} $ because $ \epsilon \in I_s $. Lastly, we have $ \mathcal{W(A)} \subseteq \Pref (\mathcal{L(A)}) $ because $ I_u \subseteq \Pref (\mathcal{L(A)}) $ for every $ u \in Q $.
        \item Fix $ \alpha \in \mathcal{W}(\mathcal{A}) $. We must prove that $ \mathcal{K}(\mathcal{W}(\mathcal{A})_\alpha) $ is finite. By definition, the set $ E $ of all edges is finite, so there exists an integer $ r $ such that every edge label has length at most $ r $. To conclude the proof, it will suffice to prove that every $ \rho \in \mathcal{K}(\mathcal{W}(\mathcal{A})_\alpha) $ has length at most $ r $ (recall that the $ \Sigma $ is also finite). Fix $ \rho \in \mathcal{K}(\mathcal{W}(\mathcal{A})_\alpha) $. Since $ \rho \in \mathcal{W}(\mathcal{A})_\alpha $, we have $ \alpha \rho \in \mathcal{W}(\mathcal{A}) $ and so $ J_{\alpha \rho} \not = \emptyset $. We know that $ \mathcal{K}(\mathcal{W}(\mathcal{A})_\alpha) $ is prefix-free, so from $ \rho \in \mathcal{W}(\mathcal{A})_\alpha $ we obtain that, for every $ \rho' \in \Sigma^+ $ being a strict nonempty prefix of $ \rho $, we have $ \alpha \rho' \not \in \mathcal{W}(\mathcal{A}) $ and so $ J_{\alpha \rho'} = \emptyset $. This means that $ \rho $ has length at most $ r $ (otherwise, $ E $ should contain an edge label whose length is bigger than $ r $). \qedhere
    \end{enumerate}
\end{proof}

\begin{rem}\label{rem:GDFAstringsreachonestate}
    Let $ \mathcal{A} = (Q, E, s, F) $ be a GDFA. Let $ \alpha \in \mathcal{W(A)} $. If $ \alpha = \epsilon $, then $ J_\epsilon = \{s \} $ because no edge is labeled with $ \epsilon $. If $ |\alpha| > 1 $, then there exists (i) a prefix $ \alpha_1 \in \Sigma^* $ of $ \alpha $ and (ii) $ u_1 \in Q $ such that $ (s, u_1, \alpha_1) \in E $, and since $ \mathcal{A} $ is a GDFA, we have (i) $ \alpha_1 \in \Sigma^+ $ and (ii) both $ \alpha_1 $ and $ u_1 $ are unique. In particular, $ \alpha_1 \in \mathcal{W(A)} $. If $ \alpha_1 $ is a strict prefix of $ \alpha $, we can repeat the argument starting from $ u_1 $. We conclude that for every $ \alpha \in \mathcal{W(A)} $ we have $ |J_\alpha| = 1 $ (that is, we have global determinism, see the informal discussion in Section~\ref{sec:introduction}). As a consequence, if $ u, v \in Q $ are distinct, then $ I_u \cap I_v = \emptyset $. In the following, if $ \mathcal{A} $ is a GDFA and $ \alpha \in \mathcal{W(A)} $, we will identify $ J_\alpha $ and the state being its unique element.
    
    Moreover, our argument shows that, if $ \mathcal{A} $ is a GDFA, then, for every $ \alpha \in \mathcal{W(A)} $ such that $ |\alpha| > 0 $, the longest strict prefix of $ \alpha $ in $ \mathcal{W(A)} $ is the unique strict prefix $ \alpha' $ of $ \alpha $ in $ \mathcal{W(A)} $ such that, letting $ \rho \in \Sigma^+ $ be the string for which $ \alpha = \alpha' \rho $, we have $ (J_{\alpha'}, J_\alpha, \rho) \in E $. This implies that if $ \mathcal{A} $ is a GDFA and $ \alpha \in \mathcal{W(A)} $, then $ \mathcal{K}(\mathcal{W}(\mathcal{A})_\alpha) = \{\rho \in \Sigma^+ \;|\; \alpha \rho \in \mathcal{W(A)}, (J_\alpha, J_{\alpha \rho}, \rho) \in E \} $.
\end{rem}

Let us give a definition.

\begin{defi}\label{def:locallybounded}
    Let $ \mathcal{A} = (Q, E, s, F) $ be a GNFA and let $ \mathcal{W} \subseteq \Sigma^* $. We say that $ \mathcal{A} $ is a \emph{$ \mathcal{W} $-GNFA} if $ \mathcal{W(A)} = \mathcal{W} $.
\end{defi}

In the classical Myhill-Nerode theorem, we consider equivalence relations defined on $ \Pref (\mathcal{L}) $. In our setting, we will need to define equivalence relations on $ \mathcal{W} $. This leads to the following general definition.

\begin{defi}\label{def:rightinvariant}
Let $ \mathcal{W} \subseteq \Sigma^* $ and let $ \sim $ be an equivalence relation on $ \mathcal{W} $. We say that $ \sim $ is \emph{right-invariant} if:
\begin{equation*}
    (\forall \alpha, \beta \in \mathcal{W})(\forall \phi \in \Sigma^*)(\alpha \sim \beta \Longrightarrow ((\alpha \phi \in \mathcal{W} \Longleftrightarrow \beta \phi \in \mathcal{W}) \land (\alpha \phi \in \mathcal{W} \Longrightarrow \alpha \phi \sim \beta \phi))).
\end{equation*}
\end{defi}

Let us give a characterization of right-invariant equivalence relations that is based on prefix-free kernels.

\begin{lem}\label{lem:characterizationrightinv}
    Let $ \mathcal{W} \subseteq \Sigma^* $ and let $ \sim $ be an equivalence relation on $ \mathcal{W} $. Then, $ \sim $ is right-invariant if and only if:
\begin{equation*}
\begin{split}
    & (\forall \alpha, \beta \in \mathcal{W})(\alpha \sim \beta \Longrightarrow ((\mathcal{K}(\mathcal{W}_\alpha) = \mathcal{K}(\mathcal{W}_\beta)) \land(\forall \rho \in \Sigma^+)(\rho \in \mathcal{K}(\mathcal{W}_\alpha) \Longrightarrow \alpha \rho \sim \beta \rho ))). \\
\end{split}
\end{equation*} 
\end{lem}

\begin{proof}
($ \Longrightarrow $) Fix $ \alpha, \beta \in \mathcal{W} $ such that $ \alpha \sim \beta $.
    \begin{itemize}
        \item Let us prove that $ \mathcal{K}(\mathcal{W}_\alpha) = \mathcal{K}(\mathcal{W}_\beta) $. To this end, we only need to prove that $ \mathcal{W}_\alpha = \mathcal{W}_\beta $. Fix $ \rho \in \Sigma^+ $. We must prove that $ \alpha \rho \in \mathcal{W} $ if and only if $ \beta \rho \in \mathcal{W} $. This follows because $ \sim $ is right-invariant and we can choose $ \phi = \rho $ in Definition~\ref{def:rightinvariant}.
        \item Fix $ \rho \in \Sigma^+ $ such that $ \rho \in \mathcal{K}(\mathcal{W}_\alpha) $. We must prove that $ \alpha \rho \sim \beta \rho $. From $ \rho \in \mathcal{K}(\mathcal{W}_\alpha) $ we obtain $ \rho \in \mathcal{W}_\alpha $, so $ \alpha \rho \in \mathcal{W} $. The conclusion follows because $ \sim $ is right-invariant and we can choose $ \phi = \rho $ in Definition~\ref{def:rightinvariant}.
    \end{itemize}
    
($ \Longleftarrow $) Let $ \alpha, \beta \in \mathcal{W} $ such that $ \alpha \sim \beta $, and let $ \phi \in \Sigma^* $ such that $ \alpha \phi \in \mathcal{W} $. We must prove that $ \beta \phi \in \mathcal{W} $ and $ \alpha \phi \sim \beta \phi $. If $ \phi = \epsilon $ the conclusion is immediate, so we can assume $ \phi \in \Sigma^+ $. Let $ \phi_1, \dots, \phi_s $ be all prefixes of $ \phi $ such that $ \alpha \phi_i \in \mathcal{W} $ for every $ 1 \le i \le s $, where $ \phi_i $ is a strict prefix of $ \phi_{i + 1} $ for every $ 1 \le i \le s - 1 $. Note that $ s \ge 2 $, $ \phi_1 = \epsilon $ and $ \phi_s = \phi $. For every $ 1 \le i \le s - 1 $, let $ \rho_i \in \Sigma^+ $ be such that $ \phi_{i + 1} = \phi_i \rho_i $. Notice that by definition we have $ \rho_i \in \mathcal{K}(T_{\alpha \phi_i}) $ for every $ 1 \le i \le s - 1 $.

Let us prove that for every $ 1 \le i \le s $ we have $ \beta \phi_i \in \mathcal{W} $ and $ \alpha \phi_i \sim \beta \phi_i $. We proceed by induction on $ i $. For $ i = 1 $, we have $ \phi_1 = \epsilon $ and we know that $ \beta \in \mathcal{W} $ and $ \alpha \sim \beta $. Now assume that $ 2 \le i \le s $. By the inductive hypothesis, we have $ \beta \phi_{i - 1} \in \mathcal{W} $ and $ \alpha \phi_{i - 1} \sim \beta \phi_{i - 1} $. We know that $ \rho_{i - 1} \in \mathcal{K}(T_{\alpha \phi_{i - 1}}) $, so by using the property described in the statement of the lemma we obtain that $ \beta \phi_i = \beta \phi_{i - 1} \rho_{i - 1} \in \mathcal{W} $ and $ \alpha \phi_i = \alpha \phi_{i - 1} \rho_{i - 1} \sim \beta \phi_{i - 1} \rho_{i - 1} = \beta \phi_i $.

In particular, for $ i = s $ we have $ \phi_s = \phi $ and we conclude that $ \beta \phi \in \mathcal{W} $ and $ \alpha \phi \sim \beta \phi $. \qedhere
\end{proof}

In general, an equivalence relation is not right-invariant. Let us show how to define a canonical right-invariant equivalence relation starting from \emph{any} equivalence relation.

\begin{lem}\label{lem:rightinvariantrefinement}
Let $ \mathcal{W} \subseteq \Sigma^* $ and let $ \sim $ be an equivalence relation on $ \mathcal{W} $. For every $ \alpha, \beta \in \mathcal{W} $, let:
\begin{equation*}
    \alpha \sim_r  \beta \Longleftrightarrow (\forall \phi \in \Sigma^*)((\alpha \phi \in \mathcal{W} \Longleftrightarrow \beta \phi \in \mathcal{W}) \land (\alpha \phi \in \mathcal{W} \Longrightarrow \alpha \phi \sim \beta \phi)).
\end{equation*}
Then $ \sim_r $ is an equivalence relation on $ \mathcal{W} $, it is right-invariant and it is the coarsest right-invariant equivalence relation on $ \mathcal{W} $ refining $ \sim $.
\end{lem}

\begin{proof}
It is immediate to check that $ \sim_r $ is an equivalence relation. Let us prove that $ \sim_r $ is right-invariant. Assume that $ \alpha, \beta \in \mathcal{W} $ and $ \phi \in \Sigma^* $ are such that $ \alpha \sim_r  \beta $ and $ \alpha \phi \in \mathcal{W} $. We must prove that $ \beta \phi \in \mathcal{W} $ and $ \alpha \phi \sim_r  \beta  \phi $. Notice that $ \beta \phi \in \mathcal{W} $ follows immediately from $ \alpha \sim_r  \beta $ and $ \alpha \phi \in \mathcal{W} $. Let us prove that $ \alpha \phi \sim_r \beta  \phi $. Fix $ \psi \in \Sigma^* $ such that $ \alpha \phi \psi \in \mathcal{W} $. We must prove that $ \beta \phi \psi \in \mathcal{W} $ and $ \alpha \phi \psi \sim \beta \phi \psi  $. This follows from $ \alpha \sim_r  \beta $ and $ \alpha \phi \psi \in \mathcal{W} $. Moreover, $ \sim_r $ refines $ \sim $ because for every $ \alpha, \beta \in \mathcal{W} $, if $ \alpha \sim_r \beta $, by letting $ \phi $ be the empty string we conclude $ \alpha \sim \beta $. Lastly, we want to prove that $ \sim_r $ is the coarsest right-invariant equivalence relation on $ \mathcal{W} $ refining $ \sim $. Let $ \sim_* $ be a right-invariant equivalence relation on $ \mathcal{W} $ refining $ \sim $. Assume that for some $ \alpha, \beta \in \mathcal{W} $ we have $ \alpha \sim_* \beta $. We must prove that $ \alpha \sim_r \beta $. Let $ \phi \in \Sigma^* $ be such that $ \alpha \phi \in \mathcal{W} $. We must prove that $ \beta \phi \in \mathcal{W} $ and $ \alpha \phi \sim \beta \phi $. Since $ \sim_* $ is right-invariant, from $ \alpha \sim_* \beta $ and $ \alpha \phi \in \mathcal{W} $ we obtain $ \beta \phi \in \mathcal{W} $ and $ \alpha \phi \sim_* \beta \phi $, which implies $ \alpha \phi \sim \beta \phi $ because $ \sim_* $ refines $ \sim $. \qedhere
\end{proof}

Lemma~\ref{lem:rightinvariantrefinement} allows us to give the following definition.

\begin{defi}
    Let $ \mathcal{W} \subseteq \Sigma^* $ and let $ \sim $ be an equivalence relation on $ \mathcal{W} $. We say that the equivalence relation $ \sim_r $ on $ \mathcal{W} $ defined in the statement of Lemma~\ref{lem:rightinvariantrefinement} is the \emph{right-invariant refinement} of $ \sim $.
\end{defi}

The \emph{Myhill-Nerode equivalence} plays a major role in the classical Myhill-Nerode theorem. Let us show how we can extend it when $ \mathcal{W} $ is not necessarily equal to $ \Pref (\mathcal{L}) $.

\begin{defi}\label{def:MyhillNerodeEquivalenceGeneralized}
Let  $ \mathcal{L}, \mathcal{W} \subseteq \Sigma^* $. The \emph{Myhill-Nerode equivalence on $ \mathcal{L} $ and $ \mathcal{W} $} is the equivalence relation $ \equiv_{\mathcal{L}, \mathcal{W}} $ on $ \mathcal{W} $ defined as the right-invariant refinement of $ \sim_{\mathcal{L}, \mathcal{W}} $, where $ \sim_{\mathcal{L}, \mathcal{W}} $ is the equivalence relation on $ \mathcal{W} $ such that for every $ \alpha, \beta \in \mathcal{W} $:
\begin{equation*}
    \alpha \sim_{\mathcal{L}, \mathcal{W}} \beta \Longleftrightarrow (\alpha \in \mathcal{L} \Longleftrightarrow \beta \in \mathcal{L}).
\end{equation*}
\end{defi}

If $ \mathcal{W} = \Pref (\mathcal{L}) $, then we retrieve the classical Myhill-Nerode equivalence relation for $ \mathcal{L} $. Let us describe some elementary properties of $ \equiv_{\mathcal{L}, \mathcal{W}} $.

\begin{lem}\label{lem:nerodegenproperties}
Let  $ \mathcal{L}, \mathcal{W} \subseteq \Sigma^* $. Then $ \equiv_{\mathcal{L}, \mathcal{W}} $ is right-invariant, and $ \mathcal{L} $ is the union of some $ \equiv_{\mathcal{L}, \mathcal{W}} $-classes.
\end{lem}

\begin{proof}
First, $ \equiv_{\mathcal{L}, \mathcal{W}} $ is right-invariant by Lemma~\ref{lem:rightinvariantrefinement} because it is a right-invariant refinement by definition. Moreover, $ \mathcal{L} $ is the union of some $ \sim_{\mathcal{L}, \mathcal{W}} $-classes, and so also of some $ \equiv_{\mathcal{L}, \mathcal{W}} $-classes because $ \equiv_{\mathcal{L}, \mathcal{W}} $ refines $ \sim_{\mathcal{L}, \mathcal{W}} $. \qedhere
\end{proof}

Let $ \mathcal{A} $ be a conventional NFA, and define the equivalence relation $ \sim_\mathcal{A} $ on $ \Pref (\mathcal{L(A)}) $ as follows: for every $ \alpha, \beta \in \Pref (\mathcal{L(A)}) $, let $ \alpha \sim_\mathcal{A} \beta $ if and only if $ J_\alpha = J_\beta $. This equivalence relation is an intermediate tool in the Myhill-Nerode theorem for conventional automata, and it can also be defined for a generalized automaton $ \mathcal{A} $ by considering the equivalence relation $ \sim_\mathcal{A} $ on $ \mathcal{W(A)} $ such that for every $ \alpha, \beta \in \mathcal{W(A)} $ we have $ \alpha \sim_\mathcal{A} \beta $ if and only if $ J_\alpha = J_\beta $. If $ \mathcal{A} $ is an NFA (or an NFA with $ \epsilon $-transitions), then $ \sim_\mathcal{A} $ is right-invariant, because for every $ \alpha \in \Pref (\mathcal{L(A)}) $ and for every prefix $ \alpha' $ of $ \alpha $, any path from the initial state to a state in $ J_\alpha $ must go through a state in $ J_{\alpha'} $. However, in general this property is not true if $ \mathcal{A} $ is a GNFA, so $ \sim_\mathcal{A} $ need not be right-invariant if $ \mathcal{A} $ is a GNFA (see Figure~\ref{fig:nonrightinvariant}). Since right-invariance is crucial in the Myhill-Nerode theorem, we will consider the right-invariant refinement of $ \sim_\mathcal{A} $.

\begin{figure}[h!]
\centering
\scalebox{0.8}{
\centering
\begin{tikzpicture}[->,>=stealth', semithick, auto, scale=1]
\node[state, initial] (1)    at (0,0)	{$ 1 $};
\node[state] (2)    at (4,2)	{$ 2 $};
\node[state, accepting] (3)    at (4, 0)	{$ 3 $};
\node[state, accepting] (4)    at (4, -2)	{$ 4 $};

\draw (1) edge [] node [] {$ a, b $} (2);
\draw (2) edge [] node [] {$ c $} (3);
\draw (1) edge [] node [] {$ ac $} (4);
\end{tikzpicture}
}
 	\caption{A GNFA $ \mathcal{A} $ such that $ \sim_\mathcal{A} $ is not right-invariant. Indeed, we have $ a, b, ac, bc \in \mathcal{W(A)} $ and $ a \sim_\mathcal{A} b $, but $ ac \not \sim_\mathcal{A} bc $.}
    \label{fig:nonrightinvariant}
\end{figure}

\begin{defi}\label{def:sim_Astates}
Let $ \mathcal{A} = (Q, E, s, F) $ be a GNFA. Let $ \equiv_\mathcal{A} $ be the right-invariant refinement of $ \sim_\mathcal{A} $, where $ \sim_\mathcal{A} $ is the equivalence relation on $ \mathcal{W(A)} $ such that for every $ \alpha, \beta \in \mathcal{W(A)} $:
\begin{equation*}
    \alpha \sim_\mathcal{A} \beta  \Longleftrightarrow J_\alpha = J_\beta.
\end{equation*}
\end{defi}

\begin{rem}\label{rem:equivalenceonGDFA}
    Let us prove that, if $ \mathcal{A} $ is GDFA, then $ \sim_\mathcal{A} $ is right-invariant. Let $ \alpha, \beta \in \mathcal{W}(\mathcal{A}) $ be such that $ J_\alpha = J_\beta $, and let $ \phi \in \mathcal{K}(\mathcal{W}(\mathcal{A})_\alpha) $. By Lemma~\ref{lem:characterizationrightinv}, we only have to prove that $ \phi \in \mathcal{K}(\mathcal{W}(\mathcal{A})_\beta) $ and $ J_{\alpha \phi} = J_{\beta \phi} $. By Remark~\ref{rem:GDFAstringsreachonestate}, we have $ \alpha \phi \in \mathcal{W(A)} $ and $ (J_\alpha, J_{\alpha \phi}, \phi) \in E $. Hence $ (J_\beta, J_{\alpha \phi}, \phi) \in E $, we obtain $ \beta \phi \in \mathcal{W(A)} $ and $ J_{\alpha \phi} = J_{\beta \phi} $, and again by Remark~\ref{rem:GDFAstringsreachonestate} we conclude $ \phi \in \mathcal{K}(\mathcal{W}(\mathcal{A})_\beta) $. Notice that, in fact, the generalized automaton $ \mathcal{A} $ in Figure~\ref{fig:nonrightinvariant} is not a GDFA.

    Since $ \equiv_\mathcal{A} $ is the right-invariant refinement of $ \sim_\mathcal{A} $, by Lemma~\ref{lem:rightinvariantrefinement} we conclude that, if $ \mathcal{A} $ is a GDFA, then $ \equiv_\mathcal{A} $ and $ \sim_\mathcal{A} $ are the same equivalence relation.
\end{rem}

Let us study the properties of $ \equiv_\mathcal{A} $.

\begin{lem}\label{lem:simAproperties}
Let $ \mathcal{A} = (Q, E, s, F) $ be a GNFA. Then, $ \equiv_\mathcal{A} $ is right-invariant, it refines $ \equiv_{\mathcal{L(A)}, \mathcal{W(A)}}$, it has finite index, and $ \mathcal{L(A)} $ is the union of some $ \equiv_\mathcal{A} $-classes. 
\end{lem}

\begin{proof}
First, $ \equiv_\mathcal{A} $ is right-invariant by Lemma~\ref{lem:rightinvariantrefinement} because it is a right-invariant refinement by definition.

Let us prove that $ \mathcal{L(A)} $ is the union of some $ \equiv_\mathcal{A} $-classes. It will suffice to show that $ \mathcal{L(A)} $ is the union of some $ \sim_\mathcal{A} $-classes, because $ \equiv_\mathcal{A} $ is a refinement of $ \sim_\mathcal{A} $. Let $ \alpha, \beta \in \mathcal{W(A)} $ such that $ \alpha \sim_\mathcal{A} \beta $, that is, $ J_\alpha = J_\beta $. We must prove that $ \alpha \in \mathcal{L(A)} $ if and only if $ \beta \in \mathcal{L(A)} $. We have $ \alpha \in \mathcal{L(A)} $ if and only if $ J_\alpha \cap F \not = \emptyset $, if and only if $ J_\beta \cap F \not = \emptyset $, if and only if $ \beta \in \mathcal{L(A)} $.

Let us prove that $ \equiv_\mathcal{A} $ refines $ \equiv_{\mathcal{L(A)}, \mathcal{W(A)}}$. Since $ \mathcal{L(A)} $ is the union of some $ \equiv_\mathcal{A} $-classes, then $ \equiv_\mathcal{A} $ refines $ \sim_{\mathcal{L(A)}, \mathcal{W(A)}} $. Then, $ \equiv_\mathcal{A} $ also refines $ \equiv_{\mathcal{L(A)}, \mathcal{W(A)}}$, because $ \equiv_\mathcal{A} $ is right-invariant and $ \equiv_{\mathcal{L(A)}, \mathcal{W(A)}}$ is the coarsest right-invariant equivalence relation refining $ \sim_{\mathcal{L(A)}, \mathcal{W(A)}} $ by Lemma~\ref{lem:rightinvariantrefinement}.

Lastly, let us prove that $ \equiv_\mathcal{A} $ has finite index. Let $ \bar{\mathcal{A}} $ be the NFA with $ \epsilon $-transitions such that $ \mathcal{L}(\bar{\mathcal{A}}) = \mathcal{L}(\mathcal{A}) $ that is constructed from $ \mathcal{A} $ by proceeding as explained in Section~\ref{sec:preliminary}. We proceed in two steps.
\begin{itemize}
\item Fix $ \alpha, \beta \in \mathcal{W(A)} $ such that $ \alpha \sim_{\bar{\mathcal{A}}} \beta $. Let us prove that we must have $ \alpha \equiv_\mathcal{A} \beta $. Since $ \equiv_\mathcal{A} $ is the right-invariant refinement of $ \sim_\mathcal{A} $, by Lemma~\ref{lem:rightinvariantrefinement} we must prove that, for every $ \phi \in \Sigma^* $ such that $ \alpha \phi \in \mathcal{W(A)} $, we have $ \beta \phi \in \mathcal{W(A)} $ and $ \alpha \phi \sim_\mathcal{A} \beta \phi $. Fix $ \phi \in \Sigma^* $ such that $ \alpha \phi \in \mathcal{W}(\mathcal{A}) $. Since $ \bar{\mathcal{A}} $ is an NFA with $ \epsilon $-transitions, we have that $ \sim_{\bar{\mathcal{A}}} $ is right-invariant (see the discussion before Definition~\ref{def:sim_Astates}). From $ \alpha \sim_{\bar{\mathcal{A}}} \beta $ and $ \alpha \phi \in \mathcal{W(A)} \subseteq \Pref (\mathcal{L(A)}) $ we obtain $ \beta \phi \in \Pref (\mathcal{L(A)}) $ and $ \alpha \phi \sim_{\bar{\mathcal{A}}} \beta \phi $, so $ J^{\bar{\mathcal{A}}}_{\alpha \phi} = J^{\bar{\mathcal{A}}}_{\beta \phi} $. By the construction of $ \bar{\mathcal{A}} $, we obtain $ J_{\alpha \phi}^\mathcal{A} = J_{\alpha \phi}^{\bar{\mathcal{A}}} \cap Q = J_{\beta \phi}^{\bar{\mathcal{A}}} \cap Q = J_{\beta \phi}^\mathcal{A} $. From $ \alpha \phi \in \mathcal{W(A)} $, we obtain $ J_{\alpha \phi}^\mathcal{A} \not = \emptyset $, so $ J_{\beta \phi}^\mathcal{A} \not = \emptyset $ and $ \beta \phi \in \mathcal{W(A)} $. Moreover, we have $ \alpha \phi \sim_\mathcal{A} \beta \phi $ because $ J^\mathcal{A}_{\alpha \phi} = J^\mathcal{A}_{\beta \phi} $. This proves that $ \alpha \equiv_\mathcal{A} \beta $.

\item Since for every $ \alpha, \beta \in \mathcal{W(A)} $ we have that $ \alpha \sim_{\bar{\mathcal{A}}} \beta $ implies $ \alpha \equiv_\mathcal{A} \beta $, in order to prove that $ \equiv_\mathcal{A} $ has finite index it will suffice to prove that $ \sim_{\bar{\mathcal{A}}} $ has finite index. To this end, observe that every $ \sim_{\bar{\mathcal{A}}} $-class can be associated with a distinct nonempty subset of the set $ \bar{Q} $ of all states in $ \bar{\mathcal{A}} $ (via the well-defined mapping $ [\alpha]_{\sim_{\bar{\mathcal{A}}}} \mapsto J^{\bar{\mathcal{A}}}_\alpha) $, so the index of $ \sim_{\bar{\mathcal{A}}} $ is bounded by  $ 2^{|\bar{Q}|} - 1 $. \qedhere
\end{itemize}
\end{proof}

Given two GNFAs $ \mathcal{A} = (Q, E, s, F)  $ and $ \mathcal{A}' = (Q', E', s', F) $, we say that $ \mathcal{A} $ and $ \mathcal{A}' $ are \emph{isomorphic} if there exists a bijection $ \phi: Q \mapsto Q' $ such that (i) for every $ u, v \in Q $ and for every $ \rho \in \Sigma^* $ we have $ (u, v, \rho) \in E $ if and only if $ (\phi (u), \phi (v), \rho) \in E' $, (ii) $ \phi (s) = s' $ and (iii) for every $ u \in Q $ we have $ u \in F $ if and only if $ \phi (u) \in F $.

The following lemma is crucial to derive our Myhill-Nerode theorem for generalized automata. 

\begin{lem}\label{lem:fromrelationtoDFA}
Let $ \mathcal{L} \subseteq \Sigma^* $ and let $ \mathcal{W} \subseteq \Sigma^* $ be a locally bounded set such that $ \mathcal{L} \cup \{\epsilon \} \subseteq \mathcal{W} \subseteq \Pref (\mathcal{L}) $. Assume that $ \mathcal{L} $ is the union of some classes of a right-invariant equivalence relation $ \sim $ on $ \mathcal{W} $ of finite index. Then, $ \mathcal{L} $ is recognized by a $ \mathcal{W} $-GDFA $ \mathcal{A_\sim} = (Q_\sim, E_\sim, s_\sim, F_\sim) $ such that:
\begin{enumerate}
\item $ |Q_\sim| $ is equal to the index of $ \sim $.
\item $ \equiv_{\mathcal{A_\sim}} $ and $ \sim $ are the same equivalence relation.
\end{enumerate}
Moreover, if $ \mathcal{B} $ is a $ \mathcal{W} $-GDFA that recognizes $ \mathcal{L} $, then $ \mathcal{A_{\equiv_\mathcal{B}}} $ is isomorphic to $ \mathcal{B} $.
\end{lem}

\begin{proof}
Define $ \mathcal{A_\sim} = (Q_\sim, E_\sim, s_\sim, F_\sim) $ as follows.
	\begin{itemize}
		\item $ Q_\sim = \{[\alpha]_\sim \;|\; \alpha \in \mathcal{W} \} $.
		\item $ E_\sim = \{ ([\alpha]_\sim, [\alpha \rho]_\sim, \rho) \;|\; \alpha \in \mathcal{W},  \rho \in \mathcal{K}(\mathcal{W}_\alpha) \} $.
        \item $ s_\sim = [\epsilon]_\sim $.
		\item $ F_\sim = \{[\alpha]_\sim \;|\; \alpha \in \mathcal{L} \} $.
	\end{itemize}
First, let us prove that $ \mathcal{A}_\sim $ is a well-defined GNFA.
\begin{itemize}
\item The set $ Q_\sim $ is finite because $ \sim $ has finite index.
\item Let us prove that the set $ E_\sim $ is well defined and finite. To prove that $ E_\sim $ is well defined, we only need to show that, if $ \alpha \in \mathcal{W} $ and $ \rho \in \mathcal{K}(\mathcal{W}_\alpha) $, then $ \alpha \rho \in \mathcal{W} $. This follows immediately from the definition of $ \mathcal{W}_\alpha $ because we have $ \rho \in \mathcal{K}(\mathcal{W}_\alpha) \subseteq \mathcal{W}_\alpha $. Now, let us prove that $ E_\sim $ is finite. Fix $ \alpha \in \mathcal{W} $. We need to show that the set of all edges leaving $ [\alpha]_\sim $ is finite. To this end, it will suffice to show that (i) $ \mathcal{K}(\mathcal{W}_\alpha) $ is finite, (ii) the set of all strings labeling some edge leaving $ [\alpha]_\sim $ is $ \mathcal{K}(\mathcal{W}_\alpha) $ and (iii) every $ \rho \in \mathcal{K}(\mathcal{W}_\alpha) $ labels exactly one edge leaving $ [\alpha]_\sim $.
\begin{itemize}
    \item $ \mathcal{K}(\mathcal{W}_\alpha) $ is finite because $ \mathcal{W} $ is locally bounded.
    \item Let us prove that the set of all strings labeling some edge leaving $ [\alpha]_\sim $ is $ \mathcal{K}(\mathcal{W}_\alpha) $. By the definition of $ E_\sim $, a string $ \rho \in \Sigma^* $ labels some edge leaving $ [\alpha]_\sim $ if and only if $ \rho \in \mathcal{K}(\mathcal{W}_{\alpha'}) $ for some $ \alpha' \in \mathcal{W} $ such that $ \alpha \sim \alpha' $. Consequently, we only need to prove that for every $ \alpha' \in \mathcal{W} $ such that $ \alpha \sim \alpha' $ we have $ \mathcal{K}(\mathcal{W}_\alpha) = \mathcal{K}(\mathcal{W}_{\alpha'}) $. Fix $ \alpha' \in \mathcal{W} $ such that $ \alpha \sim \alpha' $. Since $ \sim $ is right-invariant, we have $ \mathcal{W}_\alpha = \mathcal{W}_{\alpha'} $ and so $ \mathcal{K}(\mathcal{W}_\alpha) = \mathcal{K}(\mathcal{W}_{\alpha'}) $.
    \item Fix $ \rho \in \mathcal{K}(\mathcal{W}_\alpha) $. We must prove that $ \rho $ labels exactly one edge leaving $ [\alpha]_\sim $. We know that $ ([\alpha]_\sim, [\alpha \rho]_\sim, \rho) \in E_\sim $, so we only need to prove that every edge labeled with $ \rho $ and leaving $ [\alpha]_\sim $ is equal to $ ([\alpha]_\sim, [\alpha \rho]_\sim, \rho) $. By the definition of $ E_\sim $, every edge labeled with $ \rho $ and leaving $ [\alpha]_\sim $ is equal to $ ([\alpha']_\sim, [\alpha' \rho]_\sim, \rho) $ for some $ \alpha' \in \mathcal{W} $ such that $ \alpha \sim \alpha' $. Consequently, we only need to show that, for every $ \alpha' \in \mathcal{W} $, if $ \alpha \sim \alpha' $, then $ \alpha \rho \sim \alpha' \rho $. This follows immediately because $ \sim $ is right-invariant.
\end{itemize}
\item The initial state $ s_\sim $ is well defined because $ \epsilon \in \mathcal{W} $.
\item The set $ F_\sim $ is well defined because (i) if $ \alpha \in \mathcal{L} $, then $ \alpha \in \mathcal{W} $, so $ [\alpha]_\sim $ is well defined, and (ii) $ \mathcal{L} $ if the union of some $ \sim $-classes, so if for some $ \alpha, \alpha' \in \mathcal{W} $ we have $ \alpha \sim \alpha' $, then $ \alpha \in \mathcal{L} $ if and only if $ \alpha' \in \mathcal{L} $.
\item Let us show that every state in $ Q_\sim $ is reachable from the initial state $ s_\sim $. In other words, we must prove that, for every $ \alpha \in \mathcal{W} $, the state $ [\alpha]_\sim $ is reachable from the state $ [\epsilon]_\sim $. We proceed by induction on $ \alpha $. If $ |\alpha| = 0 $, then $ \alpha = \epsilon $ and so $ [\alpha]_\sim $ is the initial state. Now assume $ |\alpha | > 0 $. Let $ \alpha' $ be the longest strict prefix of $ \alpha $ such that $ \alpha' \in \mathcal{W} $, which must exist because $ \epsilon $ is a strict prefix of $ \alpha $ (we have $ |\alpha | > 0 $) and $ \epsilon \in \mathcal{W} $. Let $ \rho \in \Sigma^+ $ be such that $ \alpha = \alpha' \rho $. Let us prove that $ ([\alpha']_\sim, [\alpha]_\sim, \rho) \in E_\sim $. Since $ \alpha = \alpha' \rho $, we only have to show that $ \rho \in \mathcal{K}(\mathcal{W}_{\alpha'}) $. We have $ \rho \in \mathcal{W}_{\alpha'} $ because $ \alpha = \alpha' \rho \in  \mathcal{W} $, and we have $ \rho \in \mathcal{K}(\mathcal{W}_{\alpha'}) $ because if $ \rho' \in \Sigma^+ $ is a strict prefix of $ \rho $, then $ \rho' \not \in \mathcal{W}_{\alpha'} $, otherwise we would have $ \alpha' \rho' \in \mathcal{W} $, and $ \alpha' \rho' $ would be a strict prefix of $ \alpha $ longer than $ \alpha' $, which contradicts the maximality of $ \alpha' $. By the inductive hypothesis, $ [\alpha']_\sim $ is reachable from the initial state, so $ [\alpha]_\sim $ is also reachable from the initial state because $ ([\alpha']_\sim, [\alpha]_\sim, \rho) \in E_\sim $.
\item Let us prove every state is co-reachable. Fix $ \alpha \in \mathcal{W} $. We must prove that $ [\alpha]_\sim $ is either final, or it allows reaching a final state. In particular, $ \alpha \in \Pref (\mathcal{L}) $, so there exists $ \phi \in \Sigma^* $ such that $ \alpha \phi \in \mathcal{L} $, and so $ \alpha \phi \in \mathcal{W} $. Let $ \phi_1, \dots, \phi_s $ be all distinct prefixes of $ \phi $ such that $ \alpha \phi_i \in \mathcal{W} $ for every $ 1 \le i \le s $, where $ \phi_i $ is a strict prefix of $ \phi_{i + 1} $ for every $ 1 \le i \le s - 1 $. Note that $ s \ge 1 $, $ \phi_1 = \epsilon $ and $ \phi_s = \phi $. The same argument used to prove that every state is reachable from the initial state shows that $ ([\alpha \phi_i]_\sim, [\alpha \phi_{i + 1}], \rho_i) \in E_\sim $ for every $ 1 \le i \le s - 1 $, where $ \rho_i \in \Sigma^+ $ is such that $ \phi_{i + 1} = \phi_i \rho_i $. Since $ \alpha \phi_s = \alpha \phi \in \mathcal{L} $, the state $ [\alpha \phi_s]_\sim $ is final. Therefore, $ [\alpha \phi_i]_\sim $ is either final or it allows reaching a final state for every $ 1 \le i \le s $, and the conclusion follows because $ \alpha = \alpha \phi_1 $. 
\end{itemize}

Next, let us prove that $ \mathcal{A}_\sim $ is a GDFA.
\begin{itemize}
    \item To prove that no edge is labeled with $ \epsilon $, we need to show that $ \epsilon \not \in \mathcal{K}(\mathcal{W}_\alpha) $ for every $ \alpha \in \mathcal{W} $. Fix $ \alpha \in \mathcal{W} $. By definition we have $ \mathcal{W}_\alpha \subseteq \Sigma^+ $, so $ \epsilon \not \in \mathcal{K}(\mathcal{W}_\alpha) $.
    \item Fix $ \alpha \in \mathcal{W} $. We need to show that distinct edges leaving the state $ [\alpha]_\sim $ have distinct labels. This is immediate because we have already proved that the set of all strings labeling some edge leaving $ [\alpha]_\sim $ is $ \mathcal{K}(\mathcal{W}_\alpha) $ and every $ \rho \in \mathcal{K}(\mathcal{W}_\alpha) $ labels exactly one edge leaving $ [\alpha]_\sim $.
    \item Fix $ \alpha \in \mathcal{W} $. We need to show that the set of all strings labeling some edge leaving the state $ [\alpha]_\sim $ is prefix-free. We have already proved that the set of all strings labeling some edge leaving $ [\alpha]_\sim $ is $ \mathcal{K}(\mathcal{W}_\alpha) $, and $ \mathcal{K}(\mathcal{W}_\alpha) $ is prefix-free because $ \mathcal{K}(\mathcal{W}_\alpha) $ is a prefix-free kernel.
\end{itemize}

The next goal is to show that $ \mathcal{A}_\sim $ is a $ \mathcal{W} $-GDFA that recognizes $ \mathcal{L} $. To this end, let us first prove that for every $ \alpha \in \Sigma^* $ and for every $ \beta \in \mathcal{W} $ we have:
\begin{equation}\label{eq:determinization}
    (\alpha \in \mathcal{W(A_\sim)}) \land (J_\alpha = [\beta]_\sim) \Longleftrightarrow (\alpha \in \mathcal{W}) \land (\alpha \sim \beta).
\end{equation}

We proceed by induction on $ |\alpha | $. If $ |\alpha| = 0 $, then $ \alpha = \epsilon $. Notice that $ \epsilon \in \mathcal{W(A_\sim)} $ because $ \epsilon \in J_{s_\sim} $, and $ \epsilon \in \mathcal{W} $. Moreover, for every $ \beta \in \mathcal{W} $ we have $ J_\epsilon = [\beta]_\sim \Longleftrightarrow [\beta]_\sim = s_\sim \Longleftrightarrow  \epsilon \sim \beta $.

Now, assume that $ |\alpha | > 0 $. Note that the inductive hypothesis implies that for every $ \alpha' \in \Sigma^* $ such that $ |\alpha'| < |\alpha| $ we have $ \alpha' \in \mathcal{W(A_\sim)} \Longleftrightarrow \alpha' \in \mathcal{W} $, and, if $ \alpha' \in \mathcal{W(A_\sim)} $, then $ J_{\alpha'} = [\alpha']_\sim $. Indeed, ($ \Longrightarrow $) follows by choosing any $ \beta \in \mathcal{W} $ such that $ J_{\alpha'} = [\beta]_\sim $ (which exists because $ \alpha' \in \mathcal{W(A_\sim)} $) in Equation~\ref{eq:determinization}; ($ \Longleftarrow $) and the equality $ J_{\alpha'} = [\alpha']_\sim $ follow by choosing $ \beta = \alpha' $ in Equation~\ref{eq:determinization}.

Let $ \alpha' $ be the longest strict prefix of $ \alpha $ such that $ \alpha' \in \mathcal{W (A_\sim)} $, which must exist because $ \epsilon $ is a strict prefix of $ \alpha $, being $ |\alpha | > 0 $, and $ \epsilon \in \mathcal{W(A_\sim)} $. Since we know that a string shorter than $ \alpha $ is in $ \mathcal{W(A_\sim)} $ if and only if it is in $ \mathcal{W} $, then $ \alpha' $ is also the longest strict prefix of $ \alpha $ such that $ \alpha' \in \mathcal{W} $. Moreover, we know that $ J_{\alpha '} = [\alpha']_\sim $. Write $ \alpha = \alpha' \rho $, with $ \rho \in \Sigma^+ $.

$ (\Longrightarrow) $ Assume that $ (\alpha \in \mathcal{W(A_\sim)}) \land (J_\alpha = [\beta]_\sim) $. We must prove that $ (\alpha \in \mathcal{W}) \land (\alpha \sim \beta) $. Since $ \alpha', \alpha \in \mathcal{W(A_\sim)} $, $ \alpha' $ is the longest strict prefix of $ \alpha $ such such that $ \alpha' \in \mathcal{W (A_\sim)} $ and $ \alpha = \alpha' \rho $, then $ \rho \in \mathcal{K} (\mathcal{W (A_\sim)}_\alpha) $, so by Remark~\ref{rem:GDFAstringsreachonestate} we have $ (J_{\alpha'}, J_\alpha, \rho) \in E_\sim $. Since $ J_{\alpha'} = [\alpha']_\sim $ and $ J_\alpha = [\beta]_\sim $, we have $ ([\alpha']_\sim, [\beta]_\sim, \rho) \in E_\sim $. The definition of $ E_\sim $ implies $ \alpha = \alpha' \rho \in \mathcal{W} $ and $ \alpha \sim \beta $.

$ (\Longleftarrow) $ Assume that $ (\alpha \in \mathcal{W}) \land (\alpha \sim \beta) $. We must prove that $ (\alpha \in \mathcal{W(A_\sim)}) \land (J_\alpha = [\beta]_\sim) $. Let us prove that $ ([\alpha']_\sim, [\alpha]_\sim, \rho) \in E_\sim $. Since $ \alpha', \alpha \in \mathcal{W} $ and $ \alpha = \alpha' \rho $, we only have to show that $ \rho \in \mathcal{K}(\mathcal{W}_{\alpha'}) $. From  $ \alpha = \alpha' \rho $ we obtain $ \rho \in \mathcal{W}_{\alpha'} $. If $ \rho' \in \Sigma^+ $ is a strict prefix of $ \rho $, then it cannot hold $ \alpha' \rho' \in \mathcal{W} $ by the maximality of $ \alpha' $, so $ \rho \in \mathcal{K}(\mathcal{W}_{\alpha'}) $. From $ J_{\alpha'} = [\alpha']_\sim $, $ ([\alpha']_\sim, [\alpha]_\sim, \rho) \in E_\sim $ and $ \alpha = \alpha' \rho $ we obtain $ \alpha \in \mathcal{W(A_\sim)} $ and $ J_\alpha = [\alpha]_\sim $. Since $ \alpha \sim \beta $, we conclude $ J_\alpha = [\beta]_\sim $.
    
This concludes the proof of Equation~\ref{eq:determinization}.

We are now ready to show that $ \mathcal{A}_\sim $ is a $ \mathcal{W} $-GDFA. In other words, we can now show that $ \mathcal{W}(\mathcal{A}_\sim) = \mathcal{W} $. We have $ (\supseteq) $ because from $ \alpha \in \mathcal{W} $ we obtain $ \alpha \in \mathcal{W}(\mathcal{A}_\sim) $ by choosing $ \beta = \alpha $ in Equation~\ref{eq:determinization}, and we have $ (\subseteq) $ because from $ \alpha \in \mathcal{W}(\mathcal{A}_\sim) $ we obtain $ \alpha \in \mathcal{W} $ by choosing any $ \beta \in \mathcal{W} $ such that $ J_\alpha = [\beta]_\sim $ (which exists because $ \alpha \in \mathcal{W(A_\sim)} $) in Equation~\ref{eq:determinization}.

By choosing $ \beta = \alpha $ in Equation~\ref{eq:determinization}, we also obtain that for every $ \alpha \in \mathcal{W(A_\sim)} = \mathcal{W} $:
\begin{equation}\label{eq:powersetuseful}
    J_\alpha = [\alpha]_\sim.
\end{equation}

We are now ready to show that $ \mathcal{A}_\sim $ recognizes $ \mathcal{L} $, that is, $ \mathcal{L(A_\sim)} = \mathcal{L} $. We have:
\begin{equation*}
\begin{split}
    \mathcal{L(A_\sim)} & = \{\alpha \in \mathcal{W(A_\sim)} \;|\; J_\alpha = [\beta]_\sim \text{ for some $ \beta \in \mathcal{L} $} \} \\
    & = \{\alpha \in \mathcal{W} \;|\; \alpha \sim \beta \text{ for some $ \beta \in \mathcal{L} $} \} = \mathcal{L}
\end{split}
\end{equation*}
where the first equality follows from the definition of $ F_\sim $, the second equality follows from Equation~\ref{eq:determinization}, and in the last equality we have $ (\subseteq) $ because $ \mathcal{L} $ is the union of some $ \sim $-classes, and $ (\supseteq) $ because $ \mathcal{L} \subseteq \mathcal{W} $. Moreover:

\begin{enumerate}
\item The number of states of $ \mathcal{A}_\sim $ is equal to the index of $ \sim $ by the definition of $ Q_\sim $.
\item By Equation~\ref{eq:powersetuseful}, for every $ \alpha, \beta \in \mathcal{W} = \mathcal{W(A_\sim)} $ we have $ \alpha \equiv_{A_\sim} \beta \Longleftrightarrow J_\alpha = J_\beta \Longleftrightarrow [\alpha]_\sim = [\beta]_\sim \Longleftrightarrow \alpha \sim \beta $, so $ \equiv_{A_\sim} $ and $ \sim $ are the same equivalence relation.
\end{enumerate}
Lastly, suppose that $ \mathcal{B} = (Q_\mathcal{B}, E_\mathcal{B}, s_\mathcal{B}, F_\mathcal{B}) $ is a $ \mathcal{W} $-GDFA that recognizes $ \mathcal{L} $. Notice that by Lemma~\ref{lem:simAproperties} we have that $ \equiv_\mathcal{B} $ is right-invariant, $ \equiv_\mathcal{B} $ has finite index, and $ \mathcal{L} $ is the union of some $ \equiv_{\mathcal{B}} $-classes, so $ \mathcal{A_{\equiv_\mathcal{B}}} $ is well defined, and $ \mathcal{W(\mathcal{A_{\equiv_\mathcal{B}}})} = \mathcal{W} = \mathcal{W(B)} $. Let $ \phi: Q_{\equiv_\mathcal{B}} \mapsto Q_\mathcal{B} $ be the function sending the state $ [\alpha]_{\equiv_\mathcal{B}} $ of $ Q_{\equiv_\mathcal{B}} $ into the state $ J_\alpha^\mathcal{B} $ of $ Q_\mathcal{B} $, that is, $ \phi ([\alpha]_{\equiv_\mathcal{B}}) = J_\alpha^\mathcal{B} $ for every $ \alpha \in \mathcal{W} $. Notice that $ \phi $ is well defined and injective because Remark~\ref{rem:equivalenceonGDFA} implies that $ \equiv_\mathcal{B} $ and $ \sim_\mathcal{B} $ are the same equivalence relation, so we have $ [\alpha]_{\equiv_\mathcal{B}} = [\beta]_{\equiv_\mathcal{B}} \Longleftrightarrow \alpha \equiv_\mathcal{B} \beta \Longleftrightarrow \alpha \sim_\mathcal{B} \beta \Longleftrightarrow  J_\alpha^\mathcal{B} = J_\beta^\mathcal{B} $ for every $ \alpha, \beta \in \mathcal{W} $. Let us prove that $ \phi $ determines an isomorphism between $ \mathcal{A_{\equiv_\mathcal{B}}} $ and $ \mathcal{B} $. First, $ \phi $ is surjective because every state of $ \mathcal{B} $ is reachable from the initial state. Next, $ \phi(s_{\equiv_\mathcal{B}}) = \phi([\epsilon]_{\equiv_\mathcal{B}}) = J_\epsilon^\mathcal{B} = s_\mathcal{B} $. Moreover, for every $ \alpha \in \mathcal{W} $ we have $ [\alpha]_{\equiv_\mathcal{B}} \in F_{\equiv_\mathcal{B}} \Longleftrightarrow \alpha \in \mathcal{L} \Longleftrightarrow J_\alpha^\mathcal{B} \in F_\mathcal{B} $. Finally, by Remark~\ref{rem:GDFAstringsreachonestate}, for every $ \alpha, \beta \in \mathcal{W} $ and for every $ \rho \in \Sigma^+ $:
\begin{equation*}
\begin{split}
    ([\alpha]_{\equiv_\mathcal{B}}, [\beta]_{\equiv_\mathcal{B}}, \rho) \in E_{\equiv_\mathcal{B}} & \Longleftrightarrow \alpha \rho \in \mathcal{W} \land \rho \in \mathcal{K}(\mathcal{W}_{\alpha}) \land \alpha \rho \equiv_\mathcal{B} \beta \\
    & \Longleftrightarrow \alpha \rho \in \mathcal{W} \land \rho \in \mathcal{K}(\mathcal{W}_{\alpha}) \land (J_\alpha^\mathcal{B}, J_{\alpha \rho}^\mathcal{B}, \rho) \in E_\mathcal{B} \land J_{\alpha \rho}^\mathcal{B} = J_\beta^\mathcal{B} \\
    & \Longleftrightarrow (J_\alpha^\mathcal{B}, J_\beta^\mathcal{B}, \rho) \in E_\mathcal{B} \Longleftrightarrow (\phi([\alpha]_{\equiv_\mathcal{B}}), \phi([\beta]_{\equiv_\mathcal{B}}), \rho) \in E_\mathcal{B}
\end{split}
\end{equation*}
and we conclude that  $ \mathcal{A_{\equiv_\mathcal{B}}} $ and $ \mathcal{B} $ are isomorphic. \qedhere
\end{proof}

\begin{rem}\label{rem:infinitelymanyedges}
    In the statement of Lemma~\ref{lem:fromrelationtoDFA} we cannot remove the assumption that $ \mathcal{W} $ is locally bounded, because we have shown that if $ \mathcal{A} $ is a GNFA, then $ \mathcal{W(A)} $ is locally bounded (Lemma~\ref{lem:prooflocallybounded}). However, if $ \mathcal{W} $ is not locally bounded, then $ \mathcal{A}_\sim $ is still a well-defined automaton with finitely many states, but it has infinitely many edges. For example, $ \mathcal{W} = \{\epsilon \} \cup  a^*b $ is not locally bounded because (i) $ \mathcal{W}_\epsilon = a^* b $ and (ii) $ \mathcal{K}(\mathcal{W}_\epsilon) = a^* b $ is an infinite set. If $ \mathcal{L} = a^*b $, then $ \mathcal{L} \cup \{\epsilon \} \subseteq \mathcal{W} \subseteq \Pref (\mathcal{L}) $. Moreover, $ \equiv_{\mathcal{L}, \mathcal{W}} $ has finite index (the equivalence classes are $ \{\epsilon \} $ and $ a^* b $), and by Lemma~\ref{lem:nerodegenproperties} we know that $ \equiv_{\mathcal{L}, \mathcal{W}} $ is right-invariant and $ \mathcal{L} $ is the union of some $ \equiv_{\mathcal{L}, \mathcal{W}} $-classes. We conclude that $ \mathcal{A_{\equiv_{\mathcal{L}, \mathcal{W}}}} $ is well defined, but it has infinitely many edges (see Figure~\ref{fig:edgesinfinite}).

\begin{figure}[h!]
\centering
\scalebox{0.8}{
\begin{tikzpicture}[->,>=stealth', semithick, auto, scale=1]
\node[state, initial] (1)    at (0,0)	{$ 1 $};
\node[state, accepting,label=above:{}] (2)    at (6,0)	{$ 2 $};
\draw (1) edge [] node [] {$ b, ab, aab, aaab, \dots $} (2);
\end{tikzpicture}
}
 	\caption{An example where $ \mathcal{W} $ is not locally bounded.}
    \label{fig:edgesinfinite}
\end{figure}
\end{rem}

We can now prove our Myhill-Nerode theorem for generalized automata.

\begin{thm}[Myhill-Nerode theorem for generalized automata]\label{theor:nerodegeneralized}
Let $ \mathcal{L} \subseteq \Sigma^* $ and let $ \mathcal{W} \subseteq \Sigma^* $ be a locally bounded set such that $ \mathcal{L} \cup \{\epsilon \} \subseteq \mathcal{W} \subseteq \Pref (\mathcal{L}) $. The following are equivalent:
\begin{enumerate}
    \item $ \mathcal{L} $ is recognized by a $ \mathcal{W} $-GNFA.
    \item The Myhill-Nerode equivalence $ \equiv_{\mathcal{L}, \mathcal{W}} $ has finite index.
    \item There exists a right-invariant equivalence relation $ \sim $ on $ \mathcal{W} $ of finite index such that $ \mathcal{L} $ is the union of some $ \sim $-classes.
    \item $ \mathcal{L} $ is recognized by a $ \mathcal{W} $-GDFA.    
\end{enumerate}
Moreover, if one of the above statements is true (and so all the above statements are true), then there exists a unique minimal $ \mathcal{W} $-GDFA recognizing $ \mathcal{L} $ (that is, two $ \mathcal{W} $-GDFAs recognizing $ \mathcal{L} $ having the minimum number of states among all $ \mathcal{W} $-GDFAs recognizing $ \mathcal{L} $ must be isomorphic). The minimal $ \mathcal{W} $-GDFA recognizing $ \mathcal{L} $ is $ \mathcal{A_{\equiv_{\mathcal{L}, \mathcal{W}}}} $ as defined in Lemma~\ref{lem:fromrelationtoDFA}, where $ \equiv_{\mathcal{L}, \mathcal{W}} $ is the Myhill-Nerode equivalence on $ \mathcal{L} $ and $ \mathcal{W} $.
\end{thm}

\begin{proof}
    $ (1) \Longrightarrow (2) $ Let $ \mathcal{A} $ be a $ \mathcal{W} $-GDFA recognizing $ \mathcal{L} $. By Lemma~\ref{lem:simAproperties} we have that $ \equiv_\mathcal{A} $ has finite index and it refines $ \equiv_{\mathcal{L}, \mathcal{W}} $, so also $ \equiv_{\mathcal{L}, \mathcal{W}} $ has finite index.
    
    $ (2) \Longrightarrow (3) $ By Lemma~\ref{lem:nerodegenproperties} the desired equivalence relation is $ \equiv_{\mathcal{L}, \mathcal{W}} $.
    
    $ (3) \Longrightarrow (4) $ It follows from Lemma~\ref{lem:fromrelationtoDFA}.

    $ (4) \Longrightarrow (1) $ Every $ \mathcal{W} $-GDFA is a $ \mathcal{W} $-GNFA.

Now, assume that one of the above statements is true (and so all the above statements are true).  Let us prove that the minimal $ \mathcal{W} $-GDFA recognizing $ \mathcal{L} $ is $ \mathcal{A_{\equiv_{\mathcal{L}, \mathcal{W}}}} $ as defined in Lemma~\ref{lem:fromrelationtoDFA}. First, $ \mathcal{A_{\equiv_{\mathcal{L}, \mathcal{W}}}} $ is a well-defined $ \mathcal{W} $-GDFA recognizing $ \mathcal{L} $ because (i) $ \equiv_{\mathcal{L}, \mathcal{W}} $ is right-invariant and $ \mathcal{L} $ is the union of some $ \equiv_{\mathcal{L}, \mathcal{W}} $-classes by Lemma~\ref{lem:nerodegenproperties}, and (ii) $ \equiv_{\mathcal{L}, \mathcal{W}} $ has finite index by one of the statements that we assume to be true. Let $ \mathcal{B} $ be any $ \mathcal{W} $-GDFA recognizing $ \mathcal{L} $ non-isomorphic to $ \mathcal{A_{\equiv_{\mathcal{L}, \mathcal{W}}}} $. We must prove that the number of states on $ \mathcal{A_{\equiv_{\mathcal{L}, \mathcal{W}}}} $ is smaller than the number of states of $ \mathcal{B} $. By Lemma~\ref{lem:fromrelationtoDFA}, $ \mathcal{A}_{\equiv_\mathcal{B}} $ is isomorphic to $ \mathcal{B} $. We know that $ \equiv_\mathcal{B} $ is a refinement of $ \equiv_{\mathcal{L}, \mathcal{W}} $ by Lemma~\ref{lem:simAproperties}, and it must be a strict refinement of $ \equiv_{\mathcal{L}, \mathcal{W}} $, otherwise $ \mathcal{A_{\equiv_{\mathcal{L}, \mathcal{W}}}} $ would be equal to $ \mathcal{A}_{\equiv_\mathcal{B}} $, which is isomorphic to $ \mathcal{B} $, a contradiction. We conclude that the index of $ \equiv_{\mathcal{L}, \mathcal{W}} $ is smaller than the index of $ \equiv_\mathcal{B} $, so again by Lemma~\ref{lem:fromrelationtoDFA} the number of states of $ \mathcal{A_{\equiv_{\mathcal{L}, \mathcal{W}}}} $ is smaller than the number of states of $ \mathcal{A}_{\equiv_\mathcal{B}} $ and so of $ \mathcal{B} $. \qedhere
\end{proof}

The Myhill-Nerode theorem for conventional automata (Theorem~\ref{theor:MyhillNerodetextbook}) is a special case of Theorem~\ref{theor:nerodegeneralized}, because if $ \mathcal{A} $ is an NFA, then $ \mathcal{W(A)} = \Pref (\mathcal{L(A)}) $. Moreover, Theorem~\ref{theor:nerodegeneralized} is consistent with the example in Figure~\ref{fig:nouniqueminimalGDFA}, because, calling $ \mathcal{A}_1 $ and $ \mathcal{A}_2 $ the two GDFAs in Figure~\ref{fig:nouniqueminimalGDFA}, we have shown that $ \mathcal{W}(\mathcal{A}_1) \not = \mathcal{W}(\mathcal{A}_2) $.

\begin{rem}\label{rem:standardmyhill}
    The classical formulation of the Myhill-Nerode theorem is slightly different from the statement of Theorem~\ref{theor:MyhillNerodetextbook} because (i) it does not refer to arbitrary NFAs (it only refers to DFAs) and (ii) it applies to \emph{complete} DFAs.
    
    More precisely, let us consider conventional DFAs, and let us drop the requirement (see Definition~\ref{def:SNFAs}) that every state is co-reachable. Recall that a DFA is \emph{complete} if for every state $ u $ and for every character $ c $  there exists \emph{exactly} (and not at most) one state labeled $ c $ leaving $ u $. Moreover, for every $ \mathcal{L} \subseteq \Sigma^* $ let $ \equiv^\#_\mathcal{L} $ be the \emph{complete} Myhill-Nerode equivalence, that is, the equivalence relation on $ \Sigma^* $ (and not on $ \Pref (\mathcal{L}) $) such that for every $ \alpha, \beta \in \Sigma^* $ we have $ \alpha \equiv^\#_\mathcal{L} \beta $ if and only if for every $ \phi \in \Sigma^* $ we have $ \alpha \phi \in \mathcal{L} \Longleftrightarrow \beta \phi \in \mathcal{L} $. Then, according to the classical formulation of the Myhill-Nerode theorem, the following are equivalent: (i) The complete Myhill-Nerode equivalence $ \equiv_{\mathcal{L}}^\# $ has finite index; (ii) there exists a right-invariant equivalence relation $ \sim $ on $ \Sigma^* $ of finite index such that $ \mathcal{L} $ is the union of some $ \sim $-classes; (iii) $ \mathcal{L} $ is recognized by a complete DFA. Moreover, if one of the above statements is true, then up to isomorphism there exists a unique minimal complete DFA recognizing $ \mathcal{L} $.

    The proof of this result is entirely analogous to the proof of Theorem~\ref{theor:MyhillNerodetextbook}. Intuitively, since we dropped the requirement that every state is co-reachable, we can turn any non-complete DFA into a complete DFA recognizing the same language by adding a \emph{sink} state, and complete DFAs correspond to equivalence relations on $ \Sigma^* $ because, starting from the initial state, one can read \emph{every} string in $ \Sigma^* $.
\end{rem}

\section{Generalized Automata: Pattern Matching and Compression}\label{sec:WheelerGDFAs}

In this section, we prove Theorem~\ref{theor:fmindexintroduction} and some related results. In order to present the main ideas, it will suffice to consider GDFAs. The more general case of GNFAs will be considered in Appendix~\ref{app:from gdfas to gnfas}.

\subsection{Preliminary Definitions}

To introduce Wheeler generalized automata, we will first present some preliminary definitions.

Let $ V $ be a set. We say that a (binary) relation $ \le $ on $ V $ is a \emph{partial} order if $ \le $ is reflexive, antisymmetric and transitive. A partial order $ \le $ is a \emph{total} order if for every $ u, v \in V $ we have $ (u \le v) \lor (v \le u) $. We say that $ U \subseteq V $ is \emph{$ \le $-convex} if for every $ u, v, z \in V $, if $ u \le v \le z $ and $ u, z \in U $, then $ v \in U $. For every $ u, v \in V $, we write $ u < v $ if $ (u \le v) \land (u \not = v) $.

The most important data structures for solving pattern matching queries on compressed strings (such as the suffix array~\cite{manber1993suffix}, the Burrows-Wheeler Transform~\cite{burrows1994} and the FM-index~\cite{ferraginajacm2005}) are closely related to the idea of \emph{sorting} strings. Consequently, as customary in the literature on Wheeler automata~\cite{alanko2020, cotumaccio2021}, we assume that there exists a fixed total order $ \preceq $ on the alphabet $ \Sigma $ (in our examples, we always assume $ a \prec b \prec c \prec \dots $), and $ \preceq $ is extended \emph{co-lexicographically} to $ \Sigma^* $ (that is, for every $ \alpha, \beta \in \Sigma^* $ we have $ \alpha \prec \beta $ if and only if the reverse string $ \alpha^R $ is lexicographically smaller than the reverse string $ \beta^R $).

Our data structure results (including Theorem~\ref{theor:fmindexintroduction}) hold in the \emph{word RAM model} with words of size $ w \in  \Theta(\log N) $ bits, where $ N $ is the input size. When we describe our data structures in detail, we assume to be working with integer alphabets of the form $ \Sigma = \{0, 1 \dots, \sigma - 1 \} $, where $ \preceq $ is the usual total order such that $ 0 \prec 1 \prec \dots \prec \sigma - 1 $. One of our results (Theorem~\ref{theor:polynomialdecidingwheeler}) also holds in the \emph{comparison model}, which is less restrictive than the word RAM model. In the comparison model, we only assume that we can compare two elements of the alphabet $ \Sigma $ (with respect to $ \preceq $) in $ O(1) $ time, but we make no other assumption on the elements of the alphabet (in particular, $ \Sigma $ need not be an integer alphabet).

All logarithms are in base 2.

\subsection{Wheeler GDFAs}\label{sec:wheelerdefinitionsubsection}

Let us define Wheeler GDFAs. Let $ \mathcal{A} = (Q, E, s, F) $ be a GDFA. Let $ \preceq_\mathcal{A} $ be the reflexive relation on $ Q $ such that, for every $ u, v \in Q $ with $ u \not = v $, we have $ u \prec_\mathcal{A} v $ if and only if $ (\forall \alpha \in I_u)(\forall \beta \in I_v)(\alpha \prec \beta) $. Since each $ I_u $ is nonempty, it is immediate to realize that $ \preceq_\mathcal{A} $ is a partial order, but in general it is not a total order. We can then give the following definition (see Figure~\ref{fig:examwheel} for an example).

\begin{defi}\label{def:WheelerGDFA}
    Let $ \mathcal{A} = (Q, E, s, F) $ be a GDFA. We say that $ \mathcal{A} $ is \emph{Wheeler} if $ \preceq_\mathcal{A} $ is a total order.
\end{defi}

\begin{figure}[h!]
     \centering
        \scalebox{.8}{
        \begin{tikzpicture}[->,>=stealth', semithick, auto, scale=1]
\node[state, initial] (1)    at (0,0)	{$ u_1 $};
\node[state, accepting] (2)    at (1.5,1.5)	{$ u_2 $};
\node[state, accepting] (3)    at (1.5,-1.5)	{$ u_3 $};
\draw (1) edge [] node [] {$ ab, b $} (2);
\draw (1) edge [] node [] {$ ac, c $} (3);
\draw (2) edge [loop right] node [] {$ b $} (2);
\draw (3) edge [loop right] node [] {$ bc $} (3);
\end{tikzpicture}
}

	\caption{A Wheeler GDFA $ \mathcal{A} $. The states are numbered following the total order $ \preceq_\mathcal{A} $.}
 \label{fig:examwheel}
\end{figure}  

If $ \mathcal{A} $ is a conventional DFA, it is not immediately clear that Definition~\ref{def:WheelerGDFA} is equivalent to the \emph{local} definition of Wheeler DFA commonly used in the literature~\cite{alanko2020, conte2023, cotumacciospire2023}. According to the local definition, a DFA $ \mathcal{A} = (Q, E, s, F) $ is Wheeler if there exists a total order $ \le $ on $ Q $ such that (i) $ s $ comes first in the total order, (ii) for every $ (u', u, a), (v', v, b) \in E $, if $ u < v $, then $ a \preceq b $ and (iii) for every $ (u', u, a), (v', v, a) \in E $, if $ u < v $, then $ u' < v' $. Alanko et al.~\cite[Corollary 3.1]{alanko2020} proved that, if such a total order $ \le $ exists, then it is unique and it is equal to $ \preceq_\mathcal{A} $. This means that if a DFA $ \mathcal{A} $ is Wheeler according to the local definition, then $ \mathcal{A} $ is Wheeler in the sense of Definition~\ref{def:WheelerGDFA}. To show the equivalence of two definitions, we need to prove that if a DFA $ \mathcal{A} $ is Wheeler in the sense of Definition~\ref{def:WheelerGDFA}, then $ \mathcal{A} $ is Wheeler according to the local definition. In other words, we need to prove that if $ \preceq_\mathcal{A} $ is a total order, then it satisfies properties (i), (ii), (iii). This follows as a special case of the following lemma, which applies to arbitrary GDFAs (not only DFAs).

\begin{lem}\label{lem:local GDFA}
    Let $ \mathcal{A} = (Q, E, s, F) $ be a Wheeler GDFA. Then:
    \begin{enumerate}
        \item $ s $ comes first in the total order $ \preceq_\mathcal{A} $.
        \item For every $ (u', u, \rho), (v', v, \rho') \in E $, if $ u \prec_\mathcal{A} v $ and $ \rho' $ is not a strict suffix of $ \rho $, then $ \rho \preceq \rho' $.
        \item For every $ (u', u, \rho), (v', v, \rho) \in E $, if $ u \prec_\mathcal{A} v $, then $ u' \prec_\mathcal{A} v' $.
    \end{enumerate}
\end{lem}

\begin{proof}
    \begin{enumerate}
        \item Let $ u \in Q \setminus \{s \} $. We must prove that $ s \prec_\mathcal{A} u $. Since $ \preceq_\mathcal{A} $ is a total order, we only need to prove that $ u \not \prec_\mathcal{A} s $. This follows from the fact that for every $ \alpha \in I_u $ we have $ \epsilon \prec \alpha $, and $ \epsilon \in I_s $.

        \item If $ \rho = \rho' $ or $ \rho $ is a strict suffix of $ \rho' $, then the conclusion is immediate, so we can assume that $ \rho \not = \rho' $, $ \rho $ is not a strict suffix of $ \rho' $, and $ \rho' $ is not a strict suffix of $ \rho $. Let $ \alpha' \in I_{u'} $ and $ \beta' \in I_{v'} $. Then, $ \alpha' \rho \in I_u $ and $ \beta' \rho' \in I_v $. Since $ u \prec_\mathcal{A} v $, then $ \alpha' \rho \prec \beta' \rho' $, Since $ \rho \not = \rho' $, $ \rho $ is not a strict suffix of $ \rho' $, and $ \rho' $ is not a strict suffix of $ \rho $, we conclude $ \rho \prec \rho' $.

        \item Let $ \alpha' \in I_{u'} $ and $ \beta' \in I_{v'} $. We must prove that $ \alpha' \prec \beta' $. From $ (u', u, \rho), (v', v, \rho) \in E $ we obtain $ \alpha' \rho \in I_u $ and $ \beta' \rho \in I_v $, so from $ u \prec_\mathcal{A} v $ we obtain $ \alpha' \rho \prec \beta' \rho $ and thus $ \alpha' \prec \beta' $. \qedhere
    \end{enumerate}
\end{proof}

In point 2 of Lemma~\ref{lem:local GDFA} we cannot remove the assumption that $ \rho' $ is not a strict suffix of $ \rho $. For example, if in Figure~\ref{fig:basiccounterexampleswheeler} we consider the Wheeler GDFA $ \mathcal{A} = (Q, E, s, F) $ on the left, then we have that $ (u_1, u_2, ba), (u_4, u_3, a) \in E $, $ u_2 \prec_\mathcal{A} u_3 $, $ a $ is a strict suffix of $ ba $ and $ a \prec ba $. Consequently, if we aim to extend the local definition of Wheeler DFA to Wheeler GDFAs, we should keep the same assumption. However, the resulting local definition would not be equivalent to Definition~\ref{def:WheelerGDFA} (in other words, we cannot prove a converse of Lemma~\ref{lem:local GDFA} that extends the result by Alanko et al. mentioned immediately before the statement of Lemma~\ref{lem:local GDFA}). For example, if in Figure~\ref{fig:basiccounterexampleswheeler} we consider the GDFA $ \mathcal{A} = (Q, E, s, F) $ on the right, then (i) $ s $ comes first in the total order $ \le $, (ii) for every $ (u', u, \rho), (v', v, \rho') \in E $, if $ u < v $ and $ \rho' $ is not a strict suffix of $ \rho $, then $ \rho \preceq \rho' $ and (iii) for every $ (u', u, \rho), (v', v, \rho) \in E $, if $ u < v $, then $ u' < v' $; however, $ \mathcal{A} $ is not Wheeler because $ u_2 $ and $ u_3 $ are not $ \preceq_\mathcal{A} $-comparable, since $ b \prec c \prec ac $, $ c \in I_{u_3} $ and $ b, ac \in I_{u_2} $. This means that we cannot extend the local definition of Wheeler DFA to Wheeler GDFAs.

\begin{figure}
     \centering
     \begin{subfigure}[b]{0.49\textwidth}
        \centering
        \scalebox{.8}{
        \begin{tikzpicture}[->,>=stealth', semithick, auto, scale=1]
\node[state, initial] (1)    at (0,0)	{$ u_1 $};
\node[state, accepting] (2)    at (1.5,1.5)	{$ u_2 $};
\node[state, accepting] (3)    at (3,0)	{$ u_3 $};
\node[state, ] (4)    at (1.5,-1.5)	{$ u_4 $};
\draw (1) edge [] node [] {$ ba $} (2);
\draw (1) edge [] node [] {$ c $} (4);
\draw (4) edge [] node [] {$ a $} (3);
\end{tikzpicture}
}
     \end{subfigure}
     \begin{subfigure}[b]{0.49\textwidth}
        \centering
               \scalebox{.8}{
        \begin{tikzpicture}[->,>=stealth', semithick, auto, scale=1]
\node[state, initial] (1)    at (0,0)	{$ u_1 $};
\node[state, accepting] (2)    at (1.5,1.5)	{$ u_2 $};
\node[state, accepting] (3)    at (1.5,-1.5)	{$ u_3 $};
\draw (1) edge [] node [] {$ ac, b $} (2);
\draw (1) edge [] node [] {$ c $} (3);
\end{tikzpicture}
        }

     \end{subfigure}
	\caption{\emph{Left:} A Wheeler GDFA $ \mathcal{A} = (Q, E, s, F) $ showing that in point 2 of Lemma~\ref{lem:local GDFA} we cannot remove the assumption that $ \rho' $ is not a strict suffix of $ \rho $. The states are numbered following the total order $ \preceq_\mathcal{A} $. \emph{Right:} A GDFA $ \mathcal{A} = (Q, E, s, F) $ showing that we cannot extend the local definition of Wheeler DFA to Wheeler GDFAs. The states are numbered following a total order $ \le $.}
 \label{fig:basiccounterexampleswheeler}
\end{figure} 

Let us discuss one more difference between Wheeler GDFAs and conventional Wheeler DFAs: Wheeler GDFAs are more expressive than Wheeler DFAs (that is, Wheeler GDFAs recognize more regular languages). Consider the Wheeler GDFA $ \mathcal{A} $ consisting of a single state, both initial and final, with a self-loop labeled $ aa $. Then, $ \mathcal{A} $ recognizes $ \mathcal{L} = \{a^{2n} \;|\; n \ge 0 \} $, which is the classic example of a regular language that is not star-free~\cite{straubing2012finite}. However, $ \mathcal{L} $ is not recognized by any conventional Wheeler automata because every language recognized by a Wheeler automaton must be star-free~\cite{alanko2021}.

\subsection{Deciding Wheelerness}\label{sec:decidingwheelerness}

 The local definition of Wheeler DFA easily implies that the problem of deciding whether a given DFA is Wheeler can be solved in polynomial time~\cite{alanko2020}, but since we do not have a local definition of Wheeler GDFA, it is not clear whether the corresponding problem on GDFAs is also solvable in polynomial time (and we saw in the introduction that computational problems on generalized automata are usually hard). However, in Theorem~\ref{theor:polynomialdecidingwheeler} below, we prove that the problem is still solvable in polynomial time by reducing it to the problem of computing the \emph{partial} order $ \preceq_\mathcal{A^*} $ on a conventional DFA $ \mathcal{A^*} $ equivalent to a given GDFA $ \mathcal{A} $.

To present our reduction, we need to recall \emph{tries} and their properties. Let $ C \subseteq \Sigma^+ $ be a nonempty finite set, and assume that $ C $ is prefix-free. The \emph{trie} $ \mathcal{T}_C $ of $ C $ (also known as the \emph{prefix tree} of $ C $~\cite{de1959file, fredkin1960trie}, see~\cite{navarro2016} for a modern discussion) is the unique rooted directed tree such that (i) every edge is labeled with a character, (ii) two edges leaving the same node have distinct label, (iii) $ \mathcal{T}_C $ contains $ |C| $ leaves and (iv) for every leaf $ z $, the string $ \rho $ that can be read following the unique path from the root to $ z $ is in $ C $ (see Figure~\ref{fig:trieexample} for an example). For every node $ u $ of $ \mathcal{T}_C $, let $ \tau_u $ be the string that can be read by following the unique path from the root to $ u $ (in particular, $ \tau_u \in C $ if and only if $ u $ is a leaf of $ \mathcal{T}_C $). For example, in Figure~\ref{fig:trieexample} we have $ \tau_5 = ac $. If $ C = \emptyset $, we assume that $ \mathcal{T}_C $ consists of a single node and no edges.

Assume that we are given $ h $ pairwise distinct strings $ \rho_1 $, $ \rho_2 $, $ \dots $, $ \rho_h $ in \emph{lexicographic} order (not in co-lexicographic order), that is, assume that $ \rho_i $ is lexicographically smaller than $ \rho_{i + 1} $ for every $ 1 \le i \le h - 1 $. Assume also that we know that the set $ C = \{\rho_1, \rho_2, \dots, \rho_h \} $ is prefix-free. Let us show that in $ O(\sum_{i = 1}^h|\rho_i|) $ time we can (i) build (the adjacency-list representation of) $ \mathcal{T}_C $ and (ii) associate to every $ 1 \le i \le h $ the unique leaf $ u $ of $ \mathcal{T}_C $ such that $ \tau_u = \rho_i $. We can identify each node $ u $ of $ \mathcal{T}_C $ with the triple $ \phi_u = (k_u, \ell_u, r_u) $, where $ k_u $ is the distance of $ u $ from the root and $ 1 \le \ell_u \le r_u \le h $ are the two integers such that for every $ 1 \le i \le h $ we have that $ \tau_u $ is a prefix of $ \rho_i $ if and only if $ \ell_u \le i \le r_u $. For example, in Figure~\ref{fig:trieexample} we have $ \tau_2 = a $ and $ \phi_2 = (1, 1, 2) $. Note that for every node of $ u $ of $ \mathcal{T}_C $ we have $ \rho_i[1, k_u] = \tau_u $ if and only if $ \ell_u \le i \le r_u $. Moreover, for every node $ u $ of $ \mathcal{T}_C $, we have that $ u $ is a leaf if and only if (i) $ \ell_u = r_u $ and (ii) $ |\rho_{\ell_u}| = k_u $; if $ u $ is not a leaf, then $ |\rho_i| \ge k_u + 1 $ for every $ \ell_u \le i \le r_u $ (because $ C $ is prefix-free). We build $ \mathcal{T}_C $ using a queue. At any time, each element in the queue is equal to $ \phi_u $ for some node $ u $. At the beginning of the algorithm, we add $ (0, 1, h) $ to $ \mathcal{T}_C $ (which corresponds to the root of $ \mathcal{T}_C $), and we enqueue $ (0, 1, h) $. We now process all elements of the queue until it becomes empty. Assume that we pop $ (k, \ell, r) $ from the queue. If $ (k, \ell, r) $ corresponds to a leaf (which can be checked through the characterization mentioned earlier), we record that $ \ell $ is associated with the node $ (k, \ell, r) $, and we pop the next element of the queue. Now, assume that $ (k, \ell, r) $ is not a leaf. Then, we have $ |\rho_i| \ge k + 1 $ for every $ \ell \le i \le r $. Our goal is to determine the children of $ (k, \ell, r) $. We scan $ \rho_\ell [k + 1] $, $ \rho_{\ell + 1}[k + 1] $, $ \dots $, $ \rho_{r - 1} [k + 1] $, $ \rho_r [k + 1] $, and we compute the set $ D = \{\ell \} \cup \{\ell + 1 \le i \le r \;|\; \rho_i[k + 1] \not = \rho_{i - 1}[k + 1] \} $. Let $ D = \{i_1, i_2, \dots, i_q \} $, with $ q \ge 1 $ and $ i_1 < i_2 < \dots < i_q $. Then, for every $ 1 \le j \le q $, we add the new node $ (k + 1, i_j, i_{j + 1} - 1) $ to $ \mathcal{T}_C$  (where $ i_{q + 1} = r + 1 $), we add an edge from $ (k, \ell, r) $ to $ (k + 1, i_j, i_{j + 1} - 1) $ labeled $ \rho_{i_j}[k + 1] $, and we enqueue $ (k + 1, i_j, i_{j + 1} - 1) $. We can now pop the next element of the queue. When the queue is empty, we have correctly computed $ \mathcal{T}_C $ in $ O(\sum_{i = 1}^h|\rho_i|) $ time, and we have associated to every $ 1 \le i \le h $ the unique leaf $ u $ of $ \mathcal{T}_C $ such that $ \tau_u = \rho_i $.

\begin{figure}
        \centering
        \scalebox{.8}{
        \begin{tikzpicture}[->,>=stealth', semithick, auto, scale=1]
\node[state] (1)    at (0,0)	{$ 1 $};
\node[state] (2)    at (2, 1)	{$ 2 $};
\node[state] (3)    at (2, -1)	{$ 3 $};
\node[state] (4)    at (4, 2)	{$ 4 $};
\node[state] (5)    at (4, 0)	{$ 5 $};
\draw (1) edge [] node [] {$ a $} (2);
\draw (1) edge [] node [] {$ b $} (3);
\draw (2) edge [] node [] {$ a $} (4);
\draw (2) edge [] node [] {$ c $} (5);
\end{tikzpicture}
}
\caption{The trie of the prefix-free set $ C = \{aa, ac, b \} $.}
\label{fig:trieexample}
\end{figure} 

    We are now ready to prove Theorem~\ref{theor:polynomialdecidingwheeler}.

\begin{thm}\label{theor:polynomialdecidingwheeler}
    Let $ \mathcal{A} = (Q, E, s, F) $ be a GDFA, and let $ \mathfrak{e} $ be the total length of all edge labels. In $ O(\mathfrak{e} \log \mathfrak{e}) $ time we can decide whether $ \mathcal{A} $ is Wheeler and, if so, we can compute $ \preceq_\mathcal{A} $ (that is, we can sort the elements of $ Q $ with respect to $ \preceq_\mathcal{A} $). The bound $ O(\mathfrak{e} \log \mathfrak{e}) $ is also true in the comparison model.
\end{thm}

\begin{proof}
    In the rest of the proof, we can assume without loss of generality that the alphabet $ \Sigma $ is an \emph{integer} alphabet (that is, $ \Sigma = \{0, 1, \dots, \sigma - 1 \} $, with $ \sigma = |\Sigma| $) and is \emph{effective} (that is, every character in $ \Sigma $ occurs in at least one edge label, and in particular $ \sigma \le \mathfrak{e} $). Indeed, even in the more general case of the comparison model, in $ O(\mathfrak{e} \log \mathfrak{e}) $ time we can sort the multiset of size $ \mathfrak{e} $ consisting of all characters appearing in some edge label via any comparison-based sorting algorithm (e.g., merge sort, see~\cite{cormen2022introduction}) and then in $ O(\mathfrak{e}) $ time we can replace each character with its rank in the sorted list of all characters, which does not affect the partial order $ \preceq_\mathcal{A} $. 
    
    Next, we can assume without loss of generality that, for every $ u \in V $, all the edges leaving $ u $ are sorted with respect to the lexicographic order of their labels. Indeed, if the edges are not sorted, we can sort them by first comparing their start states and then comparing their labels in lexicographic order. This can be achieved in $ O(|Q| + \sigma + \mathfrak{e}) = O(\mathfrak{e}) $ time via radix sort~\cite{hopcroft1983data, paige1987three} (recall that $ \sigma \le \mathfrak{e} $ because the alphabet is effective and $ |Q| \le |E| + 1 \le \mathfrak{e} + 1 $ because (i) every state is reachable from the initial state and (ii) every edge label is a nonempty string by the definition of GDFA).
    
    The proof of the lemma consists of two main steps. In the first step, we build a conventional DFA $ \mathcal{A}^* $ starting from $ \mathcal{A} $, and in the second step we delete some edges from $ \mathcal{A}^* $, thus obtaining a simpler DFA $ \mathcal{A}' $ in which, for every state $ u $, there exists exactly one string that can be read from the initial state to $ u $. The idea is that we can use $ \mathcal{A}' $ to decide whether the original GDFA $ \mathcal{A} $ is Wheeler.
    
    For every $ u \in Q $, let $ C_u $ be the prefix-free set of all strings labeling edges leaving $ u $, and let $ \mathcal{T}_{C_u} $ be the trie of $ C_u $. In particular, if $ u $ has no outgoing edges, then $ \mathcal{T}_{C_u} $ consists of a single node and no edges. Note that the number of edges in $ \mathcal{T}_{C_u} $ is upper-bounded by the sum of the lengths of all edges leaving $ u $.

\begin{figure}
     \centering
     \begin{subfigure}[b]{0.33\textwidth}
        \centering
        \scalebox{.8}{
        \begin{tikzpicture}[->,>=stealth', semithick, auto, scale=1]
\node[state, initial] (1)    at (0,0)	{$ u_1 $};
\node[state, accepting] (2)    at (1.5,1.5)	{$ u_2 $};
\node[state, accepting] (3)    at (1.5,-1.5)	{$ u_3 $};
\draw (1) edge [] node [] {$ ab, b $} (2);
\draw (1) edge [] node [] {$ ac, c $} (3);
\draw (2) edge [loop right] node [] {$ b $} (2);
\draw (3) edge [loop right] node [] {$ bc $} (3);
\end{tikzpicture}
}

     \end{subfigure}
     \begin{subfigure}[b]{0.33\textwidth}
        \centering
        \scalebox{.8}{
        \begin{tikzpicture}[->,>=stealth', semithick, auto, scale=1]
\node[state, initial] (1)    at (0,0)	{$ u_1 $};
\node[state] (4) at (1.5,0)	{};
\node[state, accepting] (2)    at (3,1.5)	{$ u_2 $};
\node[state, accepting] (3)    at (3,-1.5)	{$ u_3 $};
\node[state] (5)    at (4.5,-1.5)	{};
\draw (1) edge [] node [] {$ a $} (4);
\draw (4) edge [] node [] {$ b $} (2);
\draw (4) edge [] node [] {$ c $} (3);
\draw (1) edge [bend right] node [] {$ c $} (3);
\draw (1) edge [bend left] node [] {$ b $} (2);
\draw (2) edge [loop right] node [] {$ b $} (2);
\draw (3) edge [bend left] node [] {$ b $} (5);
\draw (5) edge [bend left] node [] {$ c $} (3);
\end{tikzpicture}
}
     \end{subfigure}
          \begin{subfigure}[b]{0.33\textwidth}
        \centering
        \scalebox{.8}{
        \begin{tikzpicture}[->,>=stealth', semithick, auto, scale=1]
\node[state, initial, accepting] (1)    at (0,0)	{$ u_1 $};
\node[state, accepting] (4) at (1.5,0)	{};
\node[state, accepting] (2)    at (3,1.5)	{$ u_2 $};
\node[state, accepting] (3)    at (3,-1.5)	{$ u_3 $};
\node[state, accepting] (5)    at (4.5,-1.5)	{};
\draw (1) edge [] node [] {$ a $} (4);
\draw (1) edge [bend right] node [] {$ c $} (3);
\draw (1) edge [bend left] node [] {$ b $} (2);
\draw (3) edge [bend left] node [] {$ b $} (5);
\end{tikzpicture}
}
     \end{subfigure}
	\caption{\emph{Top left:} The GDFA $ \mathcal{A} $ of Figure~\ref{fig:examwheel}. \emph{Top right:} The DFA $ A^* $ built starting from $ \mathcal{A} $ in the proof of Theorem~\ref{theor:polynomialdecidingwheeler}. \emph{Bottom:} An arborescence $ \mathcal{A}' $ obtained from $ A^* $.}
 \label{fig:fromDFAtoGDFA}
\end{figure}

    We define $ \mathcal{A}^* = (Q^*, E^*, s^*, F^*) $ as follows (see Figure~\ref{fig:fromDFAtoGDFA} for an example). We first consider the trie $ \mathcal{T}_{C_u} $ for every $ u \in Q $. Then, for every $ (u', u, \rho) \in E $, we pick the unique leaf $ v $ of $ \mathcal{T}_{C_{u'}} $ such that $ \tau_v = \rho $ (that is, the leaf associated to $ \rho $), we remove the leaf $ v $, and we redirect the unique edge of $ \mathcal{T}_{C_{u'}} $ reaching $ v $ to the root of $ \mathcal{T}_{C_u} $. The initial state $ s^* $ is the root of $ \mathcal{T}_{C_s} $ and a state is in $ F^* $ if and only if it is the root of a trie $ \mathcal{T}_{C_u} $ for some $ u \in F $. By construction, $ \mathcal{A}^* $ is a DFA (in particular, every state of $ \mathcal{A}^* $ is both reachable and co-reachable). If we identify each $ u \in Q $ with the root of $ \mathcal{T}_{C_u} $ (which is a state in $ Q^* $), then we have $ I_u^\mathcal{A} = I_u^\mathcal{A^*} $ for every $ u \in Q $, so we conclude $ \mathcal{L(A^*)} = \mathcal{L(A)} $. Moreover, $ |Q^*| - 1 \le |E^*| \le \mathfrak{e} $ because (i) every state of $ Q^* $ is reachable from $ s^* $ and (ii) for every $ u \in V $ the number of edges in the trie $ \mathcal{T}_{C_u} $ of $ C_u $ is upper-bounded by the sum of the lengths of all edges leaving $ u $.
    
    Let us show that we can build $ \mathcal{A}^* $ in $ O(\mathfrak{e}) $ time. For every $ u \in Q $, all the edges leaving $ u $ are sorted with respect to the lexicographic order of their labels, so (as discussed before the statement of Theorem~\ref{theor:polynomialdecidingwheeler}) we can build all the $ \mathcal{T}_{C_u } $'s (including the $ \mathcal{T}_{C_u } $'s for which $ u $ has no outgoing edges) in $ O(|Q| + \mathfrak{e}) = O(\mathfrak{e}) $ time, and for every $ u \in V $ and for every $ \rho \in C_u $ we record the corresponding leaf of $ \mathcal{T}_{C_u} $. Through this information, we can correctly redirect the edge reaching each leaf in $ O(\mathfrak{e}) $ time, and within the same time bound we can store $ F^* $.

    Since $ I_u^\mathcal{A} = I_u^\mathcal{A^*} $ for every $ u \in Q $, we conclude that $ \mathcal{A} $ is Wheeler if and only if the restriction of the partial order $ \preceq_{\mathcal{A}^*} $ to $ Q $ is a total order.

    Consider the DFA $ \mathcal{A}' = (Q', E', s', F') $ obtained from $ \mathcal{A}^* $ in $ O(|E^*|) = O(\mathfrak{e}) $ time as follows (see Figure~\ref{fig:fromDFAtoGDFA}). We define $ Q' = Q^* $, $ s' = s^* $ and $ F' = Q' $. To define $ E' \subseteq E^* $, we navigate $ \mathcal{A}^* $ starting from $ s^* $ (e.g., via a depth-first search or a breadth-first search) and we visit all states of $ \mathcal{A}^* $, discarding all edges that reach a state that we have already visited (in other words, $ \mathcal{A}' $ is any \emph{arborescence} of $ \mathcal{A}^* $). Then, for every $ u \in Q' $ the set $ I_u^{\mathcal{A}'} $ contains exactly one string $ \gamma_u $ (because $ s' $ has no incoming edges, all the remaining states of $ Q' $ have exactly one incoming edge, and every state is reachable from $ s' $), and $ \gamma_u \in I_u^{\mathcal{A}^*} $. Notice that the $ \gamma_u $'s are pairwise distinct because $ \mathcal{A}' $ is a DFA. Since $ |I_u^{\mathcal{A}'}| = 1 $ for every $ u \in Q' $, then $ \mathcal{A}' $ is Wheeler. Moreover, we can sort the elements of $ Q' $ with respect to the $ \gamma_u $'s (which yields the total order $ \preceq_{\mathcal{A}'} $) in $ O(|Q^*|) = O(\mathfrak{e}) $ time, as shown by Ferragina et al. in their construction algorithm for the XBWT~\cite{ferraginajacm2009} (a similar reduction to the construction algorithm for the XBWT is used in the algorithm for deciding whether a conventional DFA is Wheeler, see~\cite{alanko2020}). Then, in $ O(|Q^*|) = O(\mathfrak{e}) $ time we compute the restriction of $ \preceq_{\mathcal{A}'} $ to $ Q $. In other words, we now know the enumeration $ q_1 $, $ q_2 $, $ \dots $, $ q_{|Q|} $ of $ Q $ such that $ q_1 \prec_{\mathcal{A}'} q_2 \prec_{\mathcal{A}'} q_3 \prec_{\mathcal{A}'} \dots \prec_{\mathcal{A}'} q_{|Q|} $.
    
    For every $ u \in Q^* $ (and in particular, for every $ u \in Q $) we have $ \gamma_u \in I_u^{\mathcal{A}^*} $. Consequently, for every $ u, v \in Q $, if $ u \preceq_{\mathcal{A}^*} v $, then $ u \preceq_{\mathcal{A}'} v $. Hence, the following statements are equivalent: (i) $ \mathcal{A} $ is Wheeler; (ii) the restriction of the partial order $ \preceq_{\mathcal{A}^*} $ to $ Q $ is a total order; (iii) $ q_i \prec_{\mathcal{A}^*} q_{i + 1} $ for every $ 1 \le i \le |Q| - 1 $. Moreover, if $ \mathcal{A} $ is Wheeler, then the enumeration $ q_1 $, $ q_2 $, $ \dots $, $ q_{|Q|} $ yields $ \preceq_{\mathcal{A}} $. To conclude the proof, we only have to show how to check whether $ q_i \prec_{\mathcal{A}^*} q_{i + 1} $ for every $ 1 \le i \le |Q| - 1 $.
    
    We know that $ \mathcal{A}^* $ is a conventional DFA, so we can compute the partial order $ \preceq_{\mathcal{A}^*} $ in polynomial time~\cite{cotumaccio2021, kim2023, cotumacciorecursive, beckeresa}. More precisely, the algorithm in~\cite{beckeresa} runs in $ O(|E^*| \log |Q^*|) $ time (and so $ O(\mathfrak{e} \log \mathfrak{e}) $ time), returning a data structure that in $ O(1) $ time supports the following query: given two distinct states $ u, v \in Q^* $, decide whether (i) $ u \prec_{\mathcal{A}^*} v $, (ii) $ v \prec_{\mathcal{A}^*} u $ or (iii) $ u $ and $ v $ are not $ \preceq_{\mathcal{A}^*} $-comparable. Hence, in $ O(|Q|) = O(\mathfrak{e}) $ time we can check whether $ q_i \prec_{\mathcal{A}^*} q_{i + 1} $ for every $ 1 \le i \le |Q| - 1 $. \qedhere
\end{proof}

Theorem~\ref{theor:polynomialdecidingwheeler} shows that, if $ \mathcal{A} $ is Wheeler, then we can compute $ \preceq_\mathcal{A} $ (as a linear sequence) in $ O(\mathfrak{e} \log \mathfrak{e}) $ time. In the comparison model, this bound is optimal: we cannot compute $ \preceq_\mathcal{A} $ in $ o(\mathfrak{e} \log \mathfrak{e}) $ time, otherwise we would break the well-known lower bound $ \Omega (n \log n) $  to sort $ n $ elements, which holds even if we know that the $ n $ elements are pairwise distinct~\cite{knuth1998art,cormen2022introduction}. Indeed, in $ O(n) $ time we can reduce the problem of sorting $ n $ pairwise distinct elements to the problem of computing $ \preceq_\mathcal{A} $ for a Wheeler DFA $ \mathcal{A} $ consisting of a single path. More precisely, if $ c_1 $, $ c_2 $, $ \dots $, $ c_n $ is an enumeration of the $ n $ elements to sort, we create the DFA $ \mathcal{A} = (Q, E, s, F) $, where $ Q = \{u_0, u_1, \dots, u_n \} $ has $ n + 1 $ elements, $ s = u_0 $, $ F = \{u_n \} $ and $ E = \{(u_{i - 1}, u_i, c_i) \;|\; 1 \le i \le n \} $ (for an example, see the automaton on the left in Figure~\ref{fig:lowerboundcomparison}). Then, $ \mathcal{A} $ is Wheeler, and sorting $ c_1 $, $ c_2 $, $ \dots $, $ c_n $ is equivalent to sorting the states $ u_1 $, $ u_2 $, $ \dots $, $ u_n $ with respect to $ \preceq_{\mathcal{A}} $ (because the $ n $ elements are pairwise distinct).

Observe that the problem of computing $ \preceq_{\mathcal{A}} $ is a generalization of the problem of computing the suffix array of a string (that is, the problem of lexicographically sorting all the suffixes of a string). More precisely, given a string $ \alpha $, the problem of computing the suffix array of $ \alpha $ is equivalent to the problem of computing $ \preceq_\mathcal{A} $ for a Wheeler DFA $ \mathcal{A} $ consisting of a single path obtained by considering the reverse string $ \alpha^R $ (for an example, see the automaton on the right in Figure~\ref{fig:lowerboundcomparison}). In the comparison model, building the suffix array of a string has complexity $ \Omega (n \log n) $, where $ n $ is the length of the string~\cite{farach2000jacm} (in particular, this immediately implies an alternative proof of the fact that in the comparison model we cannot compute $ \preceq_\mathcal{A} $ in $ o(\mathfrak{e} \log \mathfrak{e}) $ time). In the case of an integer alphabet in a polynomial range, the suffix array can be built in linear time~\cite{farach2000jacm}, and we leave it as an open problem to determine whether in Theorem~\ref{theor:polynomialdecidingwheeler} we can achieve $ O(\mathfrak{e}) $ time. In particular, in the case of an integer alphabet in a polynomial range, we can sort in linear time~\cite{hopcroft1983data, paige1987three}, so the proof of Theorem~\ref{theor:polynomialdecidingwheeler} implies that the bound $ O(\mathfrak{e}) $ would follow immediately if we proved that, given a DFA $ \mathcal{A} = (Q, E, s, F) $, we can compute the partial order $ \preceq_\mathcal{A} $ in $ O(|E|) $ time (determining if this is possible is also an open problem).

\begin{figure}
     \centering
     \begin{subfigure}[b]{0.49\textwidth}
        \centering
        \scalebox{.6}{
        \begin{tikzpicture}[->,>=stealth', semithick, auto, scale=1]
\node[state, initial] (1)    at (0,0)	{};
\node[state] (2)    at (2,0)	{};
\node[state] (3)    at (4,0)	{};
\node[state] (4)    at (6,0)	{};
\node[state, accepting] (5)    at (8,0)	{};
\draw (1) edge [] node [] {$ c $} (2);
\draw (2) edge [] node [] {$ d $} (3);
\draw (3) edge [] node [] {$ a $} (4);
\draw (4) edge [] node [] {$ b $} (5);
\end{tikzpicture}
}
     \end{subfigure}
     \begin{subfigure}[b]{0.49\textwidth}
        \centering
        \scalebox{.6}{
        \begin{tikzpicture}[->,>=stealth', semithick, auto, scale=1]
\node[state, initial] (1)    at (0,0)	{};
\node[state] (2)    at (2,0)	{};
\node[state] (3)    at (4,0)	{};
\node[state] (4)    at (6,0)	{};
\node[state, accepting] (5)    at (8,0)	{};
\draw (1) edge [] node [] {$ d $} (2);
\draw (2) edge [] node [] {$ f $} (3);
\draw (3) edge [] node [] {$ d $} (4);
\draw (4) edge [] node [] {$ c $} (5);
\end{tikzpicture}
}
     \end{subfigure}
	\caption{\emph{Left:} Reducing the problem of sorting $ c $, $ d $, $ a $, $ b $ to the problem of computing $ \preceq_{\mathcal{A}} $. \emph{Right:} Reducing the problem of computing the suffix array of $ \alpha = cdfd $ to the problem of computing $ \preceq_{\mathcal{A}} $ (where $ \alpha^R = dfdc $).}
 \label{fig:lowerboundcomparison}
\end{figure}

\subsection{The Burrows-Wheeler Transform of a Wheeler GDFA}

After introducing Wheeler GDFAs and discussing their properties, we now turn to our main goal, namely, proving Theorem~\ref{theor:fmindexintroduction}, which extends the FM-index from strings to Wheeler GDFAs. The FM-index of a string is a space-efficient data structure for solving pattern matching queries on the string. To achieve compressed space, pattern matching queries are not solved directly on the plain representation of the string, but on the \emph{Burrows-Wheeler Transform (BWT)} of the string~\cite{burrows1994}, which is an ingenious encoding of the considered string. Consequently, to prove Theorem~\ref{theor:fmindexintroduction}, we first need to extend the Burrows-Wheeler Transform from strings to Wheeler GDFAs.

We remark that our generalization is only the last of a chain of results showing how to extend the Burrows-Wheeler Transform from strings to more complex settings. Over the years, the Burrows-Wheeler Transform was extended to labeled trees~\cite{ferraginajacm2009}, circular strings~\cite{mantaci2007extension, hon2011succinct, cotumaccio2025improved, cotumaccio2026tcs, cotumaccio2025sorting}, de Bruijn graphs~\cite{bowe2012}, Wheeler automata~\cite{gagie2017, alanko2020, cotumaccio2025wheeler, cotumaccio2026concur} and arbitrary automata~\cite{cotumacciojacm, cotumaccio2021, cotumaccio2022, becker2025encoding, alanko2024computing}.

Let us first recall how to define the Burrows-Wheeler Transform of a string. Consider the string $ banana $. We first attach a new character $ \$ $ at the end of our string, obtaining $ banana\$ $. We assume that $ \$ $ is smaller than all the other characters. Then, we consider all circular rotations of $ banana\$ $, thus obtaining $ banana\$ $, $ anana\$b $, $ nana\$ba $, $ ana\$ban $, $ na\$bana $, $ a\$banan $, $ \$banana $. We now sort all circular suffixes in lexicographic order, obtaining the matrix in Table~\ref{table:bananamatrix}. The Burrows-Wheeler Transform of the matrix is the last column of this matrix, namely, $ annb\$aa $.

The surprising result is that, given the Burrows-Wheeler Transform of a string, one can retrieve the original string. For example, if we are only given the string $ annb\$aa $, then we can infer that $ annb\$aa $ is the Burrows-Wheeler Transform of $ banana\$ $. The advantage of storing the Burrows-Wheeler Transform of a string instead of the original string is that, if the original string is repetitive, then the Burrows-Wheeler Transform contains runs of consecutive letters. In our example, the original string $ \$banana $ does not contain consecutive letters, while the Burrows-Wheeler Transform $ annb\$aa $ contains two consecutive $ n $'s and two consecutive $ a $'s. The human DNA (which can be seen as a string on the alphabet $ \{A, C, G, T \} $, where $ A $, $ C $, $ G $ and $ T $ stand for the four nucleotides) is a long repetitive string, so when we compute its Burrows-Wheeler Transform, instead of storing (say) $ 50 $ consecutive occurrences of $ A $, we can simply store $ (A, 50) $. Remarkably, this clustering effect of the Burrows-Wheeler Transform can be described precisely in mathematical terms through the notion of \emph{entropy}~\cite{manzini2001analysis}.

    \begin{table}
        \centering
        \scalebox{0.9}{
\begin{tabularx}{0.5\textwidth}{|>{\centering\arraybackslash}X|>{\centering\arraybackslash}X|>{\centering\arraybackslash}X|>{\centering\arraybackslash}X|>{\centering\arraybackslash}X|>{\centering\arraybackslash}X|>{\centering\arraybackslash}X|}
\hline
$ \$ $ & $ b $ & $ a $ & $ n $ & $ a $ & $ n $ & \textcolor{blue}{$ a $} \\
\hline
$ a $ & $ \$ $ & $ b $ & $ a $ & $ n $ & $ a $ & \textcolor{blue}{$ n $} \\
\hline
$ a $ & $ n $ & $ a $ & $ \$ $ & $ b $ & $ a $ & \textcolor{blue}{$ n $} \\
\hline
$ a $ & $ n $ & $ a $ & $ n $ & $ a $ & $ \$ $ & \textcolor{blue}{$ b $} \\
\hline
$ b $ & $ a $ & $ n $ & $ a $ & $ n $ & $ a $ & \textcolor{blue}{$ \$ $} \\
\hline
$ n $ & $ a $ & $ \$ $ & $ b $ & $ a $ & $ n $ & \textcolor{blue}{$ a $} \\
\hline
$ n $ & $ a $ & $ n $ & $ a $ & $ \$ $ & $ b $ & \textcolor{blue}{$ a $} \\
\hline
\end{tabularx}
}
\caption{The Burrows-Wheeler Transform of $ banana\$ $.}
\label{table:bananamatrix}
\end{table}

Let us show how to define the Burrows-Wheeler Transform of a Wheeler GDFA. To this end, we need some notation. For $ i \ge 0 $, let $ \Sigma^i \subseteq \Sigma^* $ be the set of all strings of length $ i $ on the alphabet $ \Sigma $. Let $ \mathcal{A} = (Q, E, s, F) $ be a GDFA with $ n = |Q| $ states and $ e = |E| $ edges. We say that $ \mathcal{A} $ is an \emph{$ r $-GDFA} if all edge labels have length at most $ r $. Fix $ 1 \le i \le r $. Let $ E_i = \{(u', u, \rho) \in E \;|\; \rho \in \Sigma^i \} $ be the set of all edges labeled with a string of length $ i $, and let $ \Sigma_i = \{\rho \in \Sigma^i \;|\; (\exists u', u \in Q)((u', u, \rho) \in E_i) \} $ be the set of all strings of length $ i $ labeling some edge. Let $ e_i = |E_i| $ and $ \sigma_i = |\Sigma_i| $; we have $ \sigma_i \le \min \{\sigma^i, e_i \} $ (where $ \sigma^i $ is $ \sigma $ to the $ i $-th power). The \emph{$ i $-outdegree} (\emph{$ i $-indegree}, respectively) of a state is equal to the number of edges in $ E_i $ leaving (reaching, respectively) the state. The sum of the $ i $-outdegrees of all the states and the sum of the $ i $-indegrees of all the states are both equal to $ e_i $. Lastly, $ \sum_{i = 1}^r e_i = e $ and the total length of all edge labels is $ \mathfrak{e} = \sum_{i = 1}^r e_i i $.

If $ \mathcal{A} = (Q, E, s, F) $ is a Wheeler GDFA, we write $ Q = \{Q[1], Q[2], \dots, Q[n] \} $, where $ Q[1] \prec_\mathcal{A} Q[2] \prec_\mathcal{A} \dots \prec_\mathcal{A} Q[n] $. If $ 1 \le j_1 \le j_2 \le n $, let $ Q[j_1, j_2] = \{Q[j_1], Q[j_1 + 1], \dots, Q[j_2 - 1], Q[j_2] \} $, and if $ j_1 > j_2 $, let $ Q[j_1, j_2] = \emptyset $.

\begin{figure}[h!]
     \centering
     \begin{subfigure}[b]{0.6\textwidth}
        \centering
        \scalebox{.8}{
        \begin{tikzpicture}[->,>=stealth', semithick, auto, scale=1]
\node[state, initial] (1)    at (0,0)	{$ u_1 $};
\node[state, accepting] (2)    at (1.5,1.5)	{$ u_2 $};
\node[state, accepting] (3)    at (1.5,-1.5)	{$ u_3 $};
\draw (1) edge [] node [] {$ ab, b $} (2);
\draw (1) edge [] node [] {$ ac, c $} (3);
\draw (2) edge [loop right] node [] {$ b $} (2);
\draw (3) edge [loop right] node [] {$ bc $} (3);
\end{tikzpicture}
}
     \end{subfigure}
     \begin{subfigure}[b]{0.39\textwidth}
    \centering
 \begin{itemize}
    \item $ \mathtt{OUT}_1 = 001011 $
    \item $ \mathtt{OUT}_2 = 001101 $
    \item $ \mathtt{IN}_1 = 100101 $
    \item $ \mathtt{IN}_2 = 101001 $
    \item $ \mathtt{LAB}_1 = (b, c, b) $
    \item $ \mathtt{LAB}_2 = (ab, ac, bc) $
    \item $ \mathtt{FIN} = 011 $
\end{itemize}

     \end{subfigure}
	\caption{\emph{Left:} The Wheeler GDFA $ \mathcal{A} $ of Figure~\ref{fig:examwheel}. \emph{Right:} The Burrows-Wheeler Transform (BWT) of $ \mathcal{A} $.}
 \label{fig:examwheelwithbwt}
\end{figure} 

Let us define the Burrows-Wheeler Transform (BWT) of a Wheeler GDFA (see Figure~\ref{fig:examwheelwithbwt} for an example).

\begin{defi}[BWT of a Wheeler GDFA]\label{def:BWTwheelerdfa}
    Let $ \mathcal{A} = (Q, E, s, F) $ be a Wheeler $ r $-GDFA. The \emph{Burrows-Wheeler Transform} $ \BWT (\mathcal{A}) $ of $ \mathcal{A} $ consists of the following strings.
\begin{itemize}
    \item For every $ 1 \le i \le r $, the bit string $ \mathtt{OUT}_i \in \{0, 1 \}^{e_i + n} $ that stores the $ i $-outdegrees in unary. More precisely, (i) $ \mathtt{OUT}_i $ contains exactly $ n $ bits equal to $ 1 $, (ii) $ \mathtt{OUT}_i $ contains exactly $ e_i $ bits equal to $ 0 $, and (iii) for every $ 1 \le j \le n $, the number of zeros between the $(j-1)$-th bit equal to one (or the beginning of the sequence if $j=1$) and the $ j $-th bit equal to $ 1 $  yields the $ i $-outdegree of $ Q[j] $. In other words, we have:
    \begin{equation*}
        \mathtt{OUT}_i = \underbrace{0 0 \dots \dots \dots 0 0}_{\text{$ i $-outdegree of $ Q[1] $}} 1 \underbrace{0 0 \dots \dots \dots 0 0}_{\text{$ i $-outdegree of $ Q[2] $}} 1 \dots \dots \dots 1 \underbrace{0 0 \dots \dots \dots 0 0}_{\text{$ i $-outdegree of $ Q[n] $}} 1.
    \end{equation*}
    \item For every $ 1 \le i \le r $, the bit string $ \mathtt{IN}_i \in \{0, 1 \}^{e_i + n} $ that stores the $ i $-indegrees in unary. More precisely, (i) $ \mathtt{IN}_i $ contains exactly $ n $ bits equal to $ 1 $, (ii) $ \mathtt{IN}_i $ contains exactly $ e_i $ bits equal to $ 0 $, and (iii) for every $ 1 \le j \le n $, the number of zeros between the $(j-1)$-th bit equal to one (or the beginning of the sequence if $j=1$) and the $ j $-th bit equal to $ 1 $  yields the $ i $-indegree of $ Q[j] $. In other words, we have:
    \begin{equation*}
        \mathtt{IN}_i = \underbrace{0 0 \dots \dots \dots 0 0}_{\text{$ i $-indegree of $ Q[1] $}} 1 \underbrace{0 0 \dots \dots \dots 0 0}_{\text{$ i $-indegree of $ Q[2] $}} 1 \dots \dots \dots 1 \underbrace{0 0 \dots \dots \dots 0 0}_{\text{$ i $-indegree of $ Q[n] $}} 1.
    \end{equation*}
    \item For every $ 1 \le i \le r $, the string $ \mathtt{LAB}_i \in (\Sigma_i)^{e_i} $ that stores the edge labels of length $ i $ (with their multiplicities). More precisely, we sort all edges in $ E_i $ by the index of the start states (w.r.t. $ \preceq_\mathcal{A} $). Edges with the same start state are sorted by label. Then, we obtain $ \mathtt{LAB}_i $ by concatenating the labels of all edges following this edge order.  
    \item The bit string $ \mathtt{FIN} \in \{0, 1 \}^n $ that marks the final states, that is, for every $ 1 \le j \le n $, the $ j $-th bit of $ \mathtt{FIN} $ is equal to 1 if and only if $ Q[j] \in F $.
\end{itemize}
\end{defi}

The definition of the Burrows-Wheeler Transform of a string can be seen as a special case of Definition~\ref{def:BWTwheelerdfa}. For example, if we consider again the string $ banana $ (see Table~\ref{table:bananamatrix}), we can proceed as follows. We compute the reverse string $ ananab $, and we build a Wheeler DFA consisting of a cycle such that, starting from the initial state, we can read $ ananab \$ $, see Figure~\ref{fig:bananadfa} (we assume that the initial state is also the unique final state). Then, we have $ r = 1 $ (because $ \mathcal{A} $ is a conventional DFA), and the Burrows-Wheeler Transform as defined using Table~\ref{table:bananamatrix} is equal to the string $ \mathtt{LAB}_1 $ of Definition~\ref{def:BWTwheelerdfa} (in this special case the bit strings $ \mathtt{OUT}_1 $, $ \mathtt{IN}_1 $ and $ \mathtt{FIN} $ of Definition~\ref{def:BWTwheelerdfa} are uninformative because (i) every state has indegree and outdegree equal to $ 1 $ and (ii) the initial state is the unique final state).

We can now prove that the Burrows-Wheeler Transform of a Wheeler GDFA $ \mathcal{A} $ is a valid encoding of $ \mathcal{A} $. In particular, we will use this result in the proof of Theorem~\ref{theor:fmindexintroduction} to show that we can encode $ \mathcal{A} $ by using $ \mathfrak{e} \log \sigma (1 + o(1)) + O(e) $ bits.

\begin{figure}[h!]
     \centering
        \scalebox{.8}{
        \begin{tikzpicture}[->,>=stealth', semithick, auto, scale=1]
\node[state, initial, accepting] (1)    at (0,0)	{$ 1 $};
\node[state] (2)    at (2,0)	{$ 2 $};
\node[state] (3)    at (4,0)	{$ 6 $};
\node[state] (4)    at (6,0)	{$ 3 $};
\node[state] (5)    at (8, 0)	{$ 7 $};
\node[state] (6)    at (10, 0)	{$ 4 $};
\node[state] (7)    at (12, 0)	{$ 5 $};
\draw (1) edge [] node [] {$ a $} (2);
\draw (2) edge [] node [] {$ n $} (3);
\draw (3) edge [] node [] {$ a $} (4);
\draw (4) edge [] node [] {$ n $} (5);
\draw (5) edge [] node [] {$ a $} (6);
\draw (6) edge [] node [] {$ b $} (7);
\draw (7) edge [bend left] node [] {$ \$ $} (1);
\end{tikzpicture}
}

	\caption{Computing the BWT of $ banana\$ $ using a Wheeler DFA $ \mathcal{A} $. The states are numbered following the total order $ \preceq_\mathcal{A} $.}
 \label{fig:bananadfa}
\end{figure}  

\begin{thm}\label{theor:bwtGDFAwheeler}
    Let $ \mathcal{A} = (Q, E, s, F) $ be a Wheeler GDFA. If we only know $ \BWT(\mathcal{A}) $, then we can retrieve $ \mathcal{A} $ (up to isomorphism). In other words, $ \BWT(\mathcal{A}) $ is an encoding of $ \mathcal{A} $.
\end{thm}

\begin{proof}
    Note that $ \BWT(\mathcal{A}) $ consists of $ 3r + 1 $ strings, so from $ \BWT(\mathcal{A}) $ we can retrieve $ r $. From $ \BWT(\mathcal{A}) $ we can also retrieve the value $ n $ and, for every $ 1 \le i \le r $, the value $ e_i $.

    Let us show that (i) for every $ 1 \le j \le n $, we can determine whether $ Q[j] \in F $, and (ii) for every $ 1 \le j', j \le n $ and for every $ \rho \in \Sigma^+ $ such that $ |\rho| \le r $, we can determine whether $ (Q[j'], Q[j], \rho) \in E $. This is sufficient to retrieve $ \mathcal{A} $ up to isomorphism. Indeed, if we define the generalized automaton $ \mathcal{A}' = (Q', E', s', F') $ such that $ Q' = \{1, 2, \dots, n \} $, $ E' = \{(j', j, \rho) \in Q' \times Q' \times \Sigma^* \;|\; (Q[j'], Q[j], \rho) \in E \} $, $ s' = 1 $, and $ F' = \{j \in Q \;|\; Q[j] \in F \} $, then $ \mathcal{A}' $ is isomorphic to $ \mathcal{A} $, with isomorphism $ \phi: Q' \mapsto Q $ given by $ \phi(j) = Q[j] $ for every $ j \in Q' $ (note that $ \phi(s') = s $ because $ s = Q[1] $ by Lemma~\ref{lem:local GDFA}).

    Proving (i) is immediate because we can use $ F $. To prove (ii), we must show how to retrieve $ E $. It will suffice to retrieve the set $ E_i $ for every $ 1 \le i \le r $, because $ E $ is the (disjoint) union of the $ E_i $'s. Fix $ 1 \le i \le r $. From $ \mathtt{LAB}_i $ we can retrieve the labels of all edges in $ E_i $, with their multiplicities. From $ \mathtt{IN}_i $ we can retrieve the $ i $-indegree of each $ Q[j] $. By Lemma~\ref{lem:local GDFA}, for every $ \rho \in \Sigma^i $ labeling some edge reaching some state $ Q[j] $ and for every $ \rho' \in \Sigma^i $ labeling some edge reaching $ Q[j + 1] $ we must have $ \rho \preceq \rho' $. Since we know the labels of all edges in $ E_i $ with multiplicities and we know the $ i $-indegrees, then we can retrieve the labels of all edges reaching each $ Q[j] $, with multiplicities. From $ \mathtt{OUT}_i $ we can retrieve the $ i $-outdegrees of each $ Q[j] $, and the order used in the definition of $ \mathtt{LAB}_i $ implies that we can retrieve the labels of all edges leaving each $ Q[j] $. Since we know the labels of all edges reaching each $ Q[j] $ and we know the labels of all edges leaving each $ Q[j] $, then for every  $ \rho \in \Sigma^i $ we know the set of all states reached by an edge labeled $ \rho $, with multiplicities, and the set of all states having an outgoing edge labeled $ \rho $. By Lemma~\ref{lem:local GDFA}, for every $ 1 \le j_1 < j_2 \le n $, if $ Q[j_1] $ is reached by an edge labeled $ \rho $ leaving the state $ Q[j'_1] $ and $ Q[j_2] $ is reached by an edge labeled $ \rho $ leaving the state $ Q[j'_2] $, then it must be $ j'_1 < j'_2 $. As a consequence, we can retrieve the set $ E_{i, \rho} $ of all edges labeled $ \rho $ for every $ \rho \in \Sigma^i $, and so we can retrieve $ E_i $, because $ E_i $ is the (disjoint) union of the $ E_{i, \rho} $'s. \qedhere
\end{proof}

\subsection{Pattern matching on Wheeler GDFAs} The goal of this section is to show how to solve pattern matching queries on Wheeler GDFAs. Our approach is an extension of the \emph{backward search}, which is the classical algorithm for solving pattern matching queries directly on the BWT-representation of a string~\cite{ferraginajacm2005}. As a special case, we will retrieve the algorithm for solving pattern matching queries on conventional Wheeler automata~\cite{gagie2017}; however, we remark that the algorithm for arbitrary Wheeler GDFAs is significantly more complex than the algorithm for conventional Wheeler automata.

Let us formally define the pattern matching problem that we want to solve (recall the informal discussion on the SMLG problem in Section~\ref{sec:introduction}).

\begin{defi}
    Let $ \mathcal{A} $ be a GDFA. The \emph{String Matching in Labeled Graphs (SMLG)} problem for GDFAs is defined as follows: given a string $ \alpha $, compute the set of all states of $ \mathcal{A} $ reached by a path suffixed by $ \alpha $.
\end{defi}

To address the SMLG problem, we need to introduce some notation. If $ \alpha, \beta \in \Sigma^* $, we write $ \alpha \dashv \beta $ if and only if $ \alpha $ is a suffix of $ \beta $ (equivalently, if and only if there exists $ \beta' \in \Sigma^* $ such that $ \beta = \beta' \alpha $). If $ \alpha \in \Sigma^* $, let $ \alpha[i] $ be the $ i $-th character of $ \alpha $ (for $ 1 \le i \le |\alpha| $), let $ \alpha[i, j] = \alpha[i] \alpha[i + 1] \dots \alpha[j - 1] \alpha[j] $ (for $ 1 \le i \le j \le |\alpha| $), and let $ p_i(\alpha) $ and $ s_i(\alpha) $ be the prefix and the suffix of $ \alpha $ of length $ i $, respectively (for $ 0 \le i \le |\alpha| $). If $ \mathcal{A} = (Q, E, s, F) $ is a GDFA and $ u \in Q $, let $ \lambda (u) $ be the set of all strings in $ \Sigma^* $ labeling an edge reaching $ u $ (note that $ \lambda (u) \not = \emptyset $ if $ u \not = s $ because every state is reachable from the initial state).

Let us give a definition.

\begin{defi}\label{def:S_i}
Let $ \mathcal{A} = (Q, E, s, F) $ be a Wheeler GDFA, and let $ \alpha \in \Sigma^* $. Define:
\begin{itemize}
    \item $ G^\prec(\alpha) = \{  u \in Q \;|\; (\forall \beta\in I_{u})(\beta \prec \alpha)\} $;
    \item $ G_\dashv(\alpha) = \{  u \in Q \;|\; (\exists \beta\in I_{u})( \alpha \dashv \beta)\} $;
    \item $ G^\prec_\dashv(\alpha) = G^\prec(\alpha) \cup G_\dashv(\alpha) =\{  u \in Q\ \;|\; (\forall \beta\in I_{u})(\beta \prec \alpha) \vee (\exists \beta\in I_{u})( \alpha \dashv  \beta)\} $.
\end{itemize}
\end{defi}

Intuitively, the set $ G_\dashv(\alpha) $ is the set of states that the SMLG problem must return on input $ \alpha $, and $ G^\prec(\alpha) $ is the set of all states reached only by strings smaller than $ \alpha $. For example, in Figure~\ref{fig:examwheelwithbwt} we have $ G^\prec(ac) = \{u_1, u_2 \} $, $ G_\dashv(ac) = \{u_3 \} $ and $ G^\prec_\dashv(ac) = \{u_1, u_2, u_3 \} $.

\begin{rem}\label{rem:caseepsilonobvious}
    Note that $ G^\prec(\epsilon) = \emptyset $ (because $ \epsilon $ is the smallest string in $ \Sigma^* $) and $ G_\dashv(\epsilon) = Q $ (because $ \epsilon $ is a suffix of every string in $ \Sigma^* $), so $ G^\prec_\dashv(\epsilon) = Q $. In particular, $ |G^\prec(\epsilon)| = 0 $ and $ |G_\dashv(\epsilon)| = |G^\prec_\dashv(\epsilon)| = n $, where $ n = |Q | $.
\end{rem}

The following lemma shows that, as in the case of conventional Wheeler automata~\cite{alanko2021}, both $ G^\prec(\alpha) $ and $ G_\dashv(\alpha) $ are intervals with respect to the total order $ \preceq_\mathcal{A} $, and there is no state between $ G^\prec(\alpha) $ and $ G_\dashv(\alpha) $. The intuition is that Wheeler GDFAs are an extension of suffix arrays (as we have seen at the end of Section~\ref{sec:decidingwheelerness}) and, in the string setting, all suffixes that are lexicographically smaller than a string $ \alpha $ and all suffixes that start with $ \alpha $ correspond to two disjoint intervals in the list of all lexicographically-sorted suffixes.

\begin{lem}\label{lem:st}
Let $ \mathcal{A} = (Q, E, s, F) $ be a Wheeler GDFA, and let $ \alpha \in \Sigma^* $. Then:
\begin{enumerate}
    \item $ G^\prec(\alpha) \cap G_\dashv(\alpha) = \emptyset $.
    \item $ G_\dashv(\alpha) $ is $ \preceq_\mathcal{A} $-convex.
    \item If $ u, v \in Q $ are such that $ u \prec_\mathcal{A} v $ and $ v \in G^\prec(\alpha) $, then $ u \in G^\prec(\alpha) $. In other words, $ G^\prec(\alpha) = Q[1, |G^\prec(\alpha)|] $.
    \item If $ u, v \in Q $ are such that $ u \prec_\mathcal{A} v $ and $ v \in G^\prec_\dashv(\alpha) $, then $ u \in G^\prec_\dashv(\alpha) $. In other words, $ G^\prec_\dashv(\alpha) = Q[1, |G^\prec_\dashv(\alpha)|] $.
    \item $ G_\dashv(\alpha) = Q[|G^\prec(\alpha)| + 1, |G^\prec_\dashv(\alpha)|] $.
\end{enumerate}
\end{lem}

\begin{proof}
\begin{enumerate}
    \item If $ u \in G_\dashv(\alpha) $, then there exists $ \beta \in I_u $ such that $ \alpha \dashv \beta $. In particular, $ \alpha \preceq \beta $, so $ u \not \in G^\prec(\alpha) $.
    \item Assume that $ u, v, z \in Q $ are such that $ u \prec_\mathcal{A} v \prec_\mathcal{A} z $ and $ u, z \in G_\dashv(\alpha) $. We must prove that $ v \in G_\dashv(\alpha) $. Since $ u, z \in G_\dashv(\alpha) $, then there exist $ \beta \in I_u $ and $ \delta \in I_z $ such that $ \alpha \dashv \beta $ and $ \alpha \dashv \delta $. Fix any $ \gamma \in I_v $; we only have to prove that $ \alpha \dashv \gamma $. From $ u \prec_\mathcal{A} v \prec_\mathcal{A} z $ we obtain $ \beta \prec \gamma \prec \delta $. As a consequence, from $ \alpha \dashv \beta $ and $ \alpha \dashv \delta $ we conclude $ \alpha \dashv \gamma $.
    \item Let $ \beta \in I_u $. We must prove that $ \beta \prec \alpha $. Fix any $ \gamma \in I_v $. Since $ u \prec_\mathcal{A} v $, we have $ \beta \prec \gamma $. From $ v \in G^\prec(\alpha) $ we obtain $ \gamma \prec \alpha $, so we conclude $ \beta \prec \alpha $.
    \item Since $ v \in G^\prec_\dashv(\alpha) $, we have either $ v \in G^\prec(\alpha) $ or $ v \in G_\dashv(\alpha) $. If $ v \in G^\prec(\alpha) $, then $ u \in G^\prec(\alpha) $ by the previous point and so $ u \in G^\prec_\dashv(\alpha) $. Now assume that $ v \in G_\dashv(\alpha) $. If $ u \in G_\dashv(\alpha) $, then $ u \in G^\prec_\dashv(\alpha) $ and we are done, so we can assume $ u \not \in G_\dashv(\alpha) $. Let us prove that it must be $ u \in G^\prec(\alpha) $, which again implies $ u \in G^\prec_\dashv(\alpha) $. Fix $ \beta \in I_u $; we must prove that $ \beta \prec \alpha $. Since $ v \in G_\dashv(\alpha) $, then there exists $ \gamma \in \Sigma^* $ such that $ \gamma \alpha \in I_v $. From $ u \prec_\mathcal{A} v $ we obtain $ \beta \prec \gamma \alpha $. Since $ u \not \in G_\dashv(\alpha) $ implies $ \alpha \not \dashv \beta $, from  $ \beta \prec \gamma \alpha $ we conclude $\beta \prec \alpha$.
    \item By the definition of $ G^\prec_\dashv(\alpha) $ and the first point we obtain that $ G_\dashv(\alpha) $ is the disjoint union of $ G^\prec(\alpha) $ and $ G_\dashv(\alpha) $, so the conclusion follows from point 3 and point 4. \qedhere 
\end{enumerate}
\end{proof}

Since $ G_\dashv(\alpha) = Q[|G^\prec(\alpha)| + 1, |G^\prec_\dashv(\alpha)|] $, to compute $ G_\dashv(\alpha) $, it will suffice to compute $ |G^\prec(\alpha)| $ and $ |G^\prec_\dashv(\alpha)| $. To this end, we will repeatedly use point 3 of Lemma~\ref{lem:local GDFA}, which is also crucial for conventional Wheeler automata (it is a generalization of the LF-mapping used in the backward search~\cite{ferraginajacm2005}). To compute $ |G^\prec(\alpha)| $ and $ |G^\prec_\dashv(\alpha)| $, we will recursively compute $ |G^\prec (p_k(\alpha))| $ and $ |G^\prec_\dashv (p_k(\alpha))| $ for every $ 0 \le k \le |\alpha| $. For $ k = 0 $ we have $ p_0 (\alpha) = \epsilon $, and this case is easy by Remark~\ref{rem:caseepsilonobvious}. Consequently, to compute $ |G^\prec(\alpha)| $ and $ |G^\prec_\dashv(\alpha)| $, we can assume that we have already computed $ |G^\prec (p_k(\alpha))| $ and $ |G^\prec_\dashv (p_k(\alpha))| $ for every $ 0 \le k \le |\alpha| - 1 $.

First, let us show how to compute $ |G^\prec(\alpha)| $. The next lemma formalizes the following intuition: to compute $ |G^\prec(\alpha)| $, we have to consider all states whose incoming edges have a label smaller than $ \alpha $; moreover, if the label is $ s_i (\alpha) $ (for some $ i $), then the start state must be in $ G^\prec(p_{|\alpha| - i}(\alpha)) $.

\begin{lem}\label{lem:Si alpha}
Let $ \mathcal{A} = (Q, E, s, F) $ be a Wheeler GDFA, and let $ \alpha \in \Sigma^* $, with $ \alpha \not = \epsilon $. Then, $ u \in G^\prec(\alpha) $ if and only if (i) $ \rho \prec \alpha $ for every $ \rho \in \lambda (u) $ and (ii) if $ u' \in Q $ is such that $ (u', u, s_i(\alpha)) \in E $ for some $ 1 \le i \le |\alpha| - 1 $, then $ u' \in G^\prec(p_{|\alpha| - i}(\alpha)) $.
\end{lem}

\begin{proof}
$ (\Longrightarrow) $. Let us prove (i). Pick $ \rho \in \lambda (u) $. Then, there exists $ u' \in Q $ such that $ (u', u, \rho) \in E $. We must prove that $ \rho \prec \alpha $. Let $ \beta' \in I_{u'} $; from $ (u', u, \rho) \in E $ we obtain $ \beta' \rho \in I_u $, so from $u \in G^\prec(\alpha)$ we obtain $ \beta' \rho \prec \alpha $, which implies $ \rho \prec \alpha $. Let us prove (ii). Assume that $ u' \in Q $ is such that $ (u', u, s_i(\alpha)) \in E $ for some $ 1 \le i \le |\alpha| - 1 $. We must prove that $ u' \in G^\prec(p_{|\alpha| - i}(\alpha)) $. Pick $ \gamma' \in I_{u'} $. We must prove that $ \gamma' \prec p_{|\alpha| - i}(\alpha) $. Since $ (u', u, s_i(\alpha)) \in E $, we have $ \gamma' s_i(\alpha) \in I_u $, so from $u \in G^\prec(\alpha) $ we obtain $ \gamma' s_i(\alpha) \prec \alpha $, which implies $ \gamma' \prec p_{|\alpha| - i}(\alpha) $.

$ (\Longleftarrow) $ Let $ \beta \in I_u $. We must prove that $ \beta \prec \alpha $. If $ \beta = \epsilon $ the conclusion is immediate because $ \alpha \not = \epsilon $. Now, assume that $ \beta \not = \epsilon $. This means that there exists $ u' \in Q $ such that $ (u', u, s_i(\beta)) \in E $ and $ p_{|\beta| - i}(\beta) \in I_{u'} $, for some $ 1 \le i \le |\beta| $. We know that $ s_i(\beta) \prec \alpha $ by property (i). If $ s_i(\beta) \not \dashv \alpha $, then $ \beta \prec \alpha $ and we are done. Now assume that $ s_i(\beta) \dashv \alpha $.  We have $ s_i(\beta) = s_i(\alpha) $, and so $ 1 \le i \le |\alpha| - 1 $ (otherwise $ i = |\alpha| $, and $ \alpha = s_i(\alpha) =  s_i(\beta) \prec \alpha $, a contradiction). Hence, we obtain $ u' \in G^\prec(p_{|\alpha| - i}(\alpha)) $ by property (ii). Since $ p_{|\beta| - i}(\beta) \in I_{u'} $, we conclude $ p_{|\beta| - i}(\beta) \prec p_{|\alpha| - i}(\alpha) $, which implies $ \beta \prec \alpha $. \qedhere
\end{proof}

We now want to give a variant of Lemma~\ref{lem:Si alpha} which will prove to be computationally more efficient. Let us give the following definition.

\begin{defi}\label{def: inout}
Let $ \mathcal{A} = (Q, E, s, F) $ be a Wheeler GDFA. Let $ U \subseteq Q $ and $ \rho \in \Sigma^+ $. We denote by $\mathtt{out}(U,\rho)$ the number of edges labeled with $ \rho $ that leave states in $ U $, and we denote by $\mathtt{in}(U,\rho)$ the number of edges labeled with $ \rho $ that reach states in $ U $.
\end{defi}

We are now ready to give a variant of Lemma~\ref{lem:Si alpha}. We introduce two main differences:
\begin{itemize}
    \item We show that, if $ \mathcal{A} $ is an $ r $-GDFA, then in Lemma~\ref{lem:Si alpha} we do not need to check each $ 1 \le k \le |\alpha| - 1 $, but at most $ r $ values of $ k $. The intuition is that the edge $ (u', u, s_i(\alpha)) $ mentioned in the statement of Lemma~\ref{lem:Si alpha} can exist only for $ 1 \le i \le r $.
    \item We can rephrase the condition (ii) in the statement of Lemma~\ref{lem:Si alpha} by using point 3 of Lemma~\ref{lem:local GDFA}. Here is the intuition. We list all states $ Q[1] $, $ Q[2] $, $ \dots $, $ Q[|Q|] $, and then we duplicate them as in Figure~\ref{fig:intuitionleft}. Now, fix $ 1 \le i \le r $, and consider all edges labeled with $ s_i (\alpha) $. We draw all these edges as in Figure~\ref{fig:intuitionleft}: the start state is in the bottom list, and the end state is in the top list. Then, intuitively point 3 of Lemma~\ref{lem:local GDFA} ensures that the edges in Figure~\ref{fig:intuitionleft} cannot cross. Let $ f_i $ be the number of edges labeled $ s_i (\alpha) $ leaving a state in $ Q[1, |G^\prec(p_{|\alpha| - i}(\alpha))|] $. Let $ 0 \le j \le |Q| - 1 $ such that the states in $ Q[1, j + 1] $ are reached by at least $ f_i + 1 $ edges. Since the edges cannot cross, one of the $ f_i + 1 $ edges must originate from a state that is not in $ Q[1, |G^\prec(p_{|\alpha| - i}(\alpha))|] $, so $ Q[j + 1] $ cannot be a state in $ Q[1, |G^\prec(\alpha)|] $.
\end{itemize}

\begin{figure}
     \centering
        \scalebox{.8}{
        \begin{tikzpicture}[->,>=stealth', semithick, auto, scale=1]
\node[state, label=below:{$Q[1]$}] (1)    at (0,0)	{};
\node[state, label=below:{$Q[2]$}] (2)    at (2,0)	{};
\node[state, label=below:{$Q[3]$}] (3)    at (4,0)	{};
\node[state, draw=none] (4)    at (6,0)	{$ \dots $};
\node[state, label=below:{$Q[|G^\prec(p_{|\alpha| - i}(\alpha))|]$}] (5)    at (8, 0)	{};
\node[state, draw=none] (6)    at (10, 0)	{$ \dots $};
\node[state] (7)    at (12, 0)	{};
\node[state, draw=none] (8)    at (14, 0)	{$ \dots $};
\node[state, label=below:{$Q[|Q|]$}] (9)    at (16, 0)	{};

\node[state, label=above:{$Q[1]$}] (10)    at (0,4)	{};
\node[state, label=above:{$Q[2]$}] (11)    at (2,4)	{};
\node[state, draw=none] (12)    at (4,4)	{$ \dots $};
\node[state, label=above:{$Q[j]$}] (13)    at (6,4)	{$ $};
\node[state, label=above:{$Q[j + 1]$}] (14)    at (8, 4)	{};
\node[state, draw=none] (15)    at (10, 4)	{$ \dots $};
\node[state, draw=none] (16)    at (12, 4)	{$ \dots $};
\node[state, draw=none] (17)    at (14, 4)	{$ \dots $};
\node[state, label=above:{$Q[|Q|]$}] (18)    at (16, 4)	{};

\draw (1) edge [bend left] node [] {$ s_i(\alpha) $} (10);
\draw (2) edge [] node [] {$ s_i(\alpha) $} (10);
\draw (3) edge [] node [] {$ s_i(\alpha) $} (11);
\draw (5) edge [] node [] {$ s_i(\alpha) $} (13);
\draw (7) edge [] node [] {$ s_i(\alpha) $} (14);

\end{tikzpicture}
}

	\caption{The intuition behind Lemma~\ref{lem:computationalversion}.}
 \label{fig:intuitionleft}
\end{figure}

\begin{lem}\label{lem:computationalversion}
    Let $ \mathcal{A} = (Q, E, s, F) $ be a Wheeler $ r $-GDFA, and let $\alpha \in \Sigma^*$, with $ \alpha \not = \epsilon $. For $ 1 \le i \le \min\{r, |\alpha| - 1 \} $, let $ f_i = \mathtt{out}(Q[1, |G^\prec(p_{|\alpha| - i}(\alpha))|], s_i(\alpha)) $. Then, $ |G^\prec(\alpha)| $ is the largest integer $ 0 \le j \le |Q| $ such that (i) $ \rho \prec \alpha $ for every $ 1 \le t \le j $ and for every $ \rho \in \lambda (Q[t]) $, and (ii) $ \mathtt{in}(Q[1, j], s_i(\alpha)) \leq f_i $ for every $ 1 \le i \le \min\{r, |\alpha| - 1 \} $.
\end{lem}

\begin{proof}
    To prove the lemma, we will repeatedly use the equality stated in point 3 of Lemma~\ref{lem:st}. Let $ j^* $ be the largest integer $ 0 \le j \le |Q| $ such that (i) $ \rho \prec \alpha $ for every $ 1 \le t \le j $ and for every $ \rho \in \lambda (Q[t]) $, and (ii) $ \mathtt{in}(Q[1, j], s_i(\alpha)) \leq f_i $ for every $ 1 \le i \le \min\{r, |\alpha| - 1 \} $. We want to prove that $ |G^\prec(\alpha)| = j^* $.

    $ (\le) $ By the maximality of $ j^* $, it will suffice to prove that (i) $ \rho \prec \alpha $ for every $ 1 \le t \le |G^\prec(\alpha)| $ and for every $ \rho \in \lambda (Q[t]) $, and (ii) $ \mathtt{in}(Q[1, |G^\prec(\alpha)|], s_i(\alpha)) \leq f_i $ for every $ 1 \le i \le \min\{r, |\alpha| - 1 \} $.
    
    Let us prove (i). Fix $ 1 \le t \le |G^\prec(\alpha)| $ and $ \rho \in \lambda (Q[t]) $. We must prove that $ \rho \prec \alpha $. From $ 1 \le t \le |G^\prec(\alpha)| $ we obtain $ Q[t] \in G^\prec(\alpha) $, so by Lemma~\ref{lem:Si alpha} we conclude $ \rho \prec \alpha $.

    Let us prove (ii). Fix $ 1 \le i \le \min\{r, |\alpha| - 1 \} $. By Lemma~\ref{lem:Si alpha}, we know that for every $ (u', u, s_i(\alpha)) \in E $, if $ u \in G^\prec(\alpha) = Q[1, |G^\prec(\alpha)|] $, then $ u' \in G^\prec(p_{|\alpha| - i}(\alpha)) = Q[1, |G^\prec(p_{|\alpha| - i}(\alpha))|] $. The conclusion follows from the definition of $ f_i $.

    $ (\ge) $ We only have to prove that if $ |G^\prec(\alpha)| + 1 \le j \le |Q| $, then at least one of the following statements is not true:
    \begin{itemize}
        \item (a) $ \rho \prec \alpha $ for every $ 1 \le t \le j $ and for every $ \rho \in \lambda (Q[t]) $.
        \item (b) $ \mathtt{in}(Q[1, j], s_i(\alpha)) \leq f_i $ for every $ 1 \le i \le \min\{r, |\alpha| - 1 \} $.
    \end{itemize}
    Fix $ |G^\prec(\alpha)| + 1 \le j \le |Q| $, and assume that (a) is true. Then, we must prove that (b) is not true.

    Since $ |G^\prec(\alpha)| + 1 \le j \le |Q| $, we have $ Q[j] \not \in G^\prec(\alpha) $. We also know that for every $ \rho \in \lambda (Q[j]) $ we have $ \rho \prec \alpha $, so by Lemma~\ref{lem:Si alpha} we conclude that for some $ 1 \le i \le |\alpha| - 1 $ there exists $ v' \in Q $ such that $ (v', Q[j], s_i(\alpha)) \in E $ and $ v' \not \in G^\prec(p_{|\alpha| - i}(\alpha)) = Q[1, |G^\prec(p_{|\alpha| - i}(\alpha))|] $. Since $ \mathcal{A} $ is an $ r $-GDFA, we have $ 1 \le s_i(\alpha) \le r $, so $ 1 \le i \le \min\{r, |\alpha| - 1 \} $. Let us prove that $ \mathtt{in}(Q[1, j], s_i(\alpha)) > f_i $, which will imply that (b) is false. We know that $ (v', Q[j], s_i(\alpha)) \in E $ and $ v' \not \in Q[1, |G^\prec(p_{|\alpha| - i}(\alpha))|] $, so by the definition of $ f_i $ we only have to prove that, if $ u', u \in Q $ are such that $ u' \in Q[1, |G^\prec(p_{|\alpha| - i}(\alpha))|] $ and $ (u', u, s_i(\alpha)) \in E $, then $ u \in Q[1, j] $. Suppose for the sake of a contradiction that $ u \not \in Q[1, j] $. This implies that $ Q[j] \prec_\mathcal{A} u $, so from $ (v', Q[j], s_i(\alpha)), (u', u, s_i(\alpha)) \in E $ and Lemma~\ref{lem:local GDFA} we obtain $ v' \prec_\mathcal{A} u' $. Since $ u' \in Q[1, |G^\prec(p_{|\alpha| - i}(\alpha))|] $, we conclude $ v' \in Q[1, |G^\prec(p_{|\alpha| - i}(\alpha))|] $, a contradiction. \qedhere
\end{proof}

Let us show how to compute $ G^\prec_\dashv(\alpha) $. To this end, let $ G^*(\alpha) $ be the set of all states reached by an edge labeled with a string suffixed by $ \alpha $. Formally:
\begin{equation*}
        G^* (\alpha) = \{u \in Q \;|\; (\exists \rho \in \lambda (u))(\alpha \dashv \rho) \}.
\end{equation*}

Notice that $ G^* (\alpha) \subseteq G_\dashv (\alpha) $. Indeed, pick $ u \in G^* (\alpha) $. Then, there exists $ \rho \in \lambda (u) $ such that $ \alpha \dashv \rho $. In particular, there exists $ u' \in Q $ such that $ (u', u, \rho) \in E $. Pick any $ \beta \in I_{u'} $. Then, $ \beta \rho \in I_u $. From $ \alpha \dashv \rho $ we conclude $ \alpha \dashv \beta \rho $, so $ u \in G_\dashv (\alpha) $.

The following lemma shows how to compute $ G^\prec_\dashv(\alpha) $. In the proof, we will use point 3 of Lemma~\ref{lem:local GDFA} multiple times (recall that, loosely speaking, it states that edges with the same label cannot cross, see Figure~\ref{fig:intuitionleft}). Here is the intuition to prove the lemma:
\begin{itemize}
    \item For every $ 1 \le j \le |Q| $, if $ Q[j] \in G^\prec(\alpha) $, then $ Q[j] \in G^\prec_\dashv(\alpha) $.
    \item For every $ 1 \le j \le |Q| $, if $ Q[j] \in G^*(\alpha) $, then $ Q[j] \in G_\dashv (\alpha) $ and so $ Q[j] \in G^\prec_\dashv(\alpha) $.
    \item Fix $ 1 \le i \le r $. Assume that there exists at least one edge labeled $ s_i (\alpha) $ leaving a state in $ G_\dashv(p_{|\alpha| - i}(\alpha)) $. Equivalently, if $ f_i = \mathtt{out}(Q[1, |G^\prec(p_{|\alpha| - i}(\alpha))|], s_i(\alpha)) $ and $ g_i = \mathtt{out}(Q[1, |G^\prec_\dashv(p_{|\alpha| - i}(\alpha))|], s_i(\alpha)) $, assume that $ g_i > f_i $. Since the edges labeled with $ s_i (\alpha) $ cannot cross, intuitively if $ 1 \le j \le |Q| $ is such that $\mathtt{in}(Q[1, j - 1], s_i(\alpha)) < g_i $, then we must have $ Q[j] \in G^\prec_\dashv(\alpha) $ (see Figure~\ref{fig:intuitionleftplusmiddle}). Indeed, there must exists at least one edge labeled $ s_i (\alpha) $ leaving a state in $ G_\dashv(p_{|\alpha| - i}(\alpha)) $ that does not reach a state in $\mathtt{in}(Q[1, j - 1], s_i(\alpha)) $.
\end{itemize}

\begin{figure}
     \centering
        \scalebox{.8}{
        \begin{tikzpicture}[->,>=stealth', semithick, auto, scale=1]
\node[state, label=below:{$Q[1]$}] (1)    at (0,0)	{};
\node[state, label=below:{$Q[2]$}] (2)    at (2,0)	{};
\node[state, label=below:{$Q[3]$}] (3)    at (4,0)	{};
\node[state, label=below:{$Q[4]$}] (4)    at (6,0)	{};
\node[state, draw=none] (5)    at (8,0)	{$ \dots $};
\node[state, label=below:{$Q[|G^\prec_\dashv(p_{|\alpha| - i}(\alpha))|]$}] (6)    at (10, 0)	{};
\node[state, draw=none] (7)    at (12, 0)	{$ \dots $};
\node[state, label=below:{$Q[|Q|]$}] (8)    at (14, 0)	{};

\node[state, label=above:{$Q[1]$}] (9)    at (0,4)	{};
\node[state, label=above:{$Q[2]$}] (10)    at (2,4)	{};
\node[state, draw=none] (11)    at (4,4)	{$ \dots $};
\node[state, label=above:{$Q[j - 1]$}] (12)    at (6,4)	{$ $};
\node[state, label=above:{$Q[j]$}] (13)    at (8, 4)	{};
\node[state, draw=none] (14)    at (10, 4)	{$ \dots $};
\node[state, draw=none] (15)    at (12, 4)	{$ \dots $};
\node[state, label=above:{$Q[|Q|]$}] (16)    at (14, 4)	{};

\draw (1) edge [bend left] node [] {$ s_i(\alpha) $} (9);
\draw (2) edge [] node [] {$ s_i(\alpha) $} (9);
\draw (3) edge [] node [] {$ s_i(\alpha) $} (10);
\draw (6) edge [] node [] {$ s_i(\alpha) $} (13);

\end{tikzpicture}
}

	\caption{The intuition behind Lemma~\ref{lem:Ri alpha2}.}
 \label{fig:intuitionleftplusmiddle}
\end{figure}  

\begin{lem}\label{lem:Ri alpha2}
Let $ \mathcal{A} = (Q, E, s, F) $ be a Wheeler $ r $-GDFA, and let $\alpha \in \Sigma^* $, with $ \alpha \not = \epsilon $. For $ 1 \le i \le \min\{r, |\alpha| - 1 \} $, let $ f_i = \mathtt{out}(Q[1, |G^\prec(p_{|\alpha| - i}(\alpha))|], s_i(\alpha)) $ and $ g_i = \mathtt{out}(Q[1, |G^\prec_\dashv(p_{|\alpha| - i}(\alpha))|], s_i(\alpha)) $. Then, $ g_i \ge f_i $ for every $ 1 \le i \le \min\{r, |\alpha| - 1 \} $. Moreover, $ |G^\prec_\dashv(\alpha) | $ is equal to the maximum among:
\begin{itemize}
    \item $ |G^\prec(\alpha)| $.
    \item the largest integer $ 0 \le j \le |Q| $ such that, if $ j \ge 1 $, then $ Q[j] \in G^* (\alpha) $.
    \item the smallest integer $ 0 \le j \le |Q| $ such that, for every $ 1 \le i \le \min\{r, |\alpha| - 1 \} $ for which $ g_i > f_i $, we have $\mathtt{in}(Q[1, j], s_i(\alpha)) \ge g_i $.
\end{itemize}
\end{lem}

\begin{proof}
To prove the lemma, we will repeatedly use the equalities stated in points 3, 4 and 5 of Lemma~\ref{lem:st}. For every $ 1 \le i \le \min\{r, |\alpha| - 1 \} $, we have $ g_i \ge f_i $ because $ G^\prec(p_{|\alpha| - i}(\alpha)) \subseteq G^\prec_\dashv(p_{|\alpha| - i}(\alpha)) $. Let $ j_1^* $ be the largest integer $ 0 \le j \le |Q| $ such that, if $ j \ge 1 $, then $ Q[j] \in G^* (\alpha) $, and let $ j_2^* $ be the smallest integer $ 0 \le j \le |Q| $ such that $\mathtt{in}(Q[1, j], s_i(\alpha)) \ge g_i $ for every $ 1 \le i \le \min\{r, |\alpha| - 1 \} $ for which $ g_i > f_i $. Let $ l = \max \{|G^\prec(\alpha)|, j_1^*, j_2^*\} $. We want to prove that $ |G^\prec_\dashv(\alpha)| = l $.

$ (\ge) $ We have to prove that $ |G^\prec(\alpha)| \le |G^\prec_\dashv(\alpha)| $, $ j_1^* \le |G^\prec_\dashv(\alpha)| $ and $ j_2^* \le |G^\prec_\dashv(\alpha)| $. The inequality $ |G^\prec(\alpha)| \le |G^\prec_\dashv(\alpha)| $ follows from $ G^\prec(\alpha) \subseteq G^\prec_\dashv(\alpha) $. Let us prove that $ j_1^* \le |G^\prec_\dashv(\alpha)| $. If $ j_1^* = 0 $, we are done, so we can assume $ j_1^* \ge 1 $. We know that $ Q[j_1^*] \in G^* (\alpha) $, so $ Q[j_1^*] \in G_\dashv (\alpha) $ and $ Q[j_1^*] \in G^\prec_\dashv (\alpha) $, which implies $ j_1^* \le |G^\prec_\dashv(\alpha)| $.

Let us prove that $ j_2^* \le |G^\prec_\dashv(\alpha)| $. If $ j_2^* = 0 $, we are done, so we can assume $ j_2^* \ge 1 $. We know that there exists $ 1 \le i \le \min\{r, |\alpha| - 1 \} $ such that $ g_i > f_i $, $\mathtt{in}(Q[1, j_2^* - 1], s_i(\alpha)) < g_i $ and $\mathtt{in}(Q[1, j_2^*], s_i(\alpha)) \ge g_i $. We are only left with showing that there exists $ u' \in G_\dashv(p_{|\alpha| - i}(\alpha)) $ such that $ (u', Q[j_2^*], s_i(\alpha)) \in E $, because this will imply $ Q[j_2^*] \in G_\dashv(\alpha) $, so $ Q[j_2^*] \in G^\prec_\dashv(\alpha) $ and $ j_2^* \le |G^\prec_\dashv(\alpha)| $. Suppose for the sake of a contradiction that for every $ u' \in Q $ such that $ (u', Q[j_2^*], s_i(\alpha)) \in E $ we have $ u' \not \in G_\dashv(p_{|\alpha| - i}(\alpha)) $.

First, let us prove that there cannot exist two states $ u'_1 \in Q[1, |G^\prec(p_{|\alpha| - i}(\alpha))|] $ and $ u'_2 \in Q[|G^\prec_\dashv(p_{|\alpha| - i}(\alpha))| + 1, |Q|] $ such that $ (u'_1, Q[j_2^*], s_i(\alpha)) \in E $ and $ (u'_2, Q[j_2^*], s_i(\alpha)) \in E $. Suppose for the sake of a contradiction that such $ u'_1 $ and $ u'_2 $ exist. We will obtain a contradiction by showing that the existence of $ u'_1 $ and $ u'_2 $ implies that there cannot exist $ u'_3 \in G_\dashv(p_{|\alpha| - i}(\alpha)) $ and $ u_3 \in Q $ such that $ (u'_3, u_3, s_i(\alpha)) \in E $, because this would imply $ g_i = f_i $, which contradicts $ g_i > f_i $.

Suppose for the sake of a contradiction that there exist $ u'_3 \in G_\dashv(p_{|\alpha| - i}(\alpha)) $ and $ u_3 \in Q $ such that $ (u'_3, u_3, s_i(\alpha)) \in E $. We obtain a contradiction by distinguishing three cases:
\begin{itemize}
    \item we cannot have $ u_3 = Q[j_2^*] $ because for every $ u' \in Q $ such that $ (u', Q[j_2^*], s_i(\alpha)) \in E $ we have $ u' \not \in G_\dashv(p_{|\alpha| - i}(\alpha)) $.
    \item we cannot have $ u_3 \prec_\mathcal{A} Q[j_2^*] $ because, if we consider $ (u'_3, u_3, s_i(\alpha)), (u'_1, Q[j_2^*], s_i(\alpha)) \in E $, by Lemma~\ref{lem:local GDFA} we would obtain $ u'_3 \prec_\mathcal{A} u'_1 $, and since $ u'_1 \in Q[1, |G^\prec(p_{|\alpha| - i}(\alpha))|] $, we would conclude $ u'_3 \in Q[1, |G^\prec(p_{|\alpha| - i}(\alpha))|] $, which contradicts $ u'_3 \in G_\dashv(p_{|\alpha| - i}(\alpha)) $.
    \item we cannot have $ Q[j_2^*] \prec_\mathcal{A} u_3 $ because, if we consider $ (u'_2, Q[j_2^*], s_i(\alpha)), (u'_3, u_3, s_i(\alpha)) \in E $, by Lemma~\ref{lem:local GDFA} we would obtain $ u'_2 \prec_\mathcal{A} u'_3 $, and since $ u'_2 \in Q[|G^\prec_\dashv(p_{|\alpha| - i}(\alpha))| + 1, |Q|] $, we would conclude $ u'_3 \in Q[|G^\prec_\dashv(p_{|\alpha| - i}(\alpha))| + 1, |Q|] $, which contradicts $ u'_3 \in G_\dashv(p_{|\alpha| - i}(\alpha)) $.
\end{itemize}

We have proved that there cannot exist two states $ u'_1 \in Q[1, |G^\prec(p_{|\alpha| - i}(\alpha))|] $ and $ u'_2 \in Q[|G^\prec_\dashv(p_{|\alpha| - i}(\alpha))| + 1, |Q|] $ such that $ (u'_1, Q[j_2^*], s_i(\alpha)) \in E $ and $ (u'_2, Q[j_2^*], s_i(\alpha)) \in E $. Moreover, we know that for every $ u' \in Q $ such that $ (u', Q[j_2^*], s_i(\alpha)) \in E $ we have $ u' \not \in G_\dashv(p_{|\alpha| - i}(\alpha)) $. As a consequence, one of the following two cases must occur.

\begin{itemize}
    \item For every $ u'_1 \in Q $ such that $ (u'_1, Q[j_2^*], s_i(\alpha)) \in E $ we have $ u'_1 \in Q[1, |G^\prec(p_{|\alpha| - i}(\alpha))|] $. Let us prove that if $ v', v \in Q $ are such that $ v \in Q[1, j_2^*] $ and $ (v', v, s_i(\alpha)) \in E $, then $ v' \in Q[1, |G^\prec(p_{|\alpha| - i}(\alpha))|] $. This will imply $\mathtt{in}(Q[1, j_2^*], s_i(\alpha)) \le f_i < g_i $, the desired contradiction. If $ v = Q[j_2^*] $, the conclusion follows because we know that for every $ u'_1 \in Q $ such that $ (u'_1, Q[j_2^*], s_i(\alpha)) \in E $ we have $ u'_1 \in Q[1, |G^\prec(p_{|\alpha| - i}(\alpha))|] $. Now, assume that $ v \prec_\mathcal{A} Q[j_2^*] $. Since $\mathtt{in}(Q[1, j_2^* - 1], s_i(\alpha)) < g_i $ but $\mathtt{in}(Q[1, j_2^*], s_i(\alpha)) \ge g_i $, then there exists $ u^*_1 \in Q $ such that $ (u^*_1, Q[j_2^*], s_i(\alpha)) \in E $, and we know that it must be $ u^*_1 \in Q[1, |G^\prec(p_{|\alpha| - i}(\alpha))|] $. If we consider $ (v', v, s_i(\alpha)), (u^*_1, Q[j_2^*], s_i(\alpha)) \in E $, from $ v \prec_\mathcal{A} Q[j_2^*] $ and Lemma~\ref{lem:local GDFA} we obtain $ v' \prec_\mathcal{A} u^*_1 $, and since $ u^*_1 \in Q[1, |G^\prec(p_{|\alpha| - i}(\alpha))|] $, we conclude $ v' \in Q[1, |G^\prec(p_{|\alpha| - i}(\alpha))|] $.
    \item For every $ u'_2 \in Q $ such that $ (u'_2, Q[j_2^*], s_i(\alpha)) \in E $ we have $ u'_2 \in Q[|G^\prec_\dashv(p_{|\alpha| - i}(\alpha))| + 1, |Q|] $. Let us prove that if $ v', v \in Q $ are such that $ v' \in Q[1, |G^\prec_\dashv((p_{|\alpha| - i}(\alpha))|] $ and $ (v', v, s_i(\alpha)) \in E $, then $ v \in Q[1, j_2^* - 1] $. This will imply $\mathtt{in}(Q[1, j_2^* - 1], s_i(\alpha)) \ge g_k $, the desired contradiction. It cannot be $ v = Q[j_2^*] $, because this would imply $ v' \in Q[|G^\prec_\dashv(p_{|\alpha| - i}(\alpha))| + 1, |Q|] $. Now, assume for the sake of contradiction that $ Q[j_2^*] \prec_\mathcal{A} v $. Since $\mathtt{in}(Q[1, j_2^* - 1], s_i(\alpha)) < g_i $ but $\mathtt{in}(Q[1, j_2^*], s_i(\alpha)) \ge g_i $, then there exists $ u^*_2 \in Q $ such that $ (u^*_2, Q[j_2^*], s_i(\alpha)) \in E $, and we know that it must be $ u^*_2 \in Q[|G^\prec_\dashv(p_{|\alpha| - i}(\alpha))| + 1, |Q|] $. If we consider $ (u^*_2, Q[j_2^*], s_i(\alpha)), (v', v, s_i(\alpha)) \in E $, from $ Q[j_2^*] \prec_\mathcal{A} v $ and Lemma~\ref{lem:local GDFA} we obtain $ u^*_2 \prec_\mathcal{A} v' $, and since $ u^*_2 \in Q[|G^\prec_\dashv(p_{|\alpha| - i}(\alpha))| + 1, |Q|] $ , we conclude $ v' \in Q[|G^\prec_\dashv(p_{|\alpha| - i}(\alpha))| + 1, |Q|]  $, a contradiction.
\end{itemize}

$ (\le) $ We only have to prove that if $ l + 1 \le j \le |Q| $, then $ Q[j] \not \in G^\prec_\dashv(\alpha) $. Since $ j \ge l + 1 > |G^\prec(\alpha)| $, we have $ Q[j] \not \in G^\prec(\alpha) $, hence we are left with showing that $ Q[j] \not \in G_\dashv(\alpha) $. Assume for the sake of contradiction that $ Q[j] \in G_\dashv(\alpha) $. It cannot be $ Q[j] \in G^*(\alpha) $, otherwise $ j \le j_1^* \le l $, a contradiction. Hence, $ Q[j] \in G_\dashv(\alpha) \setminus G^* (\alpha) $. Since $ \mathcal{A} $ is an $ r $-GDFA, this means that for some $ 1 \le i \le \min\{r, |\alpha| -1 \} $ there exists $ u' \in Q $ such that $ (u', Q[j], s_i(\alpha)) \in E $ and $ u' \in G_\dashv(p_{|\alpha| - i}(\alpha)) $, which implies $ g_i > f_i $. Let us prove that $\mathtt{in}(Q[1, j - 1], s_i(\alpha)) < g_i $. Assume that $ (v', v, s_i(\alpha)) \in E $ is such that $ v \in Q[1, j - 1] $. If we consider $ (v', v, s_i(\alpha)), (u', Q[j], s_i(\alpha)) \in E $, from $ v \prec_\mathcal{A} Q[j] $ and Lemma~\ref{lem:local GDFA} we obtain $ v' \prec_\mathcal{A} u' $, and since $ u' \in G_\dashv(p_{|\alpha| - i}(\alpha)) $, and so $ u' \in Q[1, |G^\prec_\dashv(p_{|\alpha| - i}(\alpha))|] $, we obtain $ v' \in Q[1, |G^\prec_\dashv(p_{|\alpha| - i}(\alpha))|] $. This implies that $\mathtt{in}(Q[1, j - 1], s_i(\alpha)) \le g_i $, and it must be $\mathtt{in}(Q[1, j - 1], s_i(\alpha)) < g_i $ because $ (u', Q[j], s_i(\alpha)) \in E $ and $ u' \in G_\dashv(p_{|\alpha| - i}(\alpha)) $. We conclude $ j \le j_2^* \le l $, a contradiction. \qedhere
\end{proof}

\begin{rem}\label{rem:suffixtrivial}
    Consider the statement of Lemma~\ref{lem:Ri alpha2}. If $ |\alpha| > r $, then the largest integer $ 0 \le j \le |Q| $ such that, if $ j \ge 1 $, then $ Q[j] \in G^* (\alpha) $ is equal to $ 0 $. Intuitively, this will imply that, when we compute $ |G^\prec_\dashv(\alpha) | $ using Lemma~\ref{lem:Ri alpha2}, we will only need to perform a number of operations linear in $ r $.
\end{rem}

\subsection{The FM-index of a Wheeler GDFA} In this section, we discuss how to achieve the space bound in Theorem~\ref{theor:fmindexintroduction}. Broadly speaking, our goal is to store the Burrows-Wheeler Transform of a Wheeler GDFA (Definition~\ref{def:BWTwheelerdfa}) using compressed data structures. To this end, let us recall one of the cornerstone results in the field: we can store a string of length $ t $ over an alphabet of size $ \sigma $ using only slightly more than $ t \log \sigma $ bits in such a way that we can solve the crucial \emph{rank} and \emph{select} operations efficiently. This is remarkable because (i) there are $ \sigma^t $ strings of length $ t $ over an alphabet of size $ \sigma $, so a simple pigeonhole argument shows that we need at least $ \log (\sigma^t) = t \log \sigma $ bits to distinguish them~\cite[Section 2.1]{navarro2016}, and (ii) rank and select can be used to efficiently implement countless operations (see~\cite{navarro2016}).

\begin{lem}[~\cite{navarro2016}, Chapter 6]\label{lem:rankselectbasic}
    Let $\Sigma = \{0, 1, \dots, \sigma- 1 \} $ be an integer alphabet and let $ \alpha \in \Sigma^* $, with $ |\alpha|  = t $. If $ \sigma \le t $, then $ \alpha $ can be encoded using a data structure of $ t \log \sigma (1+o(1)) + O(t) $ bits that supports the following operations in $ O (\log \log \sigma) $ time:
    
    \begin{itemize}
        \item $ \alpha.access(i) $, for $ 1 \le i \le t $: return $ \alpha[i] $.
        \item $ \alpha.rank(i,c) $, for $ c \in \Sigma $ and $ 0 \le i \le t $: return $ |\{1 \le j \le i \ |\ \alpha[j]=c\}| $.
        \item $ \alpha.select(i,c) $, for $ c \in \Sigma $ and $ 1 \le i \le \alpha.rank(t, c) $: return the unique integer $ 1 \le j \le t $ such that (i) $ \alpha[j] = c $ and (ii) $ \alpha.rank(j,c) = i $.
    \end{itemize}
\end{lem}

For example, if $ \alpha = abaaaabaab $, then we have $ \alpha.access(2) = b $, $ \alpha.rank(5, a) = 4 $ and $ \alpha.select(4, a) = 5 $. Rank and select are closely related: for every $ c \in \Sigma $ and for every $ 1 \le i \le \alpha.rank(t, c) $, we have $ \alpha.rank(\alpha.select(i,c), c) ) = i $. Note that not only is the data structure of Lemma~\ref{lem:rankselectbasic} an encoding of $ \alpha $ (that is, from the data structure we can retrieve $ \alpha $), but we can also retrieve $ \alpha $ quickly through access operations.

\begin{rem}
    To handle some boundary cases easily, it is expedient to introduce some extensions of rank and select:
    \begin{itemize}
        \item $ \alpha.rank(i,c) $, for $ c \in \Sigma $ and $ i \ge 0 $: return $ |\{1 \le j \le \min \{i, t \} \ |\ \alpha[j]=c\}| $.
        \item $ \alpha.select(i,c) $, for $ c \in \Sigma $ and $ i \ge 1 $: if $ i \le \alpha.rank(t, c) $, return the unique integer $ 1 \le j \le t $ such that (i) $ \alpha[j] = c $ and (ii) $ \alpha.rank(j,c) = i $, and if $ i > \alpha.rank(t, c) $, return $ t + 1 $.
    \end{itemize}
    The data structure of Lemma~\ref{lem:rankselectbasic} can compute these extensions in $ O(\log \log \sigma) $ time. Indeed, to solve the extended rank, we only need to check whether $ i \le t $, and to solve the extended select, we only need to check (in $ O(\log \log \sigma) $ time) whether $ i \le \alpha.rank(t, c) $.
\end{rem}

The next lemma shows that we can store a \emph{dictionary} on $ \Sigma $ (that is, a subset of $ \Sigma $) within compressed space in such a way that we can solve a variant of rank and select efficiently. Intuitively, we need dictionaries for two reasons:
\begin{itemize}
    \item Fix $ 1 \le i \le r $. To store the Wheeler GDFA $ \mathcal{A} $, we need in particular to store every string in $ \Sigma^i $ labeling some edge. Given such a string, we might naively store each of the $ i $ characters independently. However, this is not the best approach to achieve the space bound in Theorem~\ref{theor:fmindexintroduction}: typically, only some strings in $ \Sigma^i $ label some edge (that is, $ \Sigma_i $ is strictly contained in $ \Sigma^i $, and $ \sigma_i $ could be much smaller than $ \sigma^i $). We will save space by using a dictionary to remember which strings in $ \Sigma^i $ are also in $ \Sigma_i $.
    \item We want to keep our result general without imposing any restriction on the alphabet. This means that the alphabet $ \Sigma $ need not be effective, that is, some characters in $ \Sigma $ may label no edge of $ \mathcal{A} $. Consequently, we intuitively need a dictionary to remember which characters in $ \Sigma $ label some edge.
\end{itemize}

\begin{lem}[~\cite{feigenblat2016linear}, Theorem 4.1]\label{lem:dictionary}
Let $ \Sigma = \{0, 1, \dots, \sigma - 1 \} $ be an integer alphabet, and let $ A \subseteq \Sigma $, with $ |A| = t $. Then, $ A $ can be encoded using a data structures of $ t \log (\sigma / t) + O(t) $ bits that supports the following operations in $ O(\log\log (\sigma/t))$ time:
\begin{itemize}
    \item $ A.rank(i) $, for $ 0 \le i \le \sigma - 1 $: return $ |\{j \in A\ |\ j\leq i\}| $.
    \item $ A.select(i) $, for $ 1 \le i \le t $: return the unique integer $ 0 \le j \le \sigma - 1 $ such that (i) $ j \in A $ and (ii) $ A.rank(j) = i $.
\end{itemize}
\end{lem}

\begin{rem}\label{rem:membershipquery}
The data structure of Lemma~\ref{lem:dictionary} also supports the following operations in $ O(\log\log (\sigma/t)) $ time:

\begin{itemize}
    \item $ A.memb(i) $, for $ 0 \le i \le \sigma - 1 $: decide whether $ i \in A $ (\emph{membership}).
    \item $ A.prec(i) $, for $ 0 \le i \le \sigma - 1 $: return the largest integer $ 0 \le j < i $ such that $ j \in A $, if such a $ j $ exists, otherwise return $ \bot $ (\emph{predecessor}).
    \item $ A.succ(i) $, for $ 0 \le i \le \sigma - 1 $: return the smallest integer $ i < j \le \sigma - 1 $ such that $ j \in A $, if such a $ j $ exists, otherwise return $ \bot $ (\emph{successor}).
\end{itemize}

Indeed, we can compute these operations as follows.

\begin{itemize}
    \item Let us show how to compute $ A.memb(i) $ in $ O(\log\log (\sigma/t)) $ time. We have $ i \in A $ if and only if $ A.select(A.rank(i)) = i $.
    \item Let us show how to compute $ A.prec(i) $ in $ O(\log\log (\sigma/t)) $ time. If $ i = 0 $, return $ \bot $. Now assume that $ i \ge 1 $. Compute $ A.rank (i - 1) $. If $ A.rank (i - 1) = 0 $, return $ \bot $, and if $ A.rank (i - 1) > 0 $, return $ A.select (A.rank (i - 1)) $.
    \item Let us show how to compute $ A.succ(i) $ in $ O(\log\log (\sigma/t)) $ time. Compute $ A.rank (i) $ and $ A.rank (\sigma - 1) $. If $ A.rank (i) = A.rank (\sigma - 1) $, return $ \bot $, and if $ A.rank (i) < A.rank (\sigma - 1) $, return $ A.select (A.rank (i) + 1) $.
\end{itemize}

Note that the membership operation confirms that the data structure of Lemma~\ref{lem:dictionary} is an encoding of $ A $.
\end{rem}

We are ready to extend the FM-index to Wheeler GDFAs. Recall that $ e $ is the number of edges, $ \mathfrak{e} $ is the total length of all edge labels and $ \sigma = |\Sigma| $.

\begin{thm}[FM-index of Wheeler GDFAs]\label{theor:fmindexwheelergdfa}
    Let $ \mathcal{A} = (Q, E, s, F) $ be a Wheeler $ r $-GDFA, with $ \sigma \le \mathfrak{e}^{O(1)} $ and $ r = O(1) $. Then, we can encode $ \mathcal{A} $ by using $ \mathfrak{e} \log \sigma (1 + o(1)) + O(e) $ bits so that later on, given a pattern $ \alpha \in \Sigma^* $ of length $ m $, we can solve the SMLG problem on $ \mathcal{A} $ in $ O(m \log \log \sigma) $ time. Within the same time bound, we can also decide whether $ \alpha $ is recognized by $ \mathcal{A} $.
\end{thm}

We divide the proof of Theorem~\ref{theor:fmindexwheelergdfa} into four steps:

\begin{itemize}
    \item In the first step, we describe our encoding and we prove the claimed space bound ($ \mathfrak{e} \log \sigma (1 + o(1)) + O(e) $ bits).
    \item In the second step, we show how to use our encoding to solve auxiliary operations. The auxiliary operations will be helpful in the third step.
    \item In the third step, we show how to use our encoding to solve the SMLG problem on $ \mathcal{A} $ in $ O(m \log \log \sigma) $ time.
    \item In the fourth step, we show how to decide in $ O(m \log \log \sigma) $ time whether $ \alpha $ is recognized by $ \mathcal{A} $.
\end{itemize}

Before describing the details of the proof, note that, since $ \sigma \le \mathfrak{e}^{O(1)} $ and $ r = O(1) $, for every $ 1 \le i \le r $ the elements of $ \Sigma^i $ fit in a constant number of computer words and thus can be manipulated in constant time. Let $ n = |Q| $. Recall that $ e \ge n - 1 $ because every state is reachable from the initial state, so every state different from the initial state must have an incoming edge. Moreover, recall that $ \Sigma = \{0, 1, \dots, \sigma - 1 \} $, with $ 0 \prec 1 \prec \dots \prec \sigma - 1 $. 

\subsubsection{Step 1: encoding and space bound}

Let us describe our encoding. Here is the intuition behind the encoding. First, for every $ 1 \le i \le r $, we will map each string in $ \Sigma_i $ to an integer, obtaining the set $ \Sigma_i^* $. In this way, we can manipulate each $ \Sigma_i^* $ through a dictionary (Lemma~\ref{lem:dictionary}), which will be expedient to achieve the claimed space bound and solve the SMLG problem quickly. Then, we store each component of the Burrows-Wheeler Transform of $ \mathcal{A} $ (Definition~\ref{def:BWTwheelerdfa}) through the data structure of Lemma~\ref{lem:rankselectbasic}, with two caveats: (i) we store a variant $ \mathtt{LAB}^*_i $ of each $ \mathtt{LAB}_i $, and (ii) we store some auxiliary strings $ \mathtt{AUX}^*_i $'s. We store the variant $ \mathtt{LAB}^*_i $ to reduce the space of our data structures: since we can quickly check if a string is in $ \Sigma_i^* $ (and so in $ \Sigma_i $) through a dictionary, we can map each element in $ \Sigma_i $ to an element in $ \{0, 1, \dots, \sigma_i - 1 \} $, which saves space (because each element in $ \Sigma_i $ is a string of length $ i $ on $ \Sigma $, and $ \sigma_i $ may be much smaller than $ \sigma^i $). The auxiliary strings $ \mathtt{AUX}^*_i $'s only require $ O(e) $ bits and are helpful to solve the SMLG problem quickly. We will now give a formal description of our data structures.
    
We first show how to map each string in $ \Sigma_i $ to an integer. For every $ 1 \le i \le r $, let $ \psi_i $ be the bijection from $ \Sigma^i $ to $ \{0, 1, \dots, \sigma^i - 1 \} $ that maps every element $ a_1 a_2 \dots a_{i - 1} a_i $ in $ \Sigma^i $ to the representation in base $ \sigma $ of the reverse string $ a_i a_{i - 1} \dots a_2 a_1 $. In other words, we have $ \psi_i (a_1 a_2 \dots a_{i - 1} a_i) = a_1 + a_2 \sigma + a_3 \sigma^2 + \dots a_{i - 1} \sigma^{i - 2} + a_i \sigma^{i - 1} $. In particular, for every $ \rho \in \Sigma_i $ we have $ \psi_i (\rho) \in \{0, 1, \dots, \sigma^i - 1 \} $. Note that $ \psi_i $ is \emph{monotone}: for every $ \rho, \rho' \in \Sigma^i $ we have $ \rho \prec \rho' $ if and only if $ \psi(\rho) < \psi(\rho') $. For every $ 1 \le i \le r $, define $ \Sigma^*_i = \{\psi_i (\rho) \;|\; \rho \in \Sigma_i \} $. Note that $ |\Sigma^*_i| = |\Sigma_i| = \sigma_i $ for every $ 1 \le i \le r $.

    We now describe our encoding. Let us store the following data structures.
    
    \begin{itemize}
    \item For every $ 1 \le i \le r $, the data structure of Lemma~\ref{lem:dictionary} for the set $ \Sigma^*_i \subseteq \{0, 1, \dots, \sigma^i - 1 \} $. We know that $ |\Sigma^*_i| = \sigma_i $, so the total number of required bits is $\sum_{i=1}^r (\sigma_i \log(\sigma^i/\sigma_i) + O(\sigma_i)) \leq \sum_{i=1}^r (e_i \log(\sigma^i /\sigma_i) + O(e_i)) = \sum_{i=1}^r (e_i \log \sigma^i) - \sum_{i=1}^r (e_i \log \sigma_i) + \sum_{i=1}^r O(e_i) = (\sum_{i=1}^r e_i i) \log \sigma - (\sum_{i=1}^r e_i \log \sigma_i) + O(e) = \mathfrak{e} \log \sigma - (\sum_{i = 1}^r e_i \log \sigma_i) + O(e) $. We can solve rank and select queries (and membership, predecessor and successor queries, see Remark~\ref{rem:membershipquery}) on each $ \Sigma^*_i $ in $ O(\log \log (\sigma^i / \sigma_i)) \subseteq O(\log \log \sigma^r ) \subseteq (\log r + \log \log \sigma) \subseteq O(\log \log \sigma) $ time.

    \item For every $ 1 \le i \le r $, the data structure of Lemma~\ref{lem:rankselectbasic} for the bit string $ \mathtt{OUT}_i \in \{0, 1\}^{e_i + n} $ of Definition~\ref{def:BWTwheelerdfa}. The total number of required bits is $ \sum_{i = 1}^r ((e_i + n) (1+o(1)) + O(e_i + n)) \subseteq \sum_{i = 1}^r O(e_i + n) = O(e + nr) \subseteq O(e) $. We can solve access, rank and select queries on each $ \mathtt{OUT}_i $ in $ O(1) $ time.

    \item For every $ 1 \le i \le r $, the data structure of Lemma~\ref{lem:rankselectbasic} for the bit string $ \mathtt{IN}_i \in \{0, 1\}^{e_i + n} $ of Definition~\ref{def:BWTwheelerdfa}. The total number of required bits is again $ O(e) $. We can solve access, rank and select queries on each $ \mathtt{IN}_i $ in $ O(1) $ time.

    \item For every $ 1 \le i \le r $, the data structure of Lemma~\ref{lem:rankselectbasic} for the string $ \mathtt{LAB}^*_i \in (\{0, 1, \dots, \sigma_i - 1 \})^{e_i} $ defined as follows. Consider the string $ \mathtt{LAB}_i \in (\Sigma_i)^{e_i} $ of Definition~\ref{def:BWTwheelerdfa}, and let $ \mathtt{LAB}^*_i $ be the string of length $ e_i $ such that $ \mathtt{LAB}^*_i[j] = \Sigma_i^*.rank(\psi_i (\mathtt{LAB}_i[j])) - 1 $ for every $ 1 \le j \le e_i $. In other words, (i) we compute $ \psi_i (\mathtt{LAB}_i[j]) \in \{0, 1, \dots, \sigma^i - 1 \} $ and then (ii) we compute the position of $ \psi_i (\mathtt{LAB}_i[j]) $ in the sorted list of all elements in $ \Sigma^*_i $. Note that $ \psi_i (\mathtt{LAB}_i[j]) \in \Sigma^*_i $ because $ \mathtt{LAB}_i[j] \in \Sigma_i $, hence $ 1 \le \Sigma_i^*.rank(\psi_i (\mathtt{LAB}_i[j])) \le \sigma_i $, which implies $ \mathtt{LAB}^*_i \in (\{0, 1, \dots, \sigma_i - 1 \})^{e_i} $. Moreover, $ \sigma_i \le e_i $, so the assumption required in Lemma~\ref{lem:rankselectbasic} is satisfied. Notice that for every $ 1 \le j, j' \le e_i $ we have $ \mathtt{LAB}_i^*[j] = \mathtt{LAB}_i^*[j'] $ if and only if $ \mathtt{LAB}_i[j] = \mathtt{LAB}_i[j'] $ (because $ \psi_i $ is a bijection), and we have $ \mathtt{LAB}_i^*[j] < \mathtt{LAB}_i^*[j'] $ if and only if $ \mathtt{LAB}_i[j] \prec \mathtt{LAB}_i[j'] $ (because $ \psi_i $ is monotone). The total number of required bits is $ \sum_{i = 1}^r (e_i \log \sigma_i (1 + o(1)) + O(e_i)) \le (\sum_{i = 1}^r e_i \log \sigma_i) + (\sum_{i = 1}^r e_i \log \sigma^i) \cdot o(1) + O(e) = (\sum_{i = 1}^r e_i \log \sigma_i) + (\sum_{i = 1}^r e_i i) \log \sigma \cdot o(1) + O(e) = (\sum_{i = 1}^r e_i \log \sigma_i) + \mathfrak{e} \log \sigma \cdot o(1) + O(e) $. We can solve access, rank and select queries on each $ \mathtt{LAB}^*_i $ in $ O(\log \log \sigma_i) \subseteq O(\log \log \sigma^i) \subseteq O(\log \log \sigma^r) \subseteq O(\log r + \log \log \sigma) = O(\log \log \sigma) $ time.

    \item For every $ 1 \le i \le r $, the data structure of Lemma~\ref{lem:rankselectbasic} for the (auxiliary) bit string $\mathtt{AUX}^*_i \in \{0, 1 \}^{e_i} $ defined as follows. Sort all edges in $ E_i $ by the index of the end states (w.r.t. $ \preceq_\mathcal{A} $). Edges with the same end state are sorted by label. Edges with the same end state and the same label are sorted by the index of the start states (w.r.t. $ \preceq_\mathcal{A} $). Then, we obtain $ \mathtt{AUX}_i \in (\Sigma_i)^{e_i} $ by concatenating the labels of all edges following this edge order. Note that by Lemma~\ref{lem:local GDFA}, if $ 1 \le k \le e_i - 1 $, then $ \mathtt{AUX}_i[k] \preceq \mathtt{AUX}_i[k + 1] $ (all edge labels in $ E_i $ have length $ i $, so no edge label is a strict suffix of some other edge label). The string $\mathtt{AUX}^*_i \in \{0, 1 \}^{e_i} $ is the string on $ \{0, 1 \} $ such that, for every $ 1 \le k \le e_i $, we have $\mathtt{AUX}^*[k] = 1 $ if and only if $ k = 1 $ or $ (k \ge 2) \land (\mathtt{AUX}_i[k] \neq \mathtt{AUX}_i[k-1]) $. The total number of required bits is $ \sum_{i = 1}^r (e_i (1 + o(1)) + O(e_i)) \subseteq \sum_{i = 1}^r O(e_i) = O(e) $. We can solve access, rank and select queries on each $ \mathtt{AUX}^*_i $ in $ O(1) $ time.  
    
    \item The data structure of Lemma~\ref{lem:rankselectbasic} for the bit string $ \texttt{FIN} \in \{0, 1\}^{n} $ of Definition~\ref{def:BWTwheelerdfa}. The number of required bits is  $ n (1+o(1)) + O(n) \subseteq O(n) \subseteq O(e) $. We can solve access, rank and select queries on $ \mathtt{FIN} $ in $ O(1) $ time.
\end{itemize}

    By adding up the space required of all data structures, we conclude that the total space is $ \mathfrak{e} \log\sigma(1+o(1)) + O(e) $ bits.
    
Let us show that our data structures are an encoding of $ \mathcal{A} $. By Theorem~\ref{theor:bwtGDFAwheeler}, we only need to show that our data structures are an encoding of $ \BWT (\mathcal{A}) $. By Definition~\ref{def:BWTwheelerdfa}, we need to show that our data structures are an encoding of (i) $ \mathtt{OUT}_i $, $ \mathtt{IN}_i $ and $ \mathtt{LAB}_i $, for every $ 1 \le i \le r $, and (ii) $ \mathtt{FIN} $. By Lemma~\ref{lem:rankselectbasic} and Lemma~\ref{lem:dictionary}, we know that our data structures are an encoding of (i) $ \Sigma_i^* $, $ \mathtt{OUT}_i $, $ \mathtt{IN}_i $, $ \mathtt{LAB}^*_i $ and $ \mathtt{AUX}^*_i $, for every $ 1 \le i \le r $, and (ii) $ \mathtt{FIN} $. The conclusion will follow if we show that, for every $ 1 \le i \le r $, $ \Sigma_i^* $ and $ \mathtt{LAB}^*_i $ are an encoding of $ \mathtt{LAB}_i $. Fix $ 1 \le i \le r $ and $ 1 \le j \le e_i $; it will suffice to show that we can retrieve $ \mathtt{LAB}_i[j] $ from $ \Sigma_i^* $ and $ \mathtt{LAB}^*_i[j] $. Notice that $ \psi_i (\mathtt{LAB}_i[j]) = \Sigma_i^*.select(\mathtt{LAB}^*_i[j] + 1) $, so we can retrieve $ \mathtt{LAB}_i[j] $ because $ \psi_i $ is a bijection from $ \Sigma^i $ to $ \{0, 1, \dots, \sigma^i - 1 \} $.

\subsubsection{Step 2: auxiliary operations}

Let us define four operations.

\begin{itemize}
    \item $ \mathcal{A}.op_1 (i, x, j) $, for $ 1 \le i \le r $, $ 0 \le x \le \sigma^i - 1 $ and $ 1 \le j \le n $: return $ \mathtt{out}(Q[1, j], \psi_i^{-1}(x)) $.
    \item $ \mathcal{A}.op_2 (i, x, h) $, for $ 1 \le i \le r $, $ 0 \le x \le \sigma^i - 1 $ and $ h \ge 0 $: return the largest $ 0 \le j \le n $ such that $ \mathtt{in}(Q[1, j], \psi_i^{-1}(x)) \le h $.
    \item $ \mathcal{A}.op_3 (i, x, h) $, for $ 1 \le i \le r $, $ 0 \le x \le \sigma^i - 1 $ and $ h \ge 1 $: return the smallest $ 0 \le j \le n $ such that $ \mathtt{in}(Q[1, j], \psi_i^{-1}(x)) \ge h $, or report that such a $ j $ does not exist.
    \item $ \mathcal{A}.op_4 (i, x) $, for $ 1 \le i \le r $ and $ 0 \le x \le \sigma^i - 1 $: return the largest integer $ 0 \le j \le n $ such that, for every $ 1 \le t \le j $ and for every $ \rho \in \lambda (Q[t]) \cap \Sigma^i $, we have $ \rho \preceq \psi_i^{-1}(x) $.
\end{itemize}

By reading the statements of Lemma~\ref{lem:computationalversion} and Lemma~\ref{lem:Ri alpha2}, it is easy to guess how we plan to use these four operations: we will use the first, the second and the fourth operation to implement the procedure described in Lemma~\ref{lem:computationalversion}, and we will use the first and the third operation to implement the procedure described in Lemma~\ref{lem:Ri alpha2}. Intuitively, we can solve these operations efficiently by querying the compressed data structures defined in Step 1 via rank and select queries (or related queries). In all the four queries, $ x $ is an integer between $ 0 $ and $ \sigma^i - 1 $ that corresponds to a string in $ \Sigma^i $ (namely, $ \psi_i^{-1}(x) $).

Let us show that each operation can be solved in $ O(\log \log \sigma) $ time.

\begin{itemize}
    \item $ \mathcal{A}.op_1 (i, x, j) $, for $ 1 \le i \le r $, $ 0 \le x \le \sigma^i - 1 $ and $ 1 \le j \le n $: return $ \mathtt{out}(Q[1, j], \psi_i^{-1}(x)) $. We have $ \psi_i^{-1}(x) \in \Sigma^i $. We first check whether $ \psi_i^{-1}(x) \in \Sigma_i $. We have $ \psi_i^{-1}(x) \in \Sigma_i $ if and only if $ x \in \Sigma_i^* $, so we only need to solve the query $ \Sigma_i^*.memb(x) $ in $ O(\log \log \sigma) $ time. If $ \psi_i^{-1}(x) \not \in \Sigma_i $, then $ \mathtt{out}(Q[1, j], \psi_i^{-1}(x)) = 0 $. Now assume that $ \psi_i^{-1}(x) \in \Sigma_i $. All occurrences of $ \psi_i^{-1}(x) $ in $ \mathtt{LAB}_i $ have been replaced with $ x' = \Sigma_i^*.rank(\psi_i(\psi_i^{-1}(x))) - 1 = \Sigma_i^*.rank(x) - 1 $ in $ \mathtt{LAB}_i^* $, and we can compute $ x' $ in $ O(\log \log \sigma) $ time. By the definition of $ \mathtt{OUT}_i $, the number of edges in $ E_i $ leaving a state in $ Q[1, j] $ is given by $ d = \mathtt{OUT}_i.rank(\mathtt{OUT}_i.select(j, 1), 0) $, which can be computed in $ O(1) $ time. As a consequence, in $ O(\log \log \sigma) $ time we can compute $ \mathtt{out}(Q[1, j], \psi_i^{-1}(x)) $ because by the definitions of $ \mathtt{LAB} $ and $ \mathtt{LAB}^* $ we have $ \mathtt{out}(Q[1, j], \psi_i^{-1}(x)) = \mathtt{LAB}_i^*.rank(d, x') $.
    
    \item $ \mathcal{A}.op_2 (i, x, h) $, for $ 1 \le i \le r $, $ 0 \le x \le \sigma^i - 1 $ and $ h \ge 0 $: return the largest $ 0 \le j \le n $ such that $ \mathtt{in}(Q[1, j], \psi_i^{-1}(x)) \le h $. We have $ \psi_i^{-1}(x) \in \Sigma^i $. In $ O(\log \log \sigma) $ time, we check whether $ \psi_i^{-1}(x) \in \Sigma_i $ by proceedings as in the previous point. If $ \psi_i^{-1}(x) \not \in \Sigma_i $, then we conclude $ j = n $. Now assume that $ \psi_i^{-1}(x) \in \Sigma_i $. All occurrences of $ \psi_i^{-1}(x) $ in $ \mathtt{LAB}_i $ have been replaced with $ x' = \Sigma_i^*.rank(\psi_i(\psi_i^{-1}(x))) - 1 = \Sigma_i^*.rank(x) - 1 $ in $ \mathtt{LAB}_i^* $, and we can compute $ x' $ in $ O(\log \log \sigma) $ time. Since $ \psi_i $ is a bijection, the total number of edges in $ \mathcal{A} $ labeled $ \psi_i^{-1}(x) $ is $ d = \mathtt{LAB}^*_i.rank(e_i, x') $. If $ d \le h $, we conclude $ j = n $. Now assume that $ d > h $. Since $ \psi_i $ is monotone, the number $ f $ of strings in $ \Sigma_i $ smaller than or equal to $ \psi_i^{-1}(x) $ can be computed in $ O(\log \log \sigma) $ time because $ f = \Sigma_i^*.rank(x) $. Now consider the list of all edges of $ \mathcal{A} $ sorted in the order used to define $ \mathtt{AUX}_i $. Then, the $ h + 1 $-th smallest edge labeled $ \psi_i^{-1}(x) $ in the list is the $ g $-th smallest edge of the list, where $ g = \mathtt{AUX}.select(f, 1) + h $, and we can compute $ g $ in $ O(1) $ time. By the definitions of $ \mathtt{AUX}_i $ and $ \mathtt{IN}_i $, this edge reaches state $ Q[y] $, where $ y = \mathtt{IN}_i.rank(\mathtt{IN}_i.select(g, 0), 1) + 1 $, and we can compute $ y $ in $ O(1) $ time. Hence, we conclude that the largest $ 0 \le j \le n $ such that $ \mathtt{in}(Q[1, j], \psi_i^{-1}(x)) \le h $ is $ y - 1 $.

    \item $ \mathcal{A}.op_3 (i, x, h) $, for $ 1 \le i \le r $, $ 0 \le x \le \sigma^i - 1 $ and $ h \ge 1 $: return the smallest $ 0 \le j \le n $ such that $ \mathtt{in}(Q[1, j], \psi_i^{-1}(x)) \ge h $, or report that such a $ j $ does not exist. Let $ j' $ be the largest integer for which $ 0 \le j' \le n $ and $ \mathtt{in}(Q[1, j'], \psi_i^{-1}(x)) \le h - 1 $. We can compute $ j' $ in $ O(\log \log \sigma) $ time by using $ \mathcal{A}.op_2 (i, x, h - 1) $. If $ j' < n $, then the smallest $ 0 \le j \le n $ such that $ \mathtt{in}(Q[1, j], \psi_i^{-1}(x)) \ge h $ is $ j' + 1 $ by the maximality of $ j' $, and if $ j' = n $, then such a $ j $ does not exist.

    \item $ \mathcal{A}.op_4 (i, x) $, for $ 1 \le i \le r $ and $ 0 \le x \le \sigma^i - 1 $: return the largest integer $ 0 \le j \le n $ such that, for every $ 1 \le t \le j $ and for every $ \rho \in \lambda (Q[t]) \cap \Sigma^i $, we have $ \rho \preceq \psi_i^{-1}(x) $. We compute $ \Sigma_i^*.succ(x) $ in $ O(\log \log \sigma) $ time. If $ \Sigma_i^*.succ(x) = \bot $, then for every $ y \in \Sigma^*_i $ we have $ y \le x $. Hence, for every $ \rho \in \Sigma_i $ we have $ \rho \preceq \psi_i^{-1}(x) $ (because $ \psi_i $ is monotone and $ \psi_i (\rho) \le x $). This implies that $ j = n $. Now assume that $ \Sigma_i^*.succ(x) \not = \bot $, which implies that $ \psi_i^{-1}(\Sigma_i^*.succ(x)) \in \Sigma_i $. Note that for every $ \rho \in \Sigma_i $ we have $ \rho \prec \psi_i^{-1}(\Sigma_i^*.succ(x)) $ if and only if $ \psi_i (\rho) < \Sigma_i^*.succ(x) $, if and only if $ \psi_i (\rho) \le x $, if and only if $ \rho \preceq \psi_i^{-1}(x) $. This means that we only have to compute the largest integer $ 0 \le j \le n $ such that, for every $ 1 \le t \le j $ and for every $ \rho \in \lambda(Q[t]) \cap \Sigma^i $, we have $ \rho \prec \psi_i^{-1}(\Sigma_i^*.succ(x)) $. Since $ \psi_i^{-1}(\Sigma_i^*.succ(x)) \in \Sigma_i $, by Lemma~\ref{lem:local GDFA} we conclude that we only have to compute the largest integer $ 0 \le j \le n $ such that $ \mathtt{in}(Q[1, j], \psi_i^{-1}(\Sigma_i^*.succ(x))) \le 0 $, so we only need to compute $ \mathcal{A}.op_2(i, \Sigma_i^*.succ(x), 0) $ in $ O(\log \log \sigma) $ time.
\end{itemize}

\subsubsection{Step 3: solving the SMLG problem} Let us show how to solve the SMLG problem. To handle some border cases, it is expedient to first compute an auxiliary integer $ \mathfrak{q} $. Recall that $ 0 $ is the smallest character in $ \Sigma $ (w.r.t $ \preceq $). Let $ \mathfrak{q} $ be the largest integer $ 0 \le k \le m $ such that $ p_k(\alpha) = 00 \dots 0 = 0^k $. We can compute $ \mathfrak{q} $ in $ O(m) $ time by scanning $ \alpha $ from left to right.

Recall that every string of length $ i $ corresponds to an integer between $ 0 $ and $ \sigma^i - 1 $ via the bijection $ \psi_i $. To solve the SMLG problem, we will often need to map all the suffixes of a given string into the corresponding integers through a procedure that we call \emph{suffix-mapping}. In detail, consider a string $ \rho $ of length at most $ r $. We define suffix-mapping $ \rho $ as the process of computing $ \psi_i(s_i(\rho)) $ for every $ 1 \le i \le |\rho| $. For example, if $ \sigma = 4 $ and $ \rho = 321 $, then suffix-mapping $ \rho $ means computing $ \psi_1 (1) = 1 $, $ \psi_2 (21) = 6 $ and $ \psi_3 (321) = 27 $. We can suffix-map $ \rho $ in $ O(|\rho|) \subseteq O(r) \subseteq O(1) $ time because (i) if $ |\rho| \ge 1 $, then $ \psi(s_1(\rho)) = s_1(\rho) $ and (ii) for $ 2 \le i \le |\rho| $ we have $ \psi_i(s_i(\rho)) = \rho[|\rho|- i + 1] + \psi_{i - 1}(s_{i - 1}(\rho)) \cdot \sigma $. Indeed, we have $ \psi_i(s_i(\rho)) = \rho[|\rho|- i + 1] + \rho[|\rho|- i + 2] \sigma + \rho[|\rho|- i + 3] \sigma^2 + \dots + \rho[|\rho|] \sigma^{i - 1} = \rho[|\rho|- i + 1] + (\rho[|\rho|- i + 2] + \rho[|\rho|- i + 3] \sigma + \dots + \rho[|\rho|] \sigma^{i - 2}) \cdot \sigma = \rho[|\rho|- i + 1] + \psi_{i - 1}(s_{i - 1}(\rho)) \cdot \sigma $.

We are now ready to show how to solve the SMLG problem. In other words, we can show how to compute $ G_\dashv(\alpha) $ in $ O(m \log \log \sigma) $ time. By Lemma~\ref{lem:st}, to compute $ G_\dashv(\alpha) $, we only need to compute $ |G^\prec(\alpha)| $ and $ |G^\prec_\dashv(\alpha)| $. To this end, in $ m $ steps, we will recursively compute $ |G^\prec (p_k(\alpha))| $ and $ |G^\prec_\dashv (p_k(\alpha))| $ for every $ 0 \le k \le m $, and we will obtain the conclusion by picking $ k = m $. Note that the case $ k = 0 $ is immediate because $ p_0 (\alpha) = \epsilon $, and $ |G^\prec(\epsilon)| = 0 $ and $ |G^\prec_\dashv(\epsilon)| = n $ (see Remark~\ref{rem:caseepsilonobvious}). To obtain the time bound $ O(m \log \log \sigma) $, we only need to show the following. Fix $ 1 \le k \le m $, and assume that we know $ |G^\prec (p_i(\alpha))| $ and $ |G^\prec_\dashv (p_i(\alpha))| $ for every $ 0 \le i \le k - 1 $. Then, we must prove that in $ O(\log \log \sigma) $ time we can compute $ |G^\prec (p_k(\alpha))| $ and $ |G^\prec_\dashv (p_k(\alpha))| $.

\begin{itemize}
\item Let us show how to compute $ |G^\prec (p_k(\alpha))| $ in $ O(\log \log \sigma) $ time. Let $ j_1 $ be the largest integer $ 0 \le j \le n $ such that, for every $ 1 \le t \le j $ and for every $ \rho \in \lambda (Q[t]) $, we have $ \rho \prec p_k(\alpha) $. Let $ j_2 $ be the largest integer $ 0 \le j \le n $ such that, for every $ 1 \le i \le \min \{r, k - 1 \} $, we have $ \mathtt{in}(Q[1, j], \alpha[k - i + 1, k]) \leq f_i $, where $ f_i = \mathtt{out}(Q[1, |G^\prec(p_{k - i}(\alpha))|], \alpha[k - i + 1, k]) $. By Lemma~\ref{lem:computationalversion} we have $ |G^\prec (p_k(\alpha))| = \min \{j_1, j_2 \} $, so we only need to show how to compute $ j_1 $ and $ j_2 $ in $ O(\log \log \sigma) $ time.

\begin{itemize}

\item Let us show how to compute $ j_1 $ in $ O(\log \log \sigma) $ time. For every $ 1 \le i \le r $, let $ j_{1, i} $ be the largest integer $ 0 \le j \le n $ such that, for every $ 1 \le t \le j $ and for every $ \rho \in \lambda (Q[t]) \cap \Sigma^i $, we have $ \rho \prec p_k(\alpha) $. Then, we have $ j_1 = \min \{j_{1, 1}, j_{1, 2}, \dots, j_{1, r} \} $. Hence, we only have to show that, for every $ 1 \le i \le r $, we can compute $ j_{1, i} $ in $ O(\log \log \sigma) $ time, because then we can compute $ j_1 $ in $ O(r \log \log \sigma) \subseteq O(\log \log \sigma) $ time.

We distinguish two cases: $ \mathfrak{q} \ge k $ and $ \mathfrak{q} < k $. First, assume that $ \mathfrak{q} \ge k $. This means that $ p_k(\alpha) = 0^k $. Fix $ 1 \le i \le r $. We distinguish two subcases.
\begin{itemize}
    \item Assume that $ 1 \le i \le \min \{r, k - 1 \} $. Then, $ j_{1, i} $ is the largest integer $ 0 \le j \le n $ such that, for every $ 1 \le t \le j $ and for every $ \rho \in \lambda (Q[t]) \cap \Sigma^i $, we have $ \rho \preceq 0^i $. Since $ \psi_i (0^i) = 0 $, we compute $ j_{1, i} = \mathcal{A}.op_4 (i, 0) $ in $ O(\log \log \sigma) $ time.
    \item Assume that $ k \le i \le r $. Then, $ j_{1, i} $ is the largest integer $ 0 \le j \le n $ such that, for every $ 1 \le t \le j $, there exists no $ \rho \in \lambda (Q[t]) \cap \Sigma^i $. By the definition of $ \mathtt{IN}_i $, we compute $ j_{1, i} = \mathtt{IN}_i.rank (\mathtt{IN}_i.select(1, 0), 1) $ in $ O(\log \log \sigma) $ time (note that the formula for $ j_{1, i} $ is correct even when $ e_i = 0 $ because in this case $ \mathtt{IN} = 1^n $, $ \mathtt{IN}_i.select(1, 0) = n + 1 $ and $ \mathtt{IN}_i.rank (\mathtt{IN}_i.select(1, 0), 1) = n $).
\end{itemize}

Now, assume that $ \mathfrak{q} < k $. This means that $ p_k(\alpha) \not = 0^k $, so $ p_k(\alpha) $ contains at least one character distinct from $ 0 $. For every $ 1 \le i \le r $, let $ \beta_i $ the largest string (w.r.t $ \preceq $) $ \beta \in \Sigma^i $ such that $ \beta \prec p_k(\alpha) $. Note that $ \beta_i $ is well defined because (i) $ \Sigma^i $ is a finite set and (ii) $ p_k(\alpha) \not = 0^k $ implies $ 0^i \prec p_k(\alpha) $.

Fix $ 1 \le i \le r $. Let us prove that, for every $ \rho \in \Sigma^i $, we have $ \rho \prec p_k(\alpha) $ if and only if $ \rho \preceq \beta_i $. Indeed, (i) if $ \rho \preceq \beta_i $, then $ \rho \preceq \beta_i \prec p_k(\alpha) $, and (ii) if $ \rho \prec p_k(\alpha) $, then $ \rho \preceq \beta_i $ by the maximality of $ \beta_i $. Consequently, $ j_{1, i} $ is the largest integer $ 0 \le j \le n $ such that, for every $ 1 \le t \le j $ and for every $ \rho \in \lambda (Q[t]) \cap \Sigma^i $, we have $ \rho \preceq \beta_i $. Hence, if we know $ \psi_i(\beta_i) $, then in $ O(\log \log \sigma) $ time we can compute $ j_{1, i} $ because $ j_{1, i} = \mathcal{A}.op_4(i, \psi_i(\beta_i)) $.

We are only left with showing that in $ O(1) $ time we can compute $ \psi_i(\beta_i) $ for every $ 1 \le i \le r $. Let us first determine $ \beta_i $ for every $ 1 \le i \le r $. We consider two cases:
\begin{itemize}
    \item Assume that $ 1 \le i \le \min \{r, k - 1 \} $. Then, we have $ \beta_i = \alpha[k - i + 1, k] $ because (i) $ \alpha[k - i + 1, k] \prec p_k(\alpha) $ and (ii) if $ \rho \in \Sigma^i $ satisfies $ \rho \prec p_k(\alpha) $, then $ \rho \preceq \alpha[k - i + 1, k] $. For example, if $ \sigma = 10 $ and $ p_k(\alpha) = 352 $, then $ \beta_2 = 52 $.
    \item Assume that $ k \le i \le r $. Then, we have $ \beta_i = (\sigma - 1)^{i - k + \mathfrak{q}} (\alpha[\mathfrak{q} + 1] - 1) \alpha[\mathfrak{q} + 2, k] $, where $ \alpha[\mathfrak{q} + 1] - 1 $ is the character preceding $ \alpha[\mathfrak{q} + 1] $ (we have (i) $ \mathfrak{q} < m $ because $ \mathfrak{q} < k \le m $ and (ii) $ \alpha[\mathfrak{q} + 1] \not = 0 $ by the definition of $ \mathfrak{q} $) and $ (\sigma - 1)^{i - k + \mathfrak{q}} $ is the concatenation of $ i - k + \mathfrak{q} $ occurrences of $ \sigma - 1 $, which is the largest character in $ \Sigma $. For example, if $ \sigma = 10 $ and $ p_k(\alpha) = 000752 $, then $ \beta_8 = 99999652 $. 
\end{itemize}

We conclude that in $ O(1) $ time (i) we can compute $ \psi_i(\beta_i) $ for every $ 1 \le i \le \min \{r, k - 1 \} $ by suffix-mapping the string $ \alpha[k - \min \{r, k - 1 \} + 1, k] $ and (ii) we can compute $ \psi_i(\beta_i) $ for every $ k \le i \le r $ by suffix-mapping the string $ (\sigma - 1)^{r - k + \mathfrak{q}} (\alpha[\mathfrak{q} + 1] - 1) \alpha[\mathfrak{q} + 2, k] $.

\item Let us show how to compute $ j_2 $ in $ O(\log \log \sigma) $ time. We first prove that in $ O(1) $ time we can compute $ f_i $ for every $ 1 \le i \le \min \{r, k - 1 \} $. We already know $ |G^\prec (p_i(\alpha))| $ for every $ 1 \le i \le \min \{r, k - 1 \} $. We suffix-map the string $ \alpha[k - \min \{r, k - 1 \} + 1, k] $ in $ O(1) $ time, obtaining $ \psi_i(\alpha[k - i + 1, k]) $ for every $ 1 \le i \le \min \{r, k - 1 \} $. Hence in $ O(r \log \log \sigma) \subseteq O(\log \log \sigma) $ time we compute $ f_i = \mathtt{out}(Q[1, |G^\prec(p_{k - i}(\alpha))|], \alpha[k - i + 1, k]) = \mathcal{A}.op_1(i, \psi_i(\alpha[k - i + 1, k]), |G^\prec(p_{k - i}(\alpha))|) $ for every $ 1 \le i \le \min \{r, k - 1 \} $.

For every $ 1 \le i \le \min \{r, k - 1 \} $, let $ j_{2, i} $ be the largest integer $ 0 \le j \le n $ such that $ \mathtt{in}(Q[1, j], \alpha[k - i + 1, k]) \leq f_i $. Then, we have $ j_2 = \min \{j_{2, 1}, j_{2, 2}, \dots, j_{2, \min \{r, k - 1 \} } \} $. Hence, we only have to show that, for every $ 1 \le i \le \min\{r, k - 1 \} $, we can compute $ j_{2, i} $ in $ O(\log \log \sigma) $ time, because then we can compute $ j_2 $ in $ O(r \log \log \sigma) \subseteq O(\log \log \sigma) $ time. To this end, we only need to observe that $ j_{2, i} = \mathcal{A}.op_2(i, \psi_i(\alpha[k - i + 1, k]), f_i) $.
\end{itemize}

\item Let us show how to compute $ |G^\prec_\dashv (p_k(\alpha))| $ in $ O(\log \log \sigma) $ time. Let $ j_1 $ be the largest integer $ 0 \le j \le n $ such that, if $ j \ge 1 $, then $ Q[j] \in G^*(p_k(\alpha)) $. Let $ j_2 $ be the smallest integer $ 0 \le j \le n $ such that, for every $ 1 \le i \le \min\{r, k - 1 \} $ for which $ g_i > f_i $, we have $ \mathtt{in}(Q[1, j], \alpha[k - i + 1, k]) \ge g_i $, where $ f_i = \mathtt{out}(Q[1, |G^\prec(p_{k - i}(\alpha))|], \alpha[k - i + 1, k]) $ and $ g_i = \mathtt{out}(Q[1, |G^\prec_\dashv(p_{k - i}(\alpha))|], \alpha[k - i + 1, k]) $. By Lemma~\ref{lem:Ri alpha2} we have $ |G^\prec_\dashv (p_k(\alpha))| = \max \{|G^\prec (p_k(\alpha))|, j_1, j_2 \} $, and we have already computed $ |G^\prec (p_k(\alpha))| $, so we only need to show how to compute $ j_1 $ and $ j_2 $ in $ O(\log \log \sigma) $ time.

\begin{itemize}
    \item Let us show how to compute $ j_1 $ in $ O(\log \log \sigma) $ time. If $ k > r $, we immediately conclude $ j_1 = 0 $ (see Remark~\ref{rem:suffixtrivial}), so we can assume $ k \le r $. For every $ k \le i \le r $, let $ j_{1, i} $ be the largest $ 0 \le j \le n $ such that, if $ j \ge 1 $, then there exists $ \rho \in Q[j] \cap \Sigma^i $ such that $ p_k (\alpha) \dashv \rho $. Then, we have $ j_1 = \max \{j_{1, k}, j_{1, k + 1}, \dots, j_{1, r} \} $. Hence, we only have to show that, for every $ k \le i \le r $, we can compute $ j_{1, i} $ in $ O(\log \log \sigma) $ time, because then we can compute $ j_1 $ in $ O(r \log \log \sigma) \subseteq O(\log \log \sigma) $ time.
    
    In $ O(1) $ time, we compute $ \psi_i (0^{i - k} p_k(\alpha)) $ and $ \psi_i ((\sigma - 1)^{i - k} p_k(\alpha)) $ for every $ k \le i \le r $ (recall that $ 0 $ and $ \sigma - 1 $ are the smallest and the largest character in $ \Sigma $, respectively) by suffix-mapping the strings $ 0^{r - k} p_k(\alpha) $ and $ (\sigma - 1)^{r - k} p_k(\alpha) $. Fix $ k \le i \le r $. Compute $ d_1 = \Sigma_i^*.rank(\psi_i (0^{i - k} p_k(\alpha)) - 1) $ (if $ p_k(\alpha) = 0^k $, we assume $ \Sigma_i^*.rank(\psi_i (0^{i - k} p_k(\alpha)) - 1)  = 0 $) and $ d_2 = \Sigma_i^*.rank(\psi_i ((\sigma - 1)^{i - k} p_k(\alpha))) $ in $ O(\log \log \sigma) $ time. Since $ \psi_i $ is monotone, we have $ d_1 \le d_2 $. Moreover, we have $ d_1 = d_2 $ if and only if $ j_{1, i} = 0 $, so in the rest of the proof we can assume $ d_1 < d_2 $ (and so $ j_{1, i} \ge 1 $).

    Since $ j_{1, i} \ge 1 $ and $ \Sigma^i $ is finite, we can consider the largest string $ \rho^*_i $ (w.r.t. $ \preceq $) in $ \Sigma_i $ suffixed by $ p_k(\alpha) $. By Lemma~\ref{lem:local GDFA}, $ j_{1, i} $ is the largest $ 0 \le j \le n $ such that $ \rho^*_i \in \lambda(Q[j]) $. Since $ \psi_i $ is monotone, the number $ f $ of strings in $ \Sigma_i $ smaller than or equal to $ \rho^*_i $ can be computed in $ O(\log \log \sigma) $ time because $ f = \Sigma_i^*.rank(d_2) $. Now consider the list of all edges of $ \mathcal{A} $ sorted in the order used to define $ \mathtt{AUX}_i $. Then, the largest edge labeled $ \rho^*_i $ in the list is the $ g $-th smallest edge of the list, where $ g = \mathtt{AUX}.rank(f + 1, 1) -1 $, and we can compute $ g $ in $ O(1) $ time. By the definition of $ \mathtt{AUX}_i $, this edge reaches state $ Q[j_{1, i}] $. By the definitions of $ \mathtt{AUX}_i $ and $ \mathtt{IN}_i $, we have $ j_{1, i} = \mathtt{IN}_i.rank(\mathtt{IN}_i.select(g, 0), 1) + 1 $, and we can compute $ j_{1, i} $ in $ O(1) $ time.
    
    \item Let us show how to compute $ j_2 $ in $ O(\log \log \sigma) $ time. We first prove that in $ O(1) $ time we can compute $ f_i $ and $ g_i $ for every $ 1 \le i \le \min \{r, k - 1 \} $. We already know $ |G^\prec (p_i(\alpha))| $ and $ |G^\prec_\dashv (p_i(\alpha))| $ for every $ 1 \le i \le \min \{r, k - 1 \} $. We suffix-map the string $ \alpha[k - \min \{r, k - 1 \} + 1, k] $ in $ O(1) $ time, obtaining $ \psi_i(\alpha[k - i + 1, k]) $ for every $ 1 \le i \le \min \{r, k - 1 \} $. Hence in $ O(r \log \log \sigma) \subseteq O(\log \log \sigma) $ time we compute $ f_i = \mathtt{out}(Q[1, |G^\prec(p_{k - i}(\alpha))|], \alpha[k - i + 1, k]) = \mathcal{A}.op_1(i, \psi_i(\alpha[k - i + 1, k]), |G^\prec(p_{k - i}(\alpha))|) $ and $ g_i = \mathtt{out}(Q[1, |G^\prec_\dashv(p_{k - i}(\alpha))|], \alpha[k - i + 1, k]) = \mathcal{A}.op_1(i, \psi_i(\alpha[k - i + 1, k]), |G^\prec_\dashv(p_{k - i}(\alpha))|) $ for every $ 1 \le i \le \min \{r, k - 1 \} $.

    For every $ 1 \le i \le \min \{r, k - 1 \} $, let $ j_{2, i} $ be the smallest $ 0 \le j \le n $ such that $ \mathtt{in}(Q[1, j], \alpha[k - i + 1, k]) \ge g_i $. Then, we have $ j_2 = \max \{j_{2, i} \;|\; 1 \le i \le \min\{r, k - 1 \}, g_i > f_i \} $. Hence, we only have to show that, for every $ 1 \le i \le \min\{r, k - 1 \} $ for which $ g_1 > f_i $ (and in particular $ g_i > 0 $), we can compute $ j_{2, i} $ in $ O(\log \log \sigma) $ time, because then we can compute $ j_2 $ in $ O(r \log \log \sigma) \subseteq O(\log \log \sigma) $ time. To this end, we only need to observe that $ j_{2, i} = \mathcal{A}.op_3(i, \psi_i(\alpha[k - i + 1, k]), g_i) $.
\end{itemize}

\end{itemize}

\subsubsection{Step 4: string acceptance} In the last step, we need to show that in $ O(m \log \log \sigma) $ time we can also decide whether $ \alpha $ is recognized by $ \mathcal{A} $. The key idea of the proof is to reduce the problem of deciding whether $ \alpha $ is recognized by $ \mathcal{A} $ to an appropriate instance of the SMLG problem. Let $ \mathcal{A}' $ be the GDFA obtained from $ \mathcal{A} $ as follows: (i) we add a new state $ s' $, (ii) we add the edge $ (s', s, \#) $, where $ \# \not \in \Sigma $ is a new character smaller than all characters in $ \Sigma $ (w.r.t. $ \preceq $) and $ s $ is the initial state of $ \mathcal{A} $, (iii) we let $ s $ (and not $ s' $) be the initial state of $ \mathcal{A} $. Note that $ s' $ has no incoming edges, so if we navigate $ \mathcal{A}' $ starting from $ s' $, after leaving $ s' $ we only move within $ \mathcal{A} $. For every $ \alpha \in \Sigma^* $ and for every state $ u $ of $ \mathcal{A} $, we have $ \alpha \in I^{\mathcal{A}}_u $ if and only if $ \#\alpha \in I^{\mathcal{A}'}_u $. Consequently, $ \mathcal{A}' $ is also a Wheeler GDFA (because for every $ \alpha, \beta \in \Sigma^* $ we have $ \alpha \prec \beta $ if and only if $ \# \alpha \prec \# \beta $), and $ \preceq_{\mathcal{A}'} $ is obtained from $ \preceq_{\mathcal{A}} $ by letting $ s' $ be the smallest state, without changing the mutual order of the remaining states. Checking whether $ \alpha $ is recognized by $ \mathcal{A} $ is equivalent to checking whether $ \# \alpha $ is recognized by $ \mathcal{A}' $. Since $ \mathcal{A}' $ contains exactly one edge labeled $ \# $, we can proceed as follows. We first solve the SMLG problem on $ \mathcal{A}' $ (with input $ \# \alpha $). The SMLG problem returns at most one state. If the SMLG problem returns no state, we conclude that $ \alpha $ is not recognized by $ \mathcal{A} $. If the SMLG problem returns exactly one state $ j $, then $ \alpha $ is recognized by $ \mathcal{A} $ if and only if $ j $ is a final state of $ \mathcal{A} $, so we only need to solve $ \mathtt{FIN}.access(j) $ in $ O(1) $ time. To solve the SMLG problem on $ \mathcal{A}' $ (with input $ \# \alpha $), we first process the character $ \# $. Note that $ G^\prec (\#) = \{s' \} $ and $ G^\prec_\dashv (\#) = \{s', s \} $. After processing $ \# $, to process $ \alpha $ we only move within $ \mathcal{A} $. This means that, to solve the SMLG problem on $ \mathcal{A}' $ (with input $ \# \alpha $), we can simply solve the SMLG problem of $ \mathcal{A} $ (with input $ \alpha $), as long as we artificially replace $ |G^\prec(\epsilon)| = 0 $ and $ |G^\prec_\dashv(\epsilon)| = n $ with $ |G^\prec(\epsilon)| = 0 $ and $ |G^\prec_\dashv(\epsilon)| = 1 $. The time bound $ O(m \log \log \sigma) $ follows from the bound for the SMLG problem proved in the third step.

\subsubsection{Final remarks} The proof of Theorem~\ref{theor:fmindexwheelergdfa} is now complete. We conclude this section by briefly discussing some interesting aspects of the proof.

\begin{itemize}

\item The proof of Theorem~\ref{theor:fmindexwheelergdfa} shows that our algorithm for solving the SMLG problem and deciding whether a string is recognized by a Wheeler GDFA is \emph{online}: we iteratively solve the same problems for every prefix of the pattern $ \alpha $.

\item In Theorem~\ref{theor:fmindexwheelergdfa}, we assume $ r = O(1) $. One may wonder what happens if we drop this restriction and try to derive space and time bounds parameterized by $ r $. The proof of Theorem~\ref{theor:fmindexwheelergdfa} shows that the dependency on $ r $ is very modest. For example, we can suffix-map a string (see the third step of the proof) very efficiently in $ O(r) $ time. In particular, to obtain our bounds, we do not implicitly rely on a prohibitive (say, exponential) dependency on $ r $.

    Nonetheless, if we want to derive an explicit dependency on $ r $, we need to be careful with our model of computation. Shortly after the statement of Theorem~\ref{theor:fmindexwheelergdfa}, we noticed that, if $ r = O(1) $, then for every $ 1 \le i \le r $ the elements of $ \Sigma^i $ fit in a constant number of computer words and thus can be manipulated in constant time (as typical in the word RAM model). If $ r $ is not bounded, this is in general not true, and the time required to manipulate a string in $ \Sigma^i $ depends on how $ r $ grows with respect to the size of the input. In the worst case, $ r $ can have the same asymptotic size as the input. For example, one may consider an infinite family of Wheeler GDFAs such that (i) all Wheeler GDFAs consist of a single state labeled with a self-loop and (ii) the length $ r $ of the string labeling the self-loop can be arbitrarily large. Consequently, if we want an explicit dependency on $ r $, we should be particularly careful with the limitations of the word RAM model or we should resort to a different variant of the RAM model. The proof of Theorem~\ref{theor:fmindexwheelergdfa} is already very technical, so exploring these options goes beyond the scope of the paper.

    \item In the proof of Theorem~\ref{theor:fmindexwheelergdfa}, we interpreted each string in $ \Sigma^i $ as an integer between $ 0 $ and $ \sigma^i - 1 $ (through the mapping $ \psi_i $). A natural question is whether we can prove Theorem~\ref{theor:fmindexwheelergdfa} in an easier way: we may reduce it to the corresponding result for conventional Wheeler automata by building an appropriate Wheeler automaton starting from the Wheeler GDFA $ \mathcal{A} $.

    In some special cases, this idea appears to be promising. Let $ \mathcal{A} $ be a Wheeler GDFA, and let $ \mathfrak{R} $ be the set of all strings in $ \Sigma^+ $ labeling some edge. Assume that, for every $ \rho, \rho' \in \mathfrak{R} $, $ \rho $ is not a strict suffix of $ \rho' $ and $ \rho' $ is not a strict suffix of $ \rho $. We now consider a new alphabet $ \Sigma' $ and a new total order $ \preceq' $ (on $ \Sigma' $). Let $ \Sigma' = \mathfrak{R} $ (that is, we see every edge label as a character of the new alphabet), and for every $ \rho, \rho' \in \Sigma' $ let $ \rho \prec' \rho $ (as characters of $ \mathfrak{R} $) if and only if $ \rho \prec \rho $ (as strings over $ \Sigma $). Then, we can interpret $ \mathcal{A} $ as a \emph{conventional} DFA (over the alphabet $ \mathfrak{R} $), and it is not difficult to check that $ \mathcal{A} $ is Wheeler (with respect to $ \preceq' $).

    However, if in $ \mathfrak{R} $ there exist two strings $ \rho $ and $\rho' $ such that $ \rho $ is a strict suffix of $ \rho' $, we cannot safely decide the mutual order between $ \rho $ and $ \rho' $. For example, in Figure ~\ref{fig:conversiondoesnotwork}, we have $ 2 \prec_\mathcal{A} 3 $, so we should define $ a \prec' ca $, but we also have $ 3 \prec_\mathcal{A} 4 $, so we should define $ ca \prec' a $. Intuitively, this is the same problem that we have already highlighted when we discussed point 2 of Lemma~\ref{lem:local GDFA}.
\end{itemize}

\begin{figure}
     \centering
        \scalebox{.8}{
        \begin{tikzpicture}[->,>=stealth', semithick, auto, scale=1]
\node[state, initial, accepting] (1)    at (0,0)	{$ 1 $};
\node[state, accepting] (2)    at (4,2)	{$ 2 $};
\node[state, accepting] (3)    at (4,0)	{$ 3 $};
\node[state, accepting] (4)    at (4,-2)	{$ 4 $};
\node[state] (5)    at (2,2)	{$ 5 $};
\node[state] (6)    at (2,-2)	{$ 6 $};

\draw (1) edge [] node [] {$ b $} (5);
\draw (1) edge [] node [] {$ d $} (6);
\draw (5) edge [] node [] {$ a $} (2);
\draw (1) edge [] node [] {$ ca $} (3);
\draw (6) edge [] node [] {$ a $} (4);

\end{tikzpicture}
}

	\caption{A Wheeler GDFA $ \mathcal{A} $. The states are numbered following the total order $ \preceq_\mathcal{A} $. Note that $ a $ and $ ca $ label some edges and $ a $ is a strict suffix of $ ca $.}
 \label{fig:conversiondoesnotwork}
\end{figure}

\section{Conclusions and Future Work}

In this paper, we considered the model of generalized automata, and we introduced the set $ \mathcal{W(A)} $. We showed that $ \mathcal{W(A)} $ plays the same role as $ \Pref (\mathcal{L(A)}) $ does in conventional NFAs: the set $ \mathcal{W(A)} $ can be used to derive a Myhill-Nerode theorem, and it represents the starting point for extending the FM-index to generalized automata.

Further lines of research include extending the Burrows-Wheeler Transform and the FM-index to arbitrary GNFAs. Indeed, the Burrows-Wheeler Transform and the FM-index were recently generalized from Wheeler NFAs to arbitrary NFAs through the so-called \emph{co-lex orders}~\cite{cotumacciojacm, cotumaccio2021} and \emph{co-lex relations}~\cite{cotumaccio2022}. However, we remark that the efficient time bounds for the SMLG problem that we derived in this paper cannot hold for arbitrary GNFAs due to the (conditional) lower bounds by Equi et al. that we recalled in the introduction.

Giammarresi and Montalbano described an effective procedure for computing a state-minimal GDFA equivalent to a given GDFA~\cite{giammaresi1999journal, giammaresi1995conference}, but we do not know if there exists an efficient algorithm for minimizing a GDFA. On the one hand, the Myhill-Nerode theorem for generalized automata implies that for every state-minimal GDFA $ \mathcal{A} $ there exists a set $ \mathcal{W} \subseteq \Sigma^* $ such that $ \mathcal{A} $ is isomorphic to the minimal $ \mathcal{W} $-GDFA recognizing $ \mathcal{L}(\mathcal{A})$. On the other hand, given a $ \mathcal{W} $-GDFA recognizing $ \mathcal{L} $, it should be possible to build the minimal $ \mathcal{W} $-GDFA recognizing $ \mathcal{L} $ by extending Hopcroft's algorithm~\cite{hopcroft1971} to GDFAs. If we could prove that, for every admissible $ \mathcal{W} \subseteq \Sigma^* $, the number of states of the minimal $ \mathcal{W} $-GDFA recognizing $ \mathcal{L} $ is comparable to the number of states of a minimal GDFA recognizing $ \mathcal{L} $, then we would obtain a fast algorithm that significantly reduces the number of states of a GDFA without changing the recognized language.

As discussed in Section~\ref{sec:decidingwheelerness}, we do not know whether in Theorem~\ref{theor:polynomialdecidingwheeler} we can achieve $ O(\mathfrak{e}) $ time in the case of an integer alphabet in a polynomial range. Moreover, let us outline how to investigate the expressiveness of Wheeler GDFAs. The class of \emph{Wheeler languages} is the class of all regular languages that are recognized by some Wheeler NFA~\cite{alanko2021}. Wheeler languages enjoy several properties: for example, they admit a characterization in terms of \emph{convex} equivalence relations~\cite{alanko2021}. In addition, every Wheeler language is also recognized by some DFA, and, in particular, there exists a unique state-minimal DFA recognizing a given Wheeler language~\cite{alankominimization}. The main limitation of Wheeler languages is that they capture only a small subclass of regular languages: for example, a unary language (that is, a language over an alphabet of size one) is Wheeler if and only if it is either finite or co-finite~\cite{alanko2021}. The intuitive reason why most regular languages are not Wheeler is that, if $ \mathcal{A} $ is a Wheeler NFA, then Wheelerness induces strong constraints on the set $ \Pref(\mathcal{L(A)}) $. However, when we switch to GNFAs, the role of $ \Pref(\mathcal{L(A)}) $ is played by $ \mathcal{W(A)} $, and it may hold $ \mathcal{W(A)} \subsetneqq \Pref(\mathcal{L(A)}) $, which means that now the same constraints only apply to a smaller subset. This implies (as we have seen in Section~\ref{sec:wheelerdefinitionsubsection}) that the class of all languages recognized by some Wheeler GDFA is strictly larger than the class of Wheeler languages. We can call the languages in this new class \emph{generalized Wheeler languages}: the next step is to understand which properties of Wheeler languages are still true and how it is possible to characterize this new class (see~\cite{cotumaccio2026dlt} for some recent results).

\section*{Acknowledgment}
The author is funded by the Helsinki Institute for Information Technology (HIIT).

\bibliographystyle{alphaurl}
\bibliography{bibliography-lmcs}

\appendix

\section{The FM-index: from GDFAs to GNFAs without $ \epsilon $-transitions}\label{app:from gdfas to gnfas}

This appendix aims to show that the results of Section~\ref{sec:WheelerGDFAs} can be extended to generalized \emph{nondeterministic} automata. We will consider the case of \emph{GNFAs without $ \epsilon $-transitions} (that is, GNFAs where no edge is labeled with $ \epsilon $), and we will focus on the differences between the case of GDFAs and the case of GNFAs without $ \epsilon $-transitions. Extending our results to GNFA with $ \epsilon $-transitions (within the same space and time bounds) requires additional technical machinery that goes beyond the scope of this paper. The case of GNFAs with $ \epsilon $-transitions is discussed in detail in two separate articles~\cite{cotumaccio2026acta, cotumaccio2025iwoca}.

The first step is to extend the notion of Wheelerness to GNFAs without $ \epsilon $-transitions. Let us recall how to define Wheeler NFAs (see~\cite{alanko2020, conte2023, cotumacciospire2023}). An NFA $ \mathcal{A} = (Q, E, s, F) $ is Wheeler if there exists a total order $ \le $ on $ Q $ such that (i) $ s $ comes first in the total order, (ii) for every $ (u', u, a), (v', v, b) \in E $, if $ u < v $, then $ a \preceq b $ and (iii) for every $ (u', u, a), (v', v, a) \in E $, if $ u < v $, then $ u' \le v' $.  Alanko et al.~\cite[Lemma 2.3]{alanko2020} proved that, if $ u < v $ in the total order, then $ (\forall \alpha \in I_u)(\forall \beta \in I_v)((\{\alpha, \beta \} \not \subseteq I_u \cap I_v )\Longrightarrow (\alpha \prec \beta)) $. Let us see how to extend Definition~\ref{def:WheelerGDFA} to GNFAs without $ \epsilon $-transitions. We will first generalize the definition of $ \preceq_\mathcal{A} $ from GDFAs to GNFAs without $ \epsilon $-transitions drawing inspiration from Alanko et al.'s result.

Let $ \mathcal{A} = (Q, E, s, F) $ be a GNFA without $ \epsilon $-transitions. Let $ \preceq_\mathcal{A} $ be the relation on $ Q $ such that, for every $ u, v \in Q $, we have $ u \preceq_\mathcal{A} v $ if and only if $ (\forall \alpha \in I_u)(\forall \beta \in I_v)((\{\alpha, \beta \} \not \subseteq I_u \cap I_v )\Longrightarrow (\alpha \prec \beta)) $. If $ \mathcal{A} $ is a GDFA, then $ \preceq_\mathcal{A} $ reduces to the definition given in Section~\ref{sec:wheelerdefinitionsubsection}, because for every $ u, v \in Q $ such that $ u \not = v $ we have $ I_u \cap I_v = \emptyset $ (see Remark~\ref{rem:GDFAstringsreachonestate}). We have seen that, if $ \mathcal{A} $ is a GDFA, then $ \preceq_\mathcal{A} $ is a partial order. If $ \mathcal{A} $ is an arbitrary GNFA without $ \epsilon $-transitions, in general $ \preceq_\mathcal{A} $ is only a \emph{preorder}, that is, it is a reflexive and transitive relation, but it need not be antisymmetric, see Figure~\ref{fig:counterexamplesGNFA} (this is already true for NFAs).

We can now give the following definition.

\begin{defi}\label{def:GNFA}
    Let $ \mathcal{A} = (Q, E, s, F) $ be a GNFA without $ \epsilon $-transitions. We say that $ \mathcal{A} $ is \emph{Wheeler} if there exists a total order $ \le $ on $ Q $ such that:
    \begin{itemize}
        \item \emph{(Property 1)} For every $ u, v \in Q $, if $ u \le v $, then $ u \preceq_\mathcal{A} v $.
        \item \emph{(Property 2)} For every $ (u', u, \rho), (v', v, \rho') \in E $, if $ u < v $ and $ \rho' $ is not a strict suffix of $ \rho $, then $ \rho \preceq \rho' $.
        \item \emph{(Property 3)} For every $ (u', u, \rho), (v', v, \rho) \in E $, if $ u < v $, then $ u' \le v' $.
    \end{itemize}
    We say that $ \le $ is a \emph{Wheeler order} on $ \mathcal{A} $.
\end{defi}

Note that if $ \le $ is a Wheeler order, then $ s $ comes first by Property 1, because $ \epsilon \in I_s $ and for every $ u \in Q \setminus \{s\} $ we have $ \epsilon \not \in I_u $. As a consequence, Lemma~\ref{lem:local GDFA} is true also for GNFAs without $ \epsilon $-transitions (if in the statement of Lemma~\ref{lem:local GDFA} we replace $ \preceq_\mathcal{A} $ with a Wheeler order $ \le $).

If $ \mathcal{A} $ is an NFA, then Definition~\ref{def:GNFA} reduces to the definition of a Wheeler NFA, because by Alanko et al.'s result Property 1 follows from Property 2, Property 3, and the fact that $ s $ comes first in a Wheeler order. If $ \mathcal{A} $ is a GDFA, then Definition~\ref{def:GNFA} reduces to the definition of Wheeler GDFA (Definition~\ref{def:WheelerGDFA}) by Lemma~\ref{lem:local GDFA}. 

Let us present some preliminary remarks (see Figure~\ref{fig:counterexamplesGNFA}). In general, a Wheeler order on a GNFA without $ \epsilon $-transitions is not unique (this is already true for NFAs). Moreover, Property 2 and Property 3 in Definition~\ref{def:GNFA} do not follow from Property 1 (this is already true for NFAs). Lastly, Property 1 does not follow from Properties 2, 3 and the requirement that $ s $ must come first (while we have just seen that, if $ \mathcal{A} $ is an NFA, then Property 1 follows from Properties 2, 3 and the requirement that $ s $ must come first).

The problem of deciding whether a GDFA is Wheeler can be solved in polynomial time (see Theorem~\ref{theor:polynomialdecidingwheeler}), but it becomes NP-hard on GNFAs without $ \epsilon $-transitions, because it is already NP-hard on NFAs~\cite{gibney2022}. There are natural ways of defining a Wheeler order on a GNFA without $ \epsilon $-transitions. For example, it is easy to check that a Wheeler order on a GNFA without $ \epsilon $-transitions is induced by any Wheeler order on the equivalent NFA (without $ \epsilon $-transitions) defined in Section~\ref{sec:preliminary}.

\begin{figure}
     \centering
     \begin{subfigure}[b]{0.32\textwidth}
        \centering
        \scalebox{.7}{
        \begin{tikzpicture}[->,>=stealth', semithick, auto, scale=1]
\node[state, initial] (1)    at (0,0)	{$ u_1 $};
\node[state, accepting] (2)    at (1.5,1.5)	{$ u_2 $};
\node[state, accepting] (3)    at (1.5,-1.5)	{$ u_3 $};
\draw (1) edge [] node [] {$ a $} (2);
\draw (1) edge [] node [] {$ a $} (3);
\end{tikzpicture}
}
     \end{subfigure}
     \begin{subfigure}[b]{0.32\textwidth}
        \centering
        \scalebox{.7}{
        \begin{tikzpicture}[->,>=stealth', semithick, auto, scale=1]
\node[state, initial] (1)    at (0,0)	{$ u_1 $};
\node[state] (2)    at (1.5,1.5)	{$ u_2 $};
\node[state] (3)    at (1.5,-1.5)	{$ u_3 $};
\node[state, accepting] (4)    at (4, 1.5)	{$ u_4 $};
\node[state, accepting] (5)    at (4, -1.5)	{$ u_5 $};
\draw (1) edge [] node [] {$ a, b $} (2);
\draw (1) edge [] node [] {$ a, b $} (3);
\draw (2) edge [bend left] node [] {$ c $} (5);
\draw (3) edge [bend left] node [] {$ c $} (4);
\end{tikzpicture}
}
     \end{subfigure}
     \begin{subfigure}[b]{0.32\textwidth}
        \centering
        \scalebox{.7}{
        \begin{tikzpicture}[->,>=stealth', semithick, auto, scale=1]
\node[state, initial] (1)    at (0,0)	{$ u_1 $};
\node[state] (2)    at (1.5,1.5)	{$ u_2 $};
\node[state, accepting] (3)    at (3,0)	{$ u_3 $};
\node[state, accepting] (4)    at (1.5,-1.5)	{$ u_4 $};
\draw (1) edge [] node [] {$ b $} (2);
\draw (2) edge [] node [] {$ c $} (3);
\draw (1) edge [] node [] {$ ac $} (4);
\end{tikzpicture}
}

     \end{subfigure}     
	\caption{\emph{Left:} An NFA such that $ \preceq_\mathcal{A} $ is not antisymmetric and both total orders in which $ u_1 $ comes first are Wheeler (in particular, a Wheeler order need not be unique). \emph{Center:} The total order $ \le $ given by $ u_1 < u_2 < u_3 < u_4 < u_5 $ is such that $ \le $ satisfies Property 1 of Definition~\ref{def:GNFA}, but it does not satisfy Property 2 and Property 3. \emph{Right:} The total order $ \le $ given by $ u_1 < u_2 < u_3 < u_4 $ is such that $ \le $ satisfies Property 2 and Property 3 of Definition~\ref{def:GNFA} and the initial state comes first, but it does not satisfy Property 1.}\label{fig:counterexamplesGNFA}
\end{figure} 

We will now outline how our results can be generalized from Wheeler GDFAs to Wheeler GNFAs without $ \epsilon $-transitions (see~\cite{cotumaccio2026acta, cotumaccio2025iwoca} for more details). Lemma~\ref{lem:st} is still true if we replace $ \preceq_\mathcal{A} $ with any Wheeler order $ \le $ on the GNFA without $ \epsilon $-transitions. This follows from how $ \preceq_\mathcal{A} $ is defined on GNFAs without $ \epsilon $-transitions. For example, let us prove point 3 of Lemma~\ref{lem:st}. We know that $ u, v \in Q $ are such that $ u < v $ and $ v \in G^\prec (\alpha) $, and we must prove that $ u \in G^\prec (\alpha) $. Since $ u < v $ and $ \le $ is a Wheeler order, then $ u \preceq_\mathcal{A} v $. Let $ \beta \in I_u $; we must prove that $ \beta \prec \alpha $. If $ \beta \in I_v $, then from $ v \in G^\prec (\alpha) $ we immediately conclude $ \beta \prec \alpha $. Now assume that $ \beta \not \in I_v $, and pick any $ \gamma \in I_v $. From $ u \preceq_\mathcal{A} v $, $ \beta \in I_u \setminus I_v $ and $ \gamma \in I_v $ we obtain $ \beta \prec \gamma $. From $ \gamma \in I_v $ and $ v \in G^\prec (\alpha) $ we obtain $ \gamma \prec \alpha $, so we conclude $ \beta \prec \alpha $ and we are done. Next, one can readily check that the proofs of all remaining results in Section~\ref{sec:WheelerGDFAs} still hold true (if everywhere we replace $ \preceq_\mathcal{A} $ with a Wheeler order $ \le $) because, as we have seen, Lemma~\ref{lem:local GDFA} is also true for GNFAs without $ \epsilon $-transitions. A minor tweak is needed in the (fourth step of the) proof of Theorem~\ref{theor:fmindexwheelergdfa} to decide whether a string $ \alpha $ is recognized by a GNFA without $ \epsilon $-transitions. In the case of GDFAs, the algorithm returns at most one state $ j $, and we only need to check whether $ j $ is final by solving $ \mathtt{FIN}.access(j) $ in $ O(1) $ time. In the case of GNFAs without $ \epsilon $-transitions, the algorithm returns an interval $ Q[d_1, d_2] $, and we need to determine whether at least one state is final. To this end, we only need to check in $ O(1) $ time whether $ \mathtt{FIN}.rank(d_2, 1) > \mathtt{FIN}.rank(d_1 - 1, 1) $.

\end{document}